\documentclass[aps,12pt,notitlepage,prd,nofootinbib]{revtex4-1}
\usepackage{graphicx}
\usepackage{amsmath}
\usepackage{bbm}
\usepackage{MnSymbol}
\usepackage{wasysym}
\DeclareMathOperator{\Tr}{Tr}

\begin{document}
\title{Lattice QCD: concepts, techniques and some results
\footnote{Presented at the $54^{\text{th}}$ Cracow School of Theoretical
  Physics, Zakopane, Poland}%
}
\author{Christian Hoelbling}
\affiliation{Bergische Universit\"at Wuppertal, Gausstrasse 20, D-42119
  Wuppertal, Germany}
\begin{abstract}
I give a brief introduction to lattice QCD for non-specialists. 
\end{abstract}
  
\maketitle
\section{Introduction}

Quantum chromodynamics (QCD)~\cite{Fritzsch:1973pi} is widely
recognised as being the correct fundamental theory of the strong
nuclear interaction. Its fundamental degrees of freedom are quarks and
gluons and their interactions at high energies are well described by
perturbation theory because of asymptotic
freedom~\cite{Politzer:1973fx,Gross:1973id}. There is however a
substantial discrepancy between these fundamental degrees of freedom
and the asymptotic states of the theory, which are hadrons and their
bound states. Since hadrons are thought of as strongly coupled bound
states of quarks and gluons, it is obvious that a classical
perturbative treatment is not sufficient to describe them from first
principles.

Lattice gauge theory presents a framework in which to understand
quantitatively this strongly coupled low energy sector of the theory
from first principles. There are two main motivations for this: On the
one hand, we would like to develop a full understanding of the
dynamics of QCD itself and on the other hand, we would like to
reliably subtract QCD contributions from observables designed to probe
other fundamental physics. In these lecture notes, I will explore the
first of these two motives only. In section~\ref{sec:lqcd}, I will
introduce lattice QCD and give an overview of the techniques used in
lattice QCD calculation in section~\ref{sec:cpi}. In
section~\ref{sec:had}, I will discuss the determination of the ground
state light hadron spectrum as an example of a lattice QCD calculation
and in section~\ref{sec:therm} I will briefly introduce finite
temperature lattice QCD and highlight some important results.

I would like to emphasize that these notes are not intended to be a
complete introduction to lattice QCD in any sense. The aim is rather
to provide people working in related areas with a rough overview of
lattice techniques, what they are able to provide today and what their
limitations are. Consequently, details and proofs are often omitted
and I refer the interested reader to the introductory literature on
the subject for more in depth coverage
\cite{Creutz:1984mg,Montvay:1994cy,Gupta:1997nd,Smit:2002ug,Rothe:2005nw,DeGrand:2006zz,Fodor:2009ax,Gattringer:2010zz,Fodor:2012gf}

\section{Formulation of lattice QCD}
\label{sec:lqcd}

\subsection{Continuum QCD}

QCD is an $SU(3)$ gauge theory with fermions in the fundamental
representation. The Lagrangian of QCD is
\begin{equation}
\label{eq:lqcd}
\mathcal{L}_\text{QCD}=-\frac{1}{4}G_{\mu\nu}^aG^{a \mu\nu}
+\bar\psi
\left(
 i D_\mu\gamma^\mu-m
\right)
\psi
\end{equation}
where the field strength tensor $G_{\mu\nu}^a=\partial_\mu
A_\nu^a-\partial_\nu A_\mu^a+g f^{abc}A_\mu^b A_\nu^c$ with coupling
$g$, the structure constants $f^{abc}$ of $SU(3)$ and the covariant
derivative $D_\mu=\partial_\mu+gA_\mu^a\frac{\lambda^a}{2i}$ with
$\lambda^a$ the Gell-Mann matrices. Both $\bar\psi$ and $\psi$ carry
an implicit flavour index and $m$ is a $N\times N$ matrix in flavour
space for $N$ quark flavours.

A fundamental property of the QCD Lagrangian eq.~\ref{eq:lqcd} is its
invariance under a local $SU(3)$ symmetery
\begin{equation}
\label{eq:gt}
\begin{split}
\psi(x)&\rightarrow G(x)\psi(x)\\
\bar\psi(x)&\rightarrow\bar\psi(x) G^{\dag}(x)\\
A_\mu(x)&\rightarrow G(x)
A_\mu(x)G^{\dag}(x)-\frac{i}{g}\left(\partial_\mu G(x)\right)G^{\dag}(x)
\end{split}
\end{equation}
with $G(x)\in SU(3)$ an arbitrary local gauge transformation. Since
any physical quantity can not depend on our arbitrary choice of a
gauge, only gauge invariant quantities can be physical.

In addition to gauge symmetry, the QCD Lagrangian eq.~\ref{eq:lqcd}
has a global $U(N)$ flavour symmetry
\begin{equation}
\label{eq:ft}
\begin{split}
\psi(x)&\rightarrow e^{i \tau_a\phi}\psi(x)\\
\bar\psi(x)&\rightarrow\bar\psi(x) e^{-i \tau_a\phi} \\
\end{split}
\end{equation}
where the $\tau_a$ are the $N^2$ generators of $U(N)$. In case of
vanishing quark mass $m=0$, there is an additional chiral symmetry
\begin{equation}
\label{eq:ct}
\begin{split}
\psi(x)&\rightarrow e^{i \tau_a\gamma_5\phi}\psi(x)\\
\bar\psi(x)&\rightarrow\bar\psi(x) e^{i \tau_a\gamma_5\phi} \\
\end{split}
\end{equation}
whose diagonal part $\tau_a=\mathbbm{1}$ is anomalous
\cite{Adler:1969gk,Bell:1969ts}, leaving an $SU(N)$ symmetry intact.

We quantize QCD using the Feynman path integral formalism. We can
express the expectation value of a time ordered product of operators
as
\begin{equation}
\label{eq:pathint}
\langle 0|
T(
\hat{\mathcal{O}}_1(x_1)\ldots \hat{\mathcal{O}}_n(x_n)
)
|0
\rangle=
\frac {\int  D A_\mu D \psi D \bar\psi \hat{\mathcal{O}}_1(x_1)\ldots \hat{\mathcal{O}}_n(x_n) e^{i\hat{S}[A_\mu,\psi,\bar\psi]}}
{\int  D A_\mu D \psi D \bar\psi e^{i\hat{S}[A_\mu,\psi,\bar\psi]}}
\end{equation}
where we need to integrate over all fermion and gauge fields $\psi$,
$\bar\psi$ and $A_\mu$. The integral in eq.~\ref{eq:pathint} is not
well defined unless we specify a regulator, which we will provide by
discretizing the theory on a finite space-time lattice. Before we do
so however, we perform another step: we analytically continue the
integral in eq.~\ref{eq:pathint} to imaginary time, which is possible
as long as the Hamiltonian of the theory is bounded from below. In the
resulting Euclidean path integral
\begin{equation}
\label{eq:epathint}
\langle 0|
T(
\mathcal{O}_1(x_1)\ldots \mathcal{O}_n(x_n)
)
|0
\rangle=
\frac {\int  D A_\mu D \psi D \bar\psi \mathcal{O}_1(x_1)\ldots \mathcal{O}_n(x_n) e^{-S[A_\mu,\psi,\bar\psi]}}
{\int  D A_\mu D \psi D \bar\psi e^{-S[A_\mu,\psi,\bar\psi]}}
\end{equation}
with the Euclidian QCD action
\begin{equation}
\label{eq:leqcd}
\mathcal{S}=
\int d^4x
\left(
\frac{1}{4}G_{\mu\nu}^aG^{a \mu\nu}
+\bar\psi
\left(
 i D_\mu\gamma^\mu+m
\right)
\psi
\right)
\end{equation}
the phase factor $e^{i\hat{S}}$ is replaced by a real valued
exponential $e^{-S}$. Eq.~\ref{eq:epathint} can then be interpreted as
the expectation value of the observable with respect to the positive
definite measure $D A_\mu D \psi D \bar\psi e^{-S}$.  It is
interesting to note, that the rhs. of eq.~\ref{eq:epathint} can also
be viewed as a thermodynamic expectation value with respect to a
Boltzmann factor $e^{-S}$. It is therefore customary to call the
denominator of eq.~\ref{eq:epathint} the partition function
\begin{equation}
\label{eq:partf}
\mathcal{Z}=\int  D A_\mu D \psi D \bar\psi e^{-S[A_\mu,\psi,\bar\psi]}
\end{equation}

\subsection{Lattice regularization}
\label{sect:lr}

We now proceed to introduce a UV regularization of Euclidean QCD by
discretizing the theory on a finite space-time lattice
\cite{Wilson:1974sk}. The lattice is hypercubic with a distance $a$
between nearest neighbouring points (the lattice spacing). We also
provide an IR regularization of the theory by a finite extent of the
lattice in spatial $L=N_x a$ and temporal $T=N_t a$ directions and
impose torodial boundary conditions. As is the case for any other
regularizations, we have to remove them eventually in order to obtain
physical results. In lattice terminology, the process of removing the
UV cutoff is known as the continuum limit whereas the removal of the
IR cutoff is the infinite volume limit.

Fermion fields $\psi(x)$ of the regularized theory are defined on the
lattice sites $x$ with $x_i\in a\{0,\ldots,N_x-1\}$ and $x_4\in
a\{0,\ldots,N_t-1\}$. In order to preserve exact gauge invariance,
gauge fields are treated differently however. Instead of discretizing
the gauge potential at each lattice site $A_\mu(x)$ directly, we
instead discretize the parallel transport between any site and its
nearest neighbours. In QCD, one typically uses the group element
$U_\mu(x)$ directly, where it is understood that this represents the
continuum parallel transport
\begin{equation}
\label{eq:partrans}
U_\mu(x)=\mathcal{P}e^{ig\int_x^{x+e_{\mu}}dz_\mu A_\mu(z)}
\end{equation}
where $\mathcal{P}$ denotes the path ordered product and $e_{\mu}$
is the vector of length $a$ in $\mu$ direction. The reverse parallel
transport is then given by
\begin{equation}
\label{eq:partransdag}
U_{-\mu}(x)=\mathcal{P}e^{ig\int_x^{x-e_{\mu}}dz_\mu A_\mu(z)}=U^\dag_{\mu}(x-e_{\mu})
\end{equation}
With these definitions and the gauge transformations eq.~\ref{eq:gt},
we find that the lattice fields transform as
\begin{equation}
\label{eq:gtl}
\begin{split}
\psi(x)&\rightarrow G(x)\psi(x)\\
\bar\psi(x)&\rightarrow\bar\psi(x) G^{\dag}(x)\\
U_\mu(x)&\rightarrow G(x)
U_\mu(x)G^{\dag}(x+e_{\mu})
\end{split}
\end{equation}

\begin{figure}[htb]
\centerline{
\includegraphics[width=6cm]{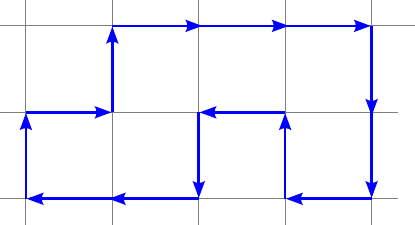}
\quad
\includegraphics[width=6cm]{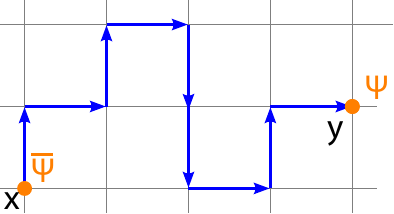}}
\caption{Two possibilities of constructing a gauge invariant object on
in a lattice gauge theory. Left: a closed loop of parallel transports
(gauge links). Right: A fermion and an antifermion connected by gauge links.}
\label{fig:giobj}
\end{figure}

In order to proceed, we need to construct gauge invariant quantities
from our lattice fields. We will need them for two distinct purposes:
First, we want to construct a lattice action and second, we need to
find gauge invariant observables. In principle, we have two choices of
constructing gauge invariant objects (see fig.~\ref{fig:giobj}).  We
can either take traces of closed loops of parallel transports (gauge
links) $\Tr(U_{\mu_1}(x)U_{\mu_2}(x+e_{\mu_1})\ldots
U^\dag_{\mu_n}(x))$ or we can take a fermion-antifermion pair that is
connected by gauge links $\bar\psi(x)U_{\mu_1}\ldots
U^\dag_{\mu_n}(y)\psi(y)$.

\begin{figure}[htb]
\centerline{
\includegraphics[height=3cm]{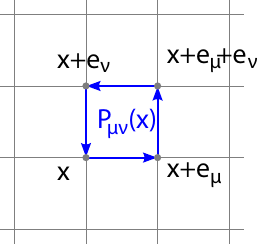}
\quad
\includegraphics[height=3cm]{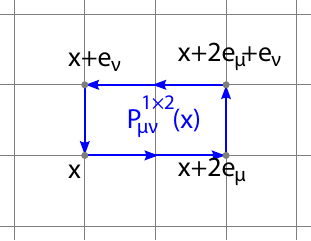}}
\caption{An elementary plaquette (left) and an extended $2\times 1$
  Wilson loop (right).}
\label{fig:plaq}
\end{figure}

The simplest\footnote{Since the lattice has a torus topology, it is
  possible to construct closed gauge loops that wind around any
  direction. This object is called Wilson line or Polyakov
  loop. Therefore it is possible on lattices with an extent of 3 or
  fewer lattice spacings in one direction to construct a gauge
  invariant object entirely consisting of even fewer gauge links}
object we can construct of gauge links alone is called the plaquette
(see fig.~\ref{fig:plaq}). It is defined as the trace of the path
ordered product of gauge links around an elementary square in the
$\mu-\nu$ plane
\begin{equation}
P_{\mu\nu}(x)=\Tr\left(U_\mu(x)U_\nu(x+e_{\mu})
U^\dag_\mu(x+e_{\nu})U^\dag_\nu(x)\right)
\end{equation}
Using eq.~\ref{eq:partrans}, we can express the plaquette in terms of
continuum gauge potentials. Taylor expanding in the lattice spacing
$a$ results in
\begin{equation}
\label{eq:plaq}
P_{\mu\nu}(x)=\Tr\left(1+iga^2G_{\mu\nu}(x)-\frac{g^2a^4}{2}G^2_{\mu\nu}(x)
\right)
+\mathcal{O}(g^6)
\end{equation}
We can use this result to construct a gauge action
\begin{equation}
\label{eq:wga}
S_G=\beta\sum_{x,\mu>\nu}
\left(
1-\frac{1}{6}\left(
P_{\mu\nu}(x)
+P^\dag_{\mu\nu}(x)
\right)
\right)
\end{equation}
with $\beta=6/g^2$ that has the correct form in the continuum limit
\begin{equation}
S_G\stackrel{a\rightarrow 0}{\longrightarrow}
\frac{1}{4}\int d^4x G^a_{\mu\nu}(x) G^a_{\mu\nu}(x)
\end{equation}

The gauge action eq.~\ref{eq:wga} is known as the Wilson plaquette
action. It has the correct continuum limit with leading corrections of
$\mathcal{O}(a^2)$. It is not the only possible discretization of the continuum
gauge action eq.~\ref{eq:leqcd} though. One could equally well take
e.g. a trace $W^{2\times 1}_{\mu\nu}$ over a closed loop of gauge
links around a $2\times 1$ rectangle (see fig.~\ref{fig:plaq}). This
object, known as the Wilson loop of size $2\times 1$, has the same
leading continuum behaviour as the Plaquette up to a trivial numerical
factor. One could therefore in principle use it instead of the
plaquette for defining a lattice gauge action, which however is not
particularly useful. What is useful however is taking a linear
combination of the elementary plaquette and the $2\times 1$ Wilson
loop \cite{Curci:1983an,Luscher:1984xn,Luscher:1984xne}. Choosing the
relative weights such that the leading order term in the continuum
limit remains unchanged, while the leading corrections of $\mathcal{O}(a^2)$
(which have the same form for both terms) cancel, we obtain an action
that has leading corrections of $\mathcal{O}(a^4)$ only. Classically, these
coefficients are easy to find. They can be read off from a Taylor
expansion of the lattice operators in terms of continuum
operators. The resulting action
\begin{equation}
\label{eq:tllwga}
S_G=\beta\sum_{x,\mu>\nu}
\left(
1-\frac{1}{6}\left(
\frac{5}{3}
P_{\mu\nu}(x)
-
\frac{1}{12}
W^{2\times 1}_{\mu\nu}(x)
\right)
+h.c.
\right)
\end{equation}
is known as the tree-level L\"uscher Weisz action. In a quantum
theory, there are radiative corrections and one can in principle
determine the relative weights by either computing them in
perturbation theory or finding them nonperturbatively
\cite{Luscher:1985zq,Iwasaki:1983ck,Takaishi:1996xj,deForcrand:1999bi}.

This construction of a gauge action that has higher order cutoff terms
is a special case of the Symanzik improvement program
\cite{Symanzik:1983dc,Symanzik:1983gh}. Generically, the idea behind
it is that the lattice theory, as an effective theory with a finite
cutoff, may contain continuum irrelevant, nonrenormalizable terms
without altering the continuum limit.  One can thus perform an
expansion of the continuum action in terms of lattice operators. The
nontrivial part of this expansion are the kinetic terms where
continuum derivative operators are expanded in discrete difference
operators. As an illustrative example, let us consider the classical
expansion of the simple derivative operator
\begin{equation}
\frac{d}{dx}f(x)=f^\prime(x)
\end{equation}
On a discrete set of points with uniform spacing $a$, we can define a
sequence of finite difference operators
\begin{equation}
\Delta_n f(x):=\frac{f(x+na)-f(x-na)}{2na}
\end{equation}
Taylor expanding this expression around $x$ one obtains
\begin{equation}
\Delta_n f(x)=\sum_{i=0}^\infty
\frac{(na)^{2i}}{(2i+1)!}f^{(2i+1)}(x)=
f^\prime(x)+\frac{1}{6}(na)^2f^{\prime\prime\prime}(x)+\mathcal{O}(a^4)
\end{equation}
and thus the finite difference operator
\begin{equation}
\label{eq:derimp}
\Delta f(x):=\left(\frac{4}{3}\Delta_1-\frac{1}{3}\Delta_2\right)f(x)
\end{equation}
has classical discretization errors
\begin{equation}
\Delta f(x)=f^\prime(x)+\mathcal{O}(a^4)
\end{equation}

\subsection{Fermion discretization}
\label{sect:ferm}

In the Euclidian continuum theory, the free fermion action reads
\begin{equation}
S_F=\int d^4x \bar\psi(x)\left(\gamma_\mu\partial_\mu+m\right)\psi(x)
\end{equation}
The most straightforward discretization of this action is
\begin{equation}
\label{eq:naiv}
S^N_F=a^4 \sum_x \bar\psi(x)\left(\gamma_\mu\Delta_\mu+m\right)\psi(x)
\end{equation}
with the simple difference operator
\begin{equation}
\Delta_\mu f(x):=\frac{f(x+e_\mu)-f(x-e_\mu)}{2a}
\end{equation}
We can diagonalize this operator in Fourier space. The
resulting inverse propagator has the form
\begin{equation}
\label{eq:kernel}
G_N^{-1}(p)=i\gamma_\mu\frac{\sin ap_\mu}{a}+m
\end{equation}
Performing the continuum limit $a\rightarrow 0$ for a {\it fixed
  physical momentum} $p_\mu$, we recover the continuum inverse
propagator
\begin{equation}
\label{eq:naii}
G_N^{-1}(p)\stackrel{a\rightarrow 0}{\longrightarrow}i\gamma_\mu p_\mu+m
\end{equation}
which has the correct physical poles at $p^2=-m^2$. Lattice
periodicity requires that these poles are repeated for
$p_\mu\rightarrow p_\mu+2\pi/a$, but eq.~\ref{eq:naii} has additional
poles within the Brillouin zone for $p_\mu\rightarrow p_\mu+\pi/a$. In
addition to the physical pole, there are $2^D-1$ of these doubler
fermion poles within the Brillouin zone, so in $4D$ the naive fermion
action eq.~\ref{eq:naiv} does in fact describe 16 species of fermions
instead of one.

\begin{figure}[htb]
\centerline{
\includegraphics[width=12cm]{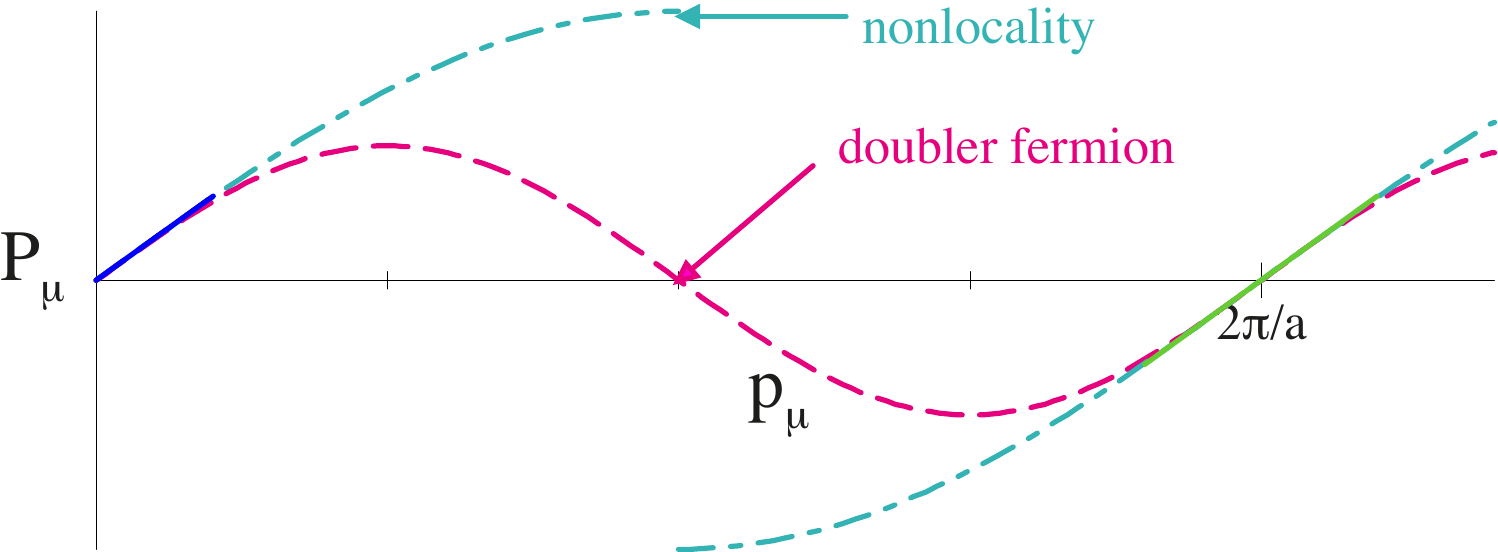}
}
\caption{Illustration of the possible behaviours of $P_\mu(ap_\mu)$.}
\label{fig:nn}
\end{figure}
 
This feature is known as the fermion doubling problem. It is not
specific to the naive fermion action, as can be seen from the
following heuristic argument. Let us try to generalize the action and
therefore the inverse propagator eq.~\ref{eq:kernel}. We may replace
$\sin(ap_\mu)/a$ with a generic function $P_\mu(ap_\mu)/a$. Around
$ap_\mu=0$, $P_\mu(ap_\mu)=ap_\mu+\mathcal{O}(a^2)$ is dictated by the
requirement of correct behaviour of physical modes in the continuum
limit (eq.~\ref{eq:naii}). Similarly, lattice periodicity requires
that $P_\mu(ap_\mu+2\pi)=ap_\mu+\mathcal{O}(a^2)$. One can therefore have an
additional zero crossing of $P_\mu(ap_\mu)$ within the Brillouin zone
or a discontinuity of the function. The former corresponds to a
doubler mode while the later, in coordinate space, corresponds to a
nonlocal operator.  This situation is depicted in fig.~\ref{fig:nn}.

Of course replacing $\sin(ap_\mu)/a$ by $P_\mu(ap_\mu)/a$ in
eq.~\ref{eq:kernel} is not the most generic ansatz. One could try to
add a term to the action that vanishes at $ap_\mu=0$ but gives a large
contribution at the doubler momenta. If we require in addition that
the chiral symmetry of the continuum action eq.~\ref{eq:ct} is
respected by the lattice action, our choices are severely
restricted. Continuum chiral symmetry requires the additional term to
anticommute with $\gamma_5$, so the only other term we may add has the
form $\gamma_\mu\gamma_5 R_\mu(ap_\mu)$. For a correct continuum
limit, this term has to vanish at $ap_\mu=2\pi n$, too. A possible
action along these lines would be
\begin{equation}
\label{eq:stup}
S_F=a^4 \sum_x \bar\psi(x)\left(\gamma_\mu\Delta_\mu+a\gamma_\mu\gamma_5\Box_\mu+m\right)\psi(x)
\end{equation}
with
\begin{equation}
\Box_\mu f(x):=\frac{f(x+e_\mu)-2f(x)+f(x-e_\mu)}{2a^2}
\end{equation}
which would lead to an inverse propagator
\begin{equation}
\label{eq:stupk}
G^{-1}(p)=i\gamma_\mu\frac{\sin ap_\mu}{a}+\gamma_\mu\gamma_5\frac{1-\cos ap_\mu}{a}+m
\end{equation}
One can check that the effect of the additional term in
eq.~\ref{eq:stupk} is just to shift the doubler poles within the
Brillouin zone. The only other possibility that is left, namely adding
cross-terms, will have a similar effect of shifting the doubler poles
within the multidimensional Brillouin zone. There is in fact a no-go
theorem by Nielsen and Ninomiya
\cite{Nielsen:1980rz,Nielsen:1981xu,Nielsen:1981hk} that states the
impossibility of a fermion discretization that simultaneously fulfills
all of the following requirements
\begin{enumerate}
\item
  The absence of doubler modes
\item
  Invariance under continuum chiral symmetry
\item
  Locality of the fermion operator
\item
  The correct continuum limit
\end{enumerate}
It was also pointed out by Karsten and Smit \cite{Karsten:1980wd},
that the emergence of doubler modes for lattice fermions is a natural
consequence of the general feature that a regulated theory is anomaly
free. The chiral anomaly of the physical mode is cancelled by the
anomaly of the doubler modes.\footnote{See also
  \cite{Chodos:1977dh,Kerler:1981tb}}

\begin{figure}[htb]
\centerline{
\includegraphics[width=12cm]{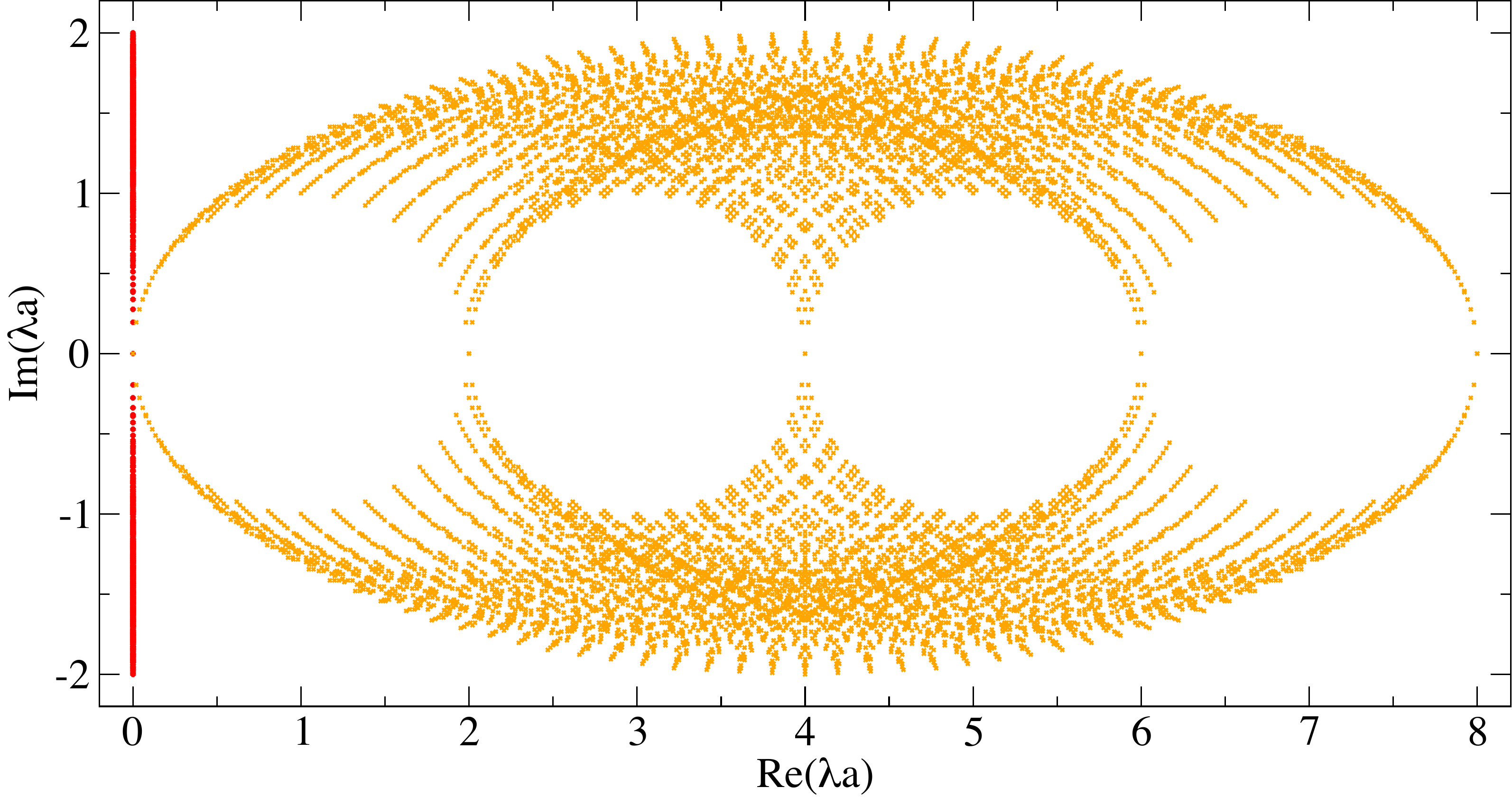}
}
\caption{\label{fig:fermop}
  Eigenvalue spectra of the free naive fermion operator
  (eq.~\ref{eq:naii}, along the imaginary axis) and the free Wilson
  fermion operator (eq.~\ref{eq:wilk}, spread into the real direction)
  for $m=0$ on a $32^4$ lattice.  The naive operator has a 16-fold
  degeneracy in each mode. Note that the spectrum of the free
  staggered fermion operator eq.~\ref{eq:stagop} is identical to the
  naive one except for the degeneracy, which is reduced by a factor of
  4.}
\end{figure}

In order to proceed, we therefore need to throw one of the desired
features of our fermion action overboard. Since we can not really
sacrifice the correct continuum limit or the locality of the operator,
the choices are to either violate continuum chiral symmetry or to live
with some doubler fermions. The former is most easily accomplished by
adding a term to the action that is very similar in spirit to
eq.~\ref{eq:stup} but violates chiral symmetry (eq.~\ref{eq:ct}). The
resulting action
\cite{Wilson:1975id}.
\begin{equation}
\label{eq:wils}
S^W_F=a^4 \sum_x \bar\psi(x)\left(\gamma_\mu\Delta_\mu+ra\Box+m\right)\psi(x)
\end{equation}
is known as the Wilson fermion action where we have used
$\Box=\sum_\mu\Box_\mu$. The Wilson parameter $r$ is usually set to
$r=1$. Fourier transforming to momentum space, we can read off the
inverse propagator as
\begin{equation}
\label{eq:wilk}
G^{-1}(p)=i\gamma_\mu\frac{\sin ap_\mu}{a}+\sum_\mu\frac{1-\cos ap_\mu}{a}+m
\end{equation}
We can see that although the Wilson term in eq.~\ref{eq:wils} is
formally suppressed by $a$ in the continuum limit, it does very
different things to physical and doubler modes. At fixed {\it
  physical} momentum $p$, the additional term in eq.~\ref{eq:wilk}
vanishes indeed $\propto a$ while at fixed {\it lattice} momentum $ap$
it gives a divergent contribution $\propto 1/a$. Notice also that
while the naive term spreads the momenta into the imaginary direction,
the Wilson term spreads them into the real one
(cf. fig.~\ref{fig:fermop}).

Introducing gluon fields in a gauge invariant manner is
straightforward. We replace the finite difference operator by the
covariant one
\begin{equation}
\label{eq:covder}
\Delta_\mu f(x):=\frac{U_\mu(x)f(x+e_\mu)-U^\dag_\mu(x-e_\mu)f(x-e_\mu)}{2a}
\end{equation}
and the discretized second derivative by the covariant
\begin{equation}
\Box_\mu f(x):=\frac{U_\mu(x)f(x+e_\mu)-2f(x)+U^\dag_\mu(x-e_\mu)f(x-e_\mu)}{2a^2}
\end{equation}

Unlike the operator of the naive fermion action eq.~\ref{eq:naiv}, the
operator
\begin{equation}
\label{eq:wop}
D_\text{W}(m)=\gamma_\mu\Delta_\mu+ra\Box+m
\end{equation}
of the Wilson action eq.~\ref{eq:wils} is no more antihermitian at
$m=0$. It does however fulfill the property
\begin{equation}
\label{eq:g5hw}
\gamma_5 D_\text{W}(m)=D^\dag_\text{W}(m) \gamma_5
\end{equation}
which is known as $\gamma_5$-hermiticity. It implies that the
eigenvalues of $D_\text{W}(m)$ are either real or appear in complex
conjugate pairs. This property renders the determinant of the operator
real, which will be important for its numerical treatment. In
addition, eq.~\ref{eq:g5hw} implies that the left eigenvector of a
complex mode $|i\rangle$ is related to the right eigenvector of the
complex conjugate mode $|\hat{i}^*\rangle$ by
$|\hat{i}^*\rangle=\gamma_5 |i\rangle$ and the left and right
eigenvectors $|j\rangle$ and $|\hat{j}\rangle$ of a real mode are
related by $|\hat{j}\rangle=\gamma_5 |j\rangle$. The latter property
is the remnant of the chirality of zero modes of the continuum
operator \cite{Atiyah:1967ih}. For a normal operator (i.e. when
$|\hat{j}\rangle=|j\rangle$), it reads $|j\rangle=\gamma_5 |j\rangle$.

Because of the explicit breaking of chiral symmetry some additional
operator mixing occurs in Wilson fermions that is absent in fermion
formulations which respect chiral symmetry. As a consequence, Wilson
fermions show some deficiencies, most notably an additive mass
renormalization and a bad $\mathcal{O}(a)$ scaling behaviour. As in the case of
gauge actions, these deficiencies can be ameliorated by following a
Symanzik improvement program.

The most straightforward idea of constructing a Wilson-like operator
with an improved continuum behaviour is including next-to-nearest
neighbour points into the derivative terms similar to
eq.~\ref{eq:derimp}. Such an operator has been proposed by Hamber and
Wu \cite{Hamber:1983qa}, but it is not used because a simpler
alternative exists. Sheikholeslami and Wohlert
\cite{Sheikholeslami:1985ij} demonstrated that $\mathcal{O}(a)$ improvement can
also be achieved by adding a Pauli term to the Wilson operator
(eq.~\ref{eq:wop})
\begin{equation}
\label{eq:clov}
D_\text{SW}=D_\text{W}(m)-a\frac{rc_{SW}}{2}\sigma_{\mu\nu}G_{\mu\nu}
\end{equation}
From eq.~\ref{eq:plaq} we see that the value of the gluon field at the
center of the plaquette can be obtained simply by taking the imaginary
part of $P_{\mu\nu}$. In order to obtain it at a lattice site, the
average over the four adjacent plaquettes is usually taken and the
term is often referred to as the clover term.

The clover term in eq.~\ref{eq:clov} comes with a coefficient
$c_\text{SW}$. For classical or tree level improvement $c_\text{SW}=1$
and the resulting action has discretization effects of $\mathcal{O}(a^2)$
classically and $\mathcal{O}(\alpha_s a)$ through quantum corrections. One can
compute the quantum corrections either perturbatively
\cite{Wohlert:1987rf,Luscher:1996vw} or nonperturbatively
\cite{Luscher:1996ug}, but the combination of tree-level clover
improvement with UV-filtering (or smearing) techniques that will be
discussed in sect.~\ref{sect:smear} provides for a very efficient
reduction of the $\mathcal{O}(\alpha_s a)$ effects \cite{Capitani:2006ni}.

We now turn towards the second option for evading the Nielsen-Ninomiya
theorem: living with doubler fermions. We saw that the naive fermion
action eq.~\ref{eq:naiv} describes a theory with $2^D=16$ poles in the
fermion propagator. The poles are located such that one can reach
another pole by adding/subtracting a momentum $\pi/a$ to any momentum
component $p_\mu$. If one then starts from the pole at $p=0$ and
defines $p_\mu^\prime=-(p_\mu\pm\pi/a)$, it is evident that the pole
at $p_\mu=\pm\pi/a$ can be reinterpreted as the physical one in the
new momenta. Note that due to the sign flip between the momenta
definitions, chirality will be reversed when going to an adjacent
pole.

In the free theory, there is an exact degeneracy between the fermions
described by each of the 16 poles. In the interacting theory, high
momentum gluons with momentum $\sim\pi/a$ will couple the different
species. As the necessary momentum diverges for $a\rightarrow 0$, one
can expect these mixing effects to disappear in the continuum
limit. We will see later that there can be subtle order of limits
effects however. 

The general strategy of living with doubler fermions is now to project
to one of the fermion species and suppress the effect of the others as
much as possible. It has been noted very early on in the development
of lattice gauge theory
\cite{Kogut:1974ag,Banks:1975gq,Susskind:1976jm} that the naive
fermion operator has an exact fourfold degeneracy even in the
interacting case that can be exposed and lifted by a simple
transformation. We start with the explicit form of the naive fermion
action eq.~\ref{eq:naiv} in the interacting case
\begin{equation}
S_N=a^4 \sum_x \bar\psi(x)\gamma_\mu\frac{U_\mu(x)\psi(x+e_\mu)-U^\dag_\mu(x-e_\mu)\psi(x-e_\mu)}{2a}+m\bar\psi(x)\psi(x)
\end{equation}
Substituting for the fermion fields
\begin{eqnarray}
 \label{eq:stagt1}
\psi(x)&=&\gamma_0^\frac{x_0}{a}\gamma_1^\frac{x_1}{a} 
\gamma_2^\frac{x_2}{a} \gamma_3^\frac{x_3}{a}\chi(x)\\
\label{eq:stagt2}
 \bar\psi(x)&=&\bar\chi(x) \gamma_3^\frac{x_3}{a}\gamma_2^\frac{x_2}{a} 
\gamma_1^\frac{x_1}{a} \gamma_0^\frac{x_0}{a}
\end{eqnarray}
we obtain
\begin{equation}
\label{eq:stagac}
S_\text{st}=a^4 \sum_x \bar\chi(x)\eta_\mu(x)\frac{U_\mu(x)\chi(x+e_\mu)-U^\dag_\mu(x-e_\mu)\chi(x-e_\mu)}{2a}+m\bar\chi(x)\chi(x)
\end{equation}
where $\eta_\mu(x)$ is a purely numerical factor
\begin{equation}
\eta_\mu(x)=(-1)^{\sum_{\nu<\mu}x_\nu}
\end{equation}
The corresponding staggered fermion operator reads
\begin{equation}
\label{eq:stagop}
D_\text{st}(m)=\eta_\mu\Delta_\mu+m
\end{equation}
We can therefore take $\bar\chi$ and $\chi$ to be single component
fields, which lifts a fourfold exact degeneracy. The individual spinor
components of the fermion field are not all present at each lattice
site anymore. We can however infer from the transformation
eqs.~\ref{eq:stagt1}-\ref{eq:stagt2} how to combine the 16 components
present in an elementary hypercube into 4 species (or tastes) of
4-component fermion spinors. As the components of the fermion field
are staggered across the lattice, the action is referred to as
staggered fermions.

Staggered fermions satisfy an equivalent of $\gamma_5$-hermiticity  eq.~\ref{eq:g5hw}
\begin{equation}
\label{eq:g5hs}
\eta_5 D_\text{st}(m)=D^\dag_\text{st}(m) \eta_5
\end{equation}
with
\begin{equation}
\label{eq:stageps}
\eta_5(x)=\left(-1\right)^{x_0+x_1+x_2+x_3}
\end{equation}
They also retain a remnant of chiral symmetry at zero mass
\begin{equation}
\label{eq:stagu1}
\left\{D_\text{st}(0),\eta_5\right\}=0
\end{equation}
which implies cutoff terms of $\mathcal{O}(a^2)$ and the absence of additive
mass renormalization. The symmetry eq.~\ref{eq:stagu1} is however very
different from the full continuum chiral symmetry. It is a $U(1)$ and
will be present, even if we want to describe a single fermion flavour
which in the continuum does not have a chiral symmetry. The
implications of this are discussed extensively in the literature
\cite{Smit:1986fn,Adams:2004mf,Durr:2004ta,Bernard:2006zw,Bernard:2007qf,Prelovsek:2005rf,Durr:2004as,Durr:2003xs,Davies:2003ik,Davies:2004hc,Aubin:2004fs,Follana:2007uv,Bazavov:2010ru,Bazavov:2009bb,Bernard:2006ee,Giedt:2006ib,Bernard:2007ma,Shamir:2006nj,Bernard:2004ab,Durr:2006ze,Creutz:2007yg,Bernard:2006vv,Hasenfratz:2006nw,Creutz:2007pr,Durr:2005ax,Sharpe:2006re,Durr:2012te}
and while no definitive conclusion has been reached, there are many
indications that staggered fermions do correctly reproduce the chiral
symmetry pattern including the anomaly if an appropriate continuum
limit is taken before going to the chiral limit.

A great number of additional fermion discretizations have been
suggested in the literature and are used to some degree in recent
lattice calculations. Among those are twisted mass fermions
\cite{Frezzotti:2000nk}, which feature an improved scaling behaviour
at the expense of flavour breaking, minimally doubled fermions
\cite{Karsten:1981gd,Wilczek:1987kw,Creutz:2007af,Borici:2007kz} with
only a single doubler pole that comes at the expense of breaking the
lattice rotational symmetry or staggered fermions with an additional
Wilson term to remove doubler modes from the physical spectrum
\cite{Adams:2010gx,deForcrand:2011ak,Hoelbling:2010jw}. The most
numerous group however are fermion formulations that to some degree
build upon the advances in understanding of chiral symmetry on the
lattice.

\subsection{Lattice chiral symmetry}

The Nielsen-Ninomiya theorem seems to forbid the existence of an
otherwise well-behaved lattice fermion with chiral symmetry. Its
definition of chiral symmetry, however, is the continuum form. It
demands that the fermion operator $D$ anticommutes with $\gamma_5$, so
$\gamma_5 D+D \gamma_5=0$. It has been realized very early on by
Ginsparg and Wilson \cite{Ginsparg:1981bj} that upon blocking from the
continuum, the continuum chiral symmetry is replaced by the relation
\begin{equation}
\label{eq:gw}
\gamma_5 D+D \gamma_5=\frac{a}{\rho}D\gamma_5 D
\end{equation}
Independently of this work, a class of lattice actions was constructed
\cite{Hasenfratz:1993sp,DeGrand:1995ji,Bietenholz:1995cy} by blocking
transformations and it was realized, that they fulfill the
Ginsparg-Wilson relation eq.~\ref{eq:gw}
\cite{Hasenfratz:1998ri}. Another lattice fermion formulation was
inspired by the discovery that in a 5-dimensional theory chiral
fermions naturally arise along a 4-dimensional defect
\cite{Callan:1984sa,Frolov:1992ck} even on a lattice with finite
cutoff \cite{Kaplan:1992bt}. The resulting domain wall fermion action
\cite{Shamir:1993zy} is still widely used today. Along similar
lines, Narayanan and Neuberger developed the overlap fermion
action\cite{Narayanan:1992wx,Narayanan:1993sk,Narayanan:1993ss,Narayanan:1994gw},
which was condensed into a 4-dimensional fermion operator by Neuberger
\cite{Neuberger:1997fp}. This operator fulfills the Ginsparg-Wilson
relation (eq.~\ref{eq:gw}) \cite{Neuberger:1998wv} and in the massless
case is given by
\begin{equation}
\label{eq:ovop}
D_\text{ov}=\frac{\rho}{a}\left(\mathbbm{1}+\frac{D_\text{W}(-\rho/a)}{\sqrt{D^\dag_\text{W}(-\rho/a)D_\text{W}(-\rho/a)}}\right)
\end{equation}
where $D_\text{W}(-\rho/a)$ is the Wilson operator (eq.~\ref{eq:wop})
at a negative bare mass $-\rho/a$. Note that the overlap operator also
fulfills $\gamma_5$-hermiticity $\gamma_5 D_\text{ov}=D^\dag_\text{ov}
\gamma_5$.

Although the Ginsparg-Wilson relation is sometimes referred to as a
minimal way of breaking chiral symmetry, it is in fact the correct
chiral symmetry relation of the regulated theory
\cite{Luscher:1998pqa}. This is somewhat more apparent after a trivial
rewriting of eq.~\ref{eq:gw}
\begin{equation}
\label{eq:gwr}
\gamma_5 D+D \hat\gamma_5=0
\qquad
\hat\gamma_5=\gamma_5\left(1-\frac{a}{\rho}D\right)
\end{equation}
It implies that the fermion action
\begin{equation}
\label{eq:gwa}
S=\bar\psi D\psi
\end{equation}
is invariant under an infinitesimal chiral transformation
\begin{equation}
\label{eq:gwt}
\bar\psi\rightarrow\bar\psi (\mathbbm{1}+i\epsilon\gamma_5)
\qquad
\psi\rightarrow (\mathbbm{1}+i\epsilon\hat\gamma_5)\psi
\end{equation}
which acts differently on the fermion and antifermion fields. Note
however that eq.~\ref{eq:gwr} implies that in the continuum limit the
continuum form of chiral symmetry is restored
\begin{equation}
\label{eq:g5ct}
\hat\gamma_5=\gamma_5\left(1-\frac{a}{\rho}D\right)\stackrel{a\rightarrow
0}{\longrightarrow}\gamma_5
\end{equation}

\begin{figure}[htb]
\centerline{
\includegraphics[width=12cm]{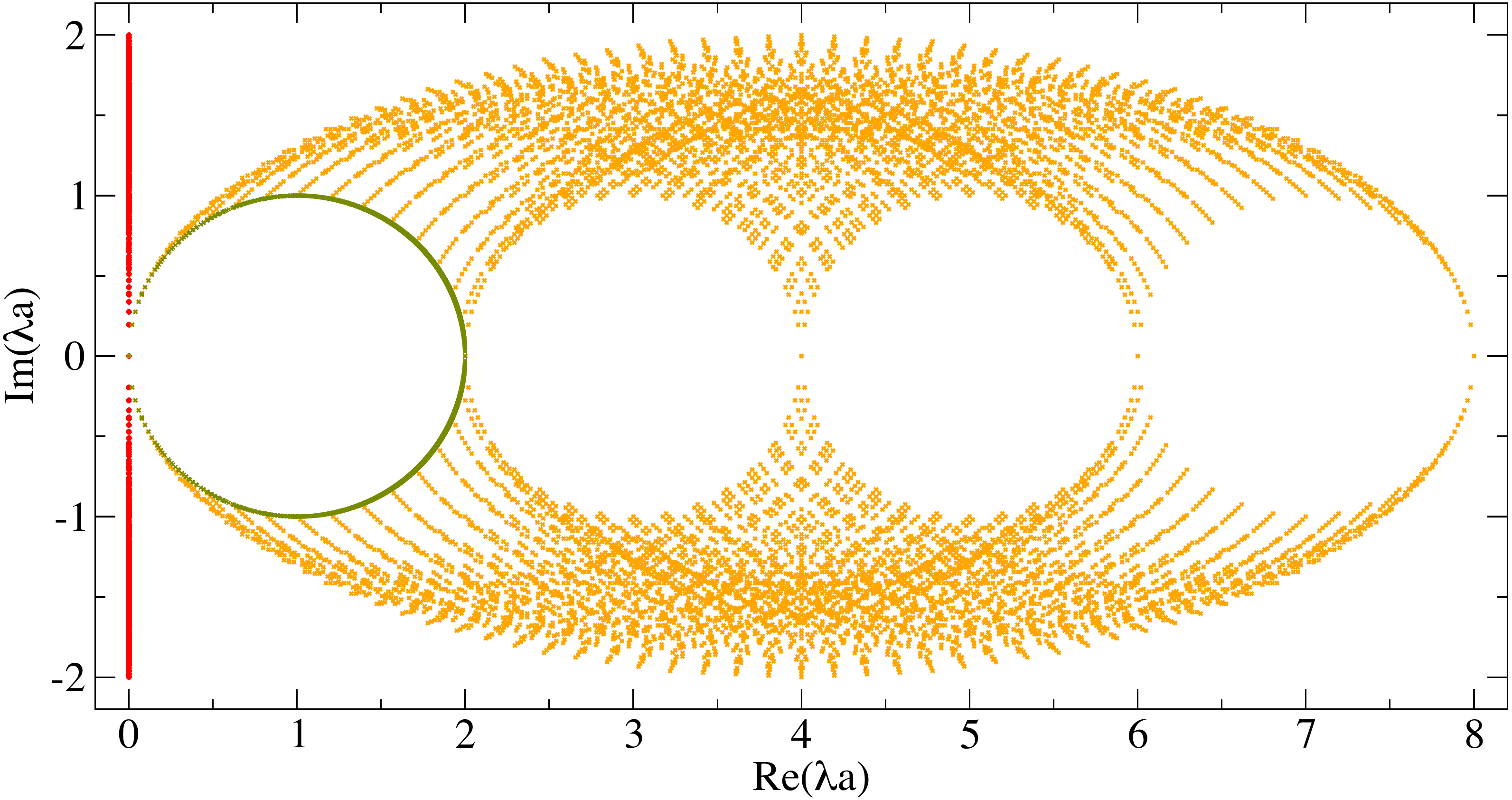}
}
\caption{\label{fig:fermop2} Eigenvalue spectrum of the free overlap
  operator with $\rho=1$ (eq.~\ref{eq:ovop}, circle) compared to the
  free Wilson (eq.~\ref{eq:wilk}, spread into the real direction) and
  staggered (eq.~\ref{eq:stagop}, along the imaginary axis)
  operators on a $32^4$ lattice.}
\end{figure}

Together with $\gamma_5$ hermiticity $D\gamma_5=\gamma_5 D$ the
Ginsparg-Wilson relation eq.~\ref{eq:gw} implies
\begin{equation}
D^\dag+D=\frac{a}{\rho}DD^\dag
\end{equation}
which means that $\frac{a}{\rho}D-1$ is a unitary operator and the
eigenvalues of $D$ lie on a circle of radius $\rho/a$ touching the
imaginary axis at the origin (see fig.~\ref{fig:fermop2}). The real
modes of $D$ are therefore located at either $0$ or $2\rho/a$ and it
can easily be shown that they are chiral. The modes at $2\rho/a$
correspond to all unphysical doubler branches and they can in fact be
removed by a simple transformation of the fermion fields
\begin{equation}
\label{eq:gwtild}
\tilde\psi=\tilde{\mathbbm{1}}\psi
\qquad
\tilde{\mathbbm{1}}=\mathbbm{1}-\frac{a}{2\rho}D
\end{equation}
The (massless) fermion action can now be written as
\begin{equation}
\label{eq:gwtilda}
S=\bar\psi\frac{D}{1-\frac{a}{2\rho}D}\tilde\psi
\end{equation}
which is an antihermitian operator. The chiral modes are at the origin
and the doublers have been removed off to infinity. In fact, in the
new field variables the action now obeys the continuum form of chiral
symmetry, i.e. it is invariant under
\begin{equation}
\label{eq:gwtt}
\bar\psi\rightarrow\bar\psi (\mathbbm{1}+i\epsilon\gamma_5)
\qquad
\tilde\psi\rightarrow (\mathbbm{1}+i\epsilon\gamma_5)\tilde\psi
\end{equation}
All consequences of chiral symmetry, such as conserved axial currents,
the correct anomaly, the absence of additive mass renormalization and
discretization effects that start at $\mathcal{O}(a^2)$ only are
therefore present for Ginsparg-Wilson fermions provided that the
correctly rotated field variables are used for constructing the
observables. One can now also add a mass term to Ginsparg-Wilson
fermions that behaves exactly like a continuum mass term
\begin{equation}
D(m)=D+\tilde{\mathbbm{1}}m
\end{equation}

The main disadvantage of chirally symmetric fermions for numerical
computations is their cost. Simple fermion discretizations, such as
the Wilson or staggered ones, typically have fermion operators with a
limited number of couplings to neighbour sites. Numerically, this
translates into them being sparse matrices. In contrast, chirally
symmetric operators, such as e.g. the overlap operator
eq.~\ref{eq:ovop}, tend to be full matrices which are much more
demanding computationally. In fact, there is a theorem
\cite{Horvath:1998cm,Bietenholz:1999dg} that chirally symmetric
operators can not be realized with a finite number of couplings to
their nearest neighbours in four dimensions, a property that is
referred to as ultralocality in the lattice literature. One can still
have an ultralocal chiral fermion operator if one includes an extra
fifth dimension as is done in the case of domain wall fermions. The
price to pay there however is that the extent of the fifth dimension
has to be made infinitely large in principle to have exact chiral
symmetry. Obviously this strategy implies considerable additional
computational effort, too.

\subsection{Effects of UV modes}
\label{sect:smear}

It is evident from fig.~\ref{fig:fermop2} that the physical modes of a
lattice fermion operator which are located around the origin are by
far outnumbered by UV modes. Modes that are close to the cutoff do not
carry a lot of physical information in the interacting case
though. With increasing cutoff, these modes are ever deeper in the
perturbative regime that is dominated by asymptotic freedom and their
fluctuations mainly contribute towards enlarging cutoff effects.

\begin{figure}[htb]
\centerline{
\includegraphics[width=6cm]{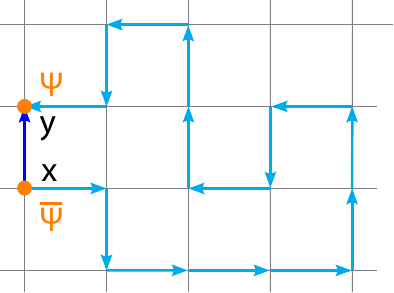}
}
\caption{\label{fig:smear}Illustration of two different lattice paths,
  one of them minimal, between neighbouring points $x$ and $y$.}
\end{figure}

It is possible to eliminate a large part of these fluctuations by a
simple modification of the fermion operator. Remember that gauge
interactions were introduced into the lattice theory by the covariant
derivative (eq.~\ref{eq:covder}) which contains the parallel transport
$U_\mu(x)$ between the lattice points $x$ and $x+e_\mu$. Note that we
have chosen the gauge connection $U_\mu(x)$ along the minimal path
connecting the two neighbouring ponts. Although this choice seems
reasonable, it is not unique. In principle, one can choose any path
connecting the two sites (see fig.~\ref{fig:smear}).

\begin{figure}[htb]
\centerline{
\includegraphics[width=8cm]{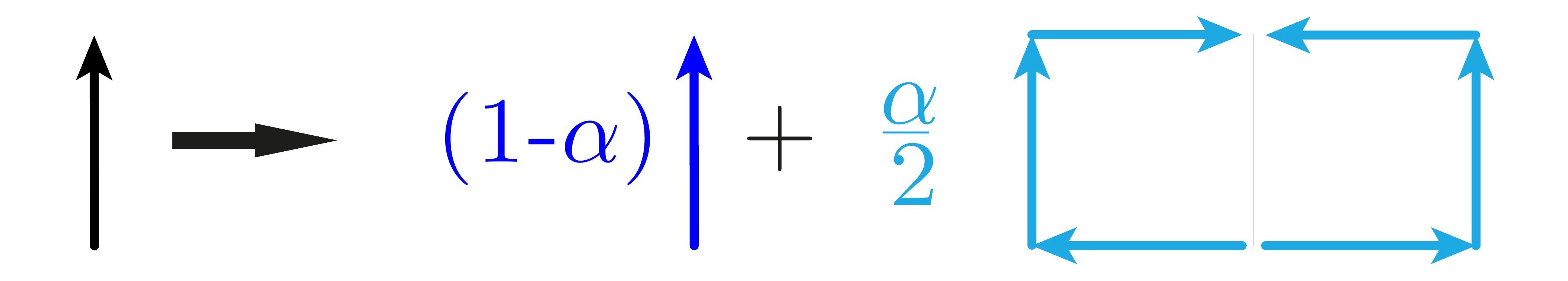}
}
\caption{\label{fig:apesmear}Illustration of the APE smearing
  procedure \cite{Albanese:1987ds} in the 2-dimensional case. The link
connecting nearest neighbouring sites in the fermion operator is
replaced by a weighted average of the ``thin link'' (weight
$1-\alpha$) and the ``staple'' (weight $\alpha/2(D-1)$).}
\end{figure}

This ambiguity can be used to effectively damp the coupling of gluons
with momenta close to the cutoff to the fermions. The procedure is
generically termed smearing, link fattening or UV-filtering and the
first instance was proposed by the APE collaboration
\cite{Albanese:1987ds}. In the APE smearing recipe, a fat or smeared
gauge link $U^\text{(APE)}_\mu(x)$ in $D$ dimensions is defined as a
weighted average
\begin{equation}
\label{eq:apesmear}
U^\text{(APE)}_\mu(x)=
(\alpha-1) U_\mu(x)+
\frac{\alpha}{2(D-1)}\Omega_\mu(x)
\end{equation}
where
\begin{equation}
\Omega_\mu(x)=
\sum_{\mu\neq\mu}U_\nu(x)
U_\mu(x+e_\nu)
U_\nu^\dag(x+e_\mu)
+
U^\dag_\nu(x-e_\nu)
U_\mu(x-e_\nu)
U_\nu(x-e_\nu+e_\mu)
\end{equation}
is the sum over staples (see fig.~\ref{fig:apesmear}) with the smearing
parameter $\alpha$ typically chosen to be $\alpha\sim 0.6$. The
smeared link $U^\text{(APE)}_\mu(x)$ can then be used in the fermion
operator instead of the original ``thin link'' $U_\mu(x)$. For
sufficiently smooth gauge configurations, i.e. for gauge
configurations where the gluonic fields do not carry substantial
momenta components at the cutoff scale, the difference between thin
links and staples is irrelevant in the continuum limit. Consequently,
the continuum limit is not affected by replacing thin links with
smeared ones in the fermionic operator.

This simple recipe has a shortcoming though. The new link variable
$U^\text{(APE)}_\mu(x)$ was obtained by averaging elements of the
gauge group and therefore is not in general an element of the gauge
group itself. This can be remedied by a simple unitary backprojection
\begin{equation}
\label{eq:as2}
U^\prime
=
\frac{U^\text{(APE)}}{\sqrt{
{U^\text{(APE)}}^\dag
U^\text{(APE)}
}}
\end{equation}
followed by dividing out the phase of the determinant
\begin{equation}
\label{eq:as3}
\hat{U}=
\frac{U^\prime}
{\left(\det(U^\prime)\right)^{1/3}}
\end{equation}
This procedure is continuum irrelevant on sufficiently smooth gauge
fields, too. One can therefore replace the thin links $U$ in the
fermion action with $\hat{U}$.

While the backprojection eqs.~\ref{eq:as2},\ref{eq:as3} produces an
element of the gauge group, it is however not differentiable. This
turns out to be an obstacle for dynamical fermion algorithms as they
require the derivative of the fermionic action with respect to the
original gauge field $U$. Morningstar and Peardon have suggested a
modification of the APE smearing procedure \cite{Morningstar:2003gk}
that is both differentiable and equivalent to APE smearing for small
smearing parameters $\alpha$. They start by constructing the
antihermitian part of the plaquettes spanned by the staples
\begin{equation}
\label{eq:mp1}
A_\mu(x)=\frac{\Omega_\mu(x)U^\dag_\mu(x)-U_\mu(x)\Omega^\dag_\mu(x)}{2}
\end{equation}
and making it traceless
\begin{equation}
\label{eq:mp2}
S_\mu(x)=A_\mu(x)-\frac{1}{3}\Tr A_\mu(x)
\end{equation}
Exponentiating the result with a smearing parameter $\rho$ and
multiplying it on the original link
\begin{equation}
\label{eq:mp3}
V_\mu(x)=e^{\rho S_\mu(x)}U_\mu(x)
\end{equation}
gives the so-called stout link $V_\mu(x)$. For small smearing
parameters, stout link smearing with a smearing parameter
$\rho=\alpha/2(D-1)$ is equaivalent to APE smearing.

There are many variants of the smearing procedure that are commonly
used. The simplest one is the repeated application of the smearing
procedure. One can e.g. use the $V_\mu(x)$ from eq.~\ref{eq:mp3} as an
input to eq.~\ref{eq:mp1} instead of the original thin link
$U_\mu(x)$. If the number of steps is kept finite, the procedure still
amounts to a continuum irrelevant redefinition of the fermion action.

Instead of repeating the entire smearing procedure $n$ times, it is
also possible to change the staples used and the smearing parameter
upon each application. Hasenfratz and Kenchtli
\cite{Hasenfratz:2001hp} suggested a smearing procedure along these
lines consisting of $D-1$ steps. They construct an APE-smeared link
out of staples that are smeared themselves. This nested smearing is
a variant of the APE smearing leaving out all directions that would
cause a link to be outside the adjacent elementary hypercubes of the
original target link. This nesting is then repeated until thin links are used
in the $(D-1)^\text{st}$ step. This smearing procedure is therefore
known as hypercubic or HYP smearing. An analytic variant of this
procedure, the hypercubic exponential (HEX) smearing
\cite{Capitani:2006ni} is also in use today. 

\begin{figure}[htb]
\centerline{
\includegraphics[width=12cm]{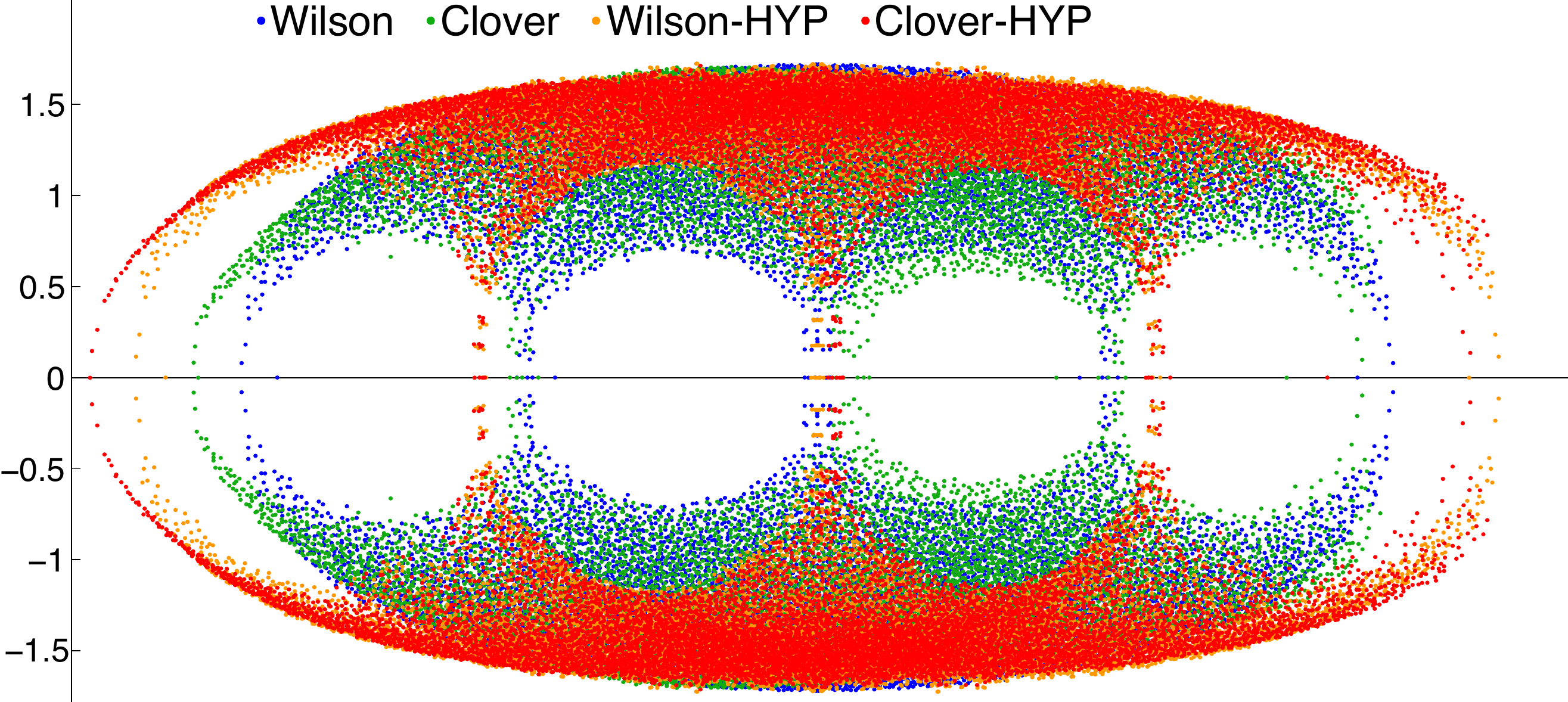}
}
\caption{\label{fig:smearsp}Eigenvalue spectrum of Wilson-type
  operators on one single gauge field background on a $6^4$
  lattice. Data courtesy S. D\"urr \cite{Durr:2010ch,Durr:2013gp}.}
\end{figure}

To Illustrate the effect that smearing has on the eigenmode spectrum
of the Wilson operator, fig.~\ref{fig:smearsp} is plotting the entire
spectrum of massless Wilson operators on a single gauge configuration
of topological charge $1$. Remember that the physically relevant low
momentum modes are in the vicinity of the origin and that for the free
case and for Ginsparg-Wilson and staggered fermions the physical
eigenmodes spread from the origin along the imaginary axis resp. a
circle touching it (cf. fig.~\ref{fig:fermop2}). For the interacting
case we see, that the eigenmodes of the Wilson operator do not touch
the imaginary axis at all indicating a large additive mass
renormalization. In addition, the would-be chiral mode on the real
axis is far away from the low lying complex modes indicating a large
mixing with doubler modes.

Adding a clover term (eq.~\ref{eq:clov}) with the tree-level
$c_\text{SW}=1$ does mitigate both these effects somewhat as does one
step of HYP-smearing. The combined effect of clover improvement and
smearing however does result in a significantly improved operator
spectrum in the relevant region.

Similarly, beneficial effects of smearing can be observed for
staggered fermions. In their case, a gluon field with a momentum
component $\sim\pi/a$ near the cutoff can transform between the
staggered ``tastes''. A suppression of these spurious interactions
that are absent in the continuum therefore improves the degeneracy
between the physical and remaining doubler branches and leads to a
smaller breaking of the taste symmetry.

\begin{figure}[htb]
\centerline{
\includegraphics[width=12cm]{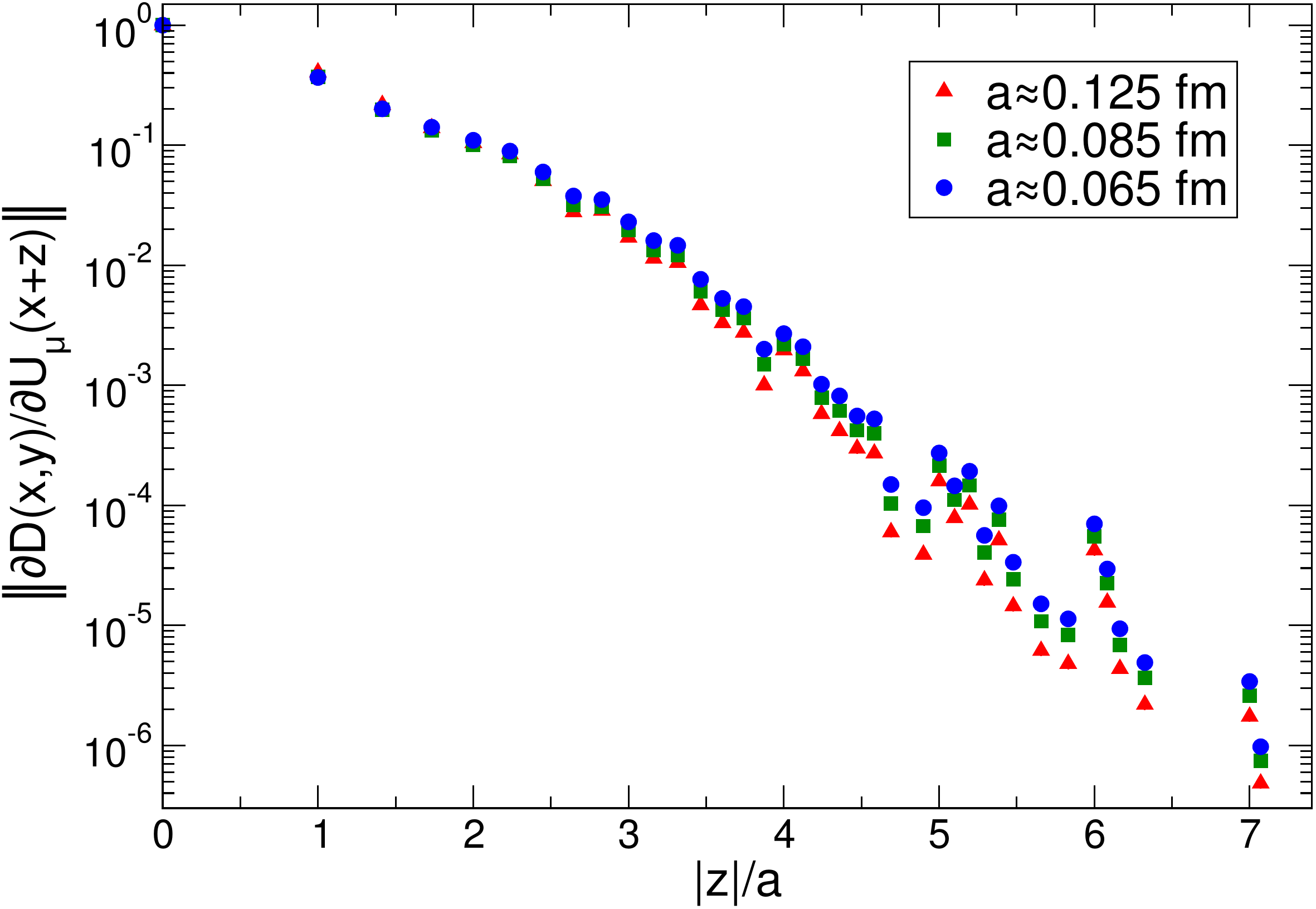}
}
\caption{\label{fig:loc}Locality of the gauge field to fermion
  coupling for a 6-step stout smeared action. As one can clearly see,
  the exponential decay of the coupling with distance in lattice units
  has an envelope that is independent of lattice spacing. Results and
  figure from \cite{Durr:2008zz}.}
\end{figure}

One might be worried about the effect of iterated smearing on the
locality of the fermion operator. In fact, the locality of the fermion
operator itself is not affected by the smearing at all. Changing the
link variables in the fermion operator does not alter the sites
connected to each other via these links. What is affected by link
smearing is the fermion to gauge field coupling: it acquires a
momentum dependent form factor. For small $\alpha$, perturbation
theory tells us that the gauge field coupling is smeared out after N
steps over an effective radius squared of \cite{Bernard:1999kc}
\begin{equation}
\label{eq:spread}
\langle r^2\rangle_\text{eff}=\frac{a^2N\alpha}{D-1}
\end{equation}
For fixed $\alpha$ and $N$ therefore the coupling should be local in
the continuum limit. This assertion has been tested numerically for
the case of 6-step stout smearing \cite{Durr:2008zz}. As one can see
in fig.~\ref{fig:loc}, the sensitivity of the fermion operator towards
a variation of the gauge field is bounded from above by an exponential
in lattice units. On top of that, the coupling is still ultralocal. It
is exactly $0$ outside of the smearing radius.

\section{Computing the path integral}
\label{sec:cpi}

\subsection{Fermion fields and observables}
\label{sect:obs}

Until now we have discussed how to discretize both gauge field and
fermion actions on a lattice. As a next step, we would like to compute
the expectation values of fermionic and gauge field observables. For
the gauge fields this seems straightforward, but the classical limit
of fermions are anticommuting Grassmann fields. Assuming that we have
a single staggered fermion field (with one component per lattice site)
on a lattice with $N$ points, the implementation of the full Grassmann
algebra would require an object with $2^N$ components. As $N\sim 10^6$
for a $32^4$ lattice, which is not large by todays standards, this is
absolutely prohibitive.

\begin{figure}[htb]
\centerline{
\includegraphics[width=12cm]{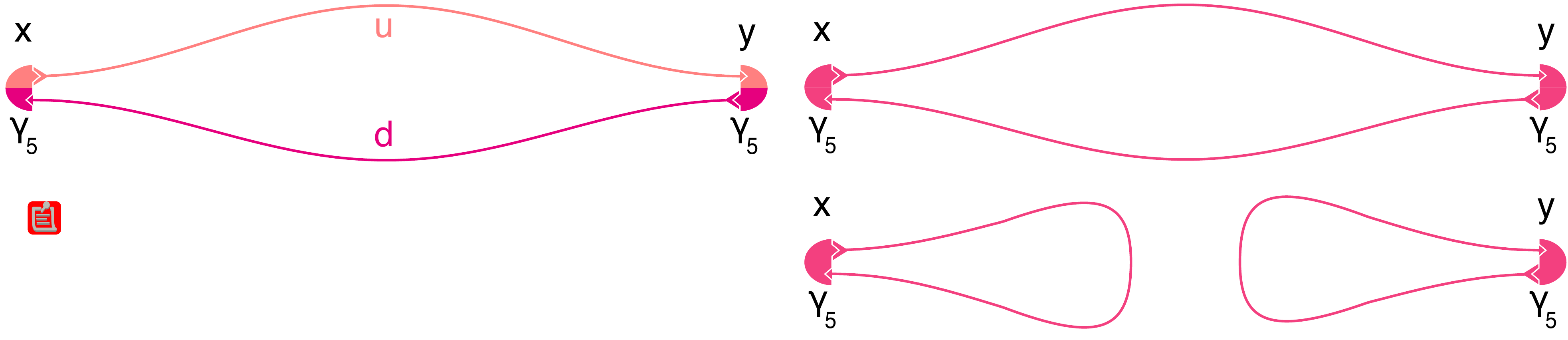}
}
\caption{\label{fig:meprops}Diagrammatic representation of the Wick
  contraction of a flavour non-singlet propagator (left) and the two
  contractions of a flavour singlet propagator (right)}
\end{figure}

In order to proceed, we note that in general the fermion action is
bilinear in the fermion fields
\begin{equation}
\label{eq:fermac}
S_F=\bar\psi D\psi
\end{equation}
Denoting the gauge action by $S_G$, the partition function
eq.~\ref{eq:partf} takes the form
\begin{equation}
\label{eq:z}
\mathcal{Z}=\int\prod_{x,\mu}[dU_\mu(x)][d\bar\psi(x)][d\psi(x)]
e^{-S_G-\bar\psi D\psi}
\end{equation}
Using the rules of Gassman integration, we can formally integrate out
the $\bar\psi$ and $\psi$ fields in eq.~\ref{eq:z} to obtain
\begin{equation}
\label{eq:zint}
\mathcal{Z}=\int\prod_{x,\mu}[dU_\mu(x)]
\det D[U]
e^{-S_G}
\end{equation}
In order to obtain expectation values of observables, we also need to
integrate out the fermion fields in the numerator of
eq.~\ref{eq:epathint}. For gluonic observables this is
straightforward. For fermionic observables, we take as an example the
generic fermion bilinear $\psi_\alpha(x)\bar\psi_\beta(y)$ where
$\alpha$ and $\beta$ generically denote all spinor and flavour
indices. We obtain
\begin{equation}
\label{eq:pfint}
\begin{split}
&\int\prod_{x,\mu}[dU_\mu(x)][d\bar\psi(x)][d\psi(x)]
\psi_\alpha(x)\bar\psi_\beta(y)
e^{-S_G-\bar\psi D\psi}=\\
&\int\prod_{x,\mu}[dU_\mu(x)]
\det D[U]
D^{-1}_{\alpha,\beta}(x,y)
e^{-S_G}
\end{split}
\end{equation}
so here too the path integral over the fermionic fields may be
replaced by a simple factor $\det D[U]$ in the gluonic path integral
or, said differently, by adding an effective gluonic action of the
form $-\ln\det D[U]$ to $S_G$. We now introduce the shorthand notation
\begin{equation}
\label{eq:avgdef}
\langle O[U]\rangle:=
\frac{1}{\mathcal{Z}}
\int\prod_{x,\mu}[dU_\mu(x)]
O[U]
e^{-(S_G -\ln\det D[U])}
\end{equation}
For more complex fermionic observables, one can show that the Wick
theorem is reobtained with the contractions given by the corresponding
inverse of the fermion matrix. We can e.g. obtain
\begin{equation}
\label{eq:mesprop}
\langle 0|
T(
(\bar\psi_\text{u}\gamma_5\psi_\text{d})_x
(\bar\psi_\text{d}\gamma_5\psi_\text{u})_y
)
|0
\rangle=
\left\langle
\Tr\left(
D^{-1}_\text{u}(x,y)
\gamma_5
D^{-1}_\text{d}(y,x)
\gamma_5
\right)
\right\rangle
\end{equation}
which by using $\gamma_5$-hermiticity (eq.~\ref{eq:g5hw}) can be
rewritten into
\begin{equation}
\label{eq:mespropg5h}
\langle 0|
T(
(\bar\psi_\text{u}\gamma_5\psi_\text{d})_x
(\bar\psi_\text{d}\gamma_5\psi_\text{u})_y
)
|0
\rangle=
\left\langle
\Tr\left(
D^{-1}_\text{u}(x,y)
{D^{-1}_\text{d}}^\dag(x,y)
\right)
\right\rangle
\end{equation}
where u and d denote the quark flavours and the trace is taken over
color and spin indices. The observable eq.~\ref{eq:mesprop} can be
diagrammatically represented as shown on the left panel of
fig.~\ref{fig:meprops}.

Similarly, for a flavour-singlet observable we obtain
\begin{equation}
\label{eq:mespropfs}
\begin{split}
\langle 0|
T(
(\bar\psi\gamma_5\psi)_x
(\bar\psi\gamma_5\psi)_y
)
|0
\rangle=
&
\left\langle
\Tr\left(
D^{-1}(x,y)
\gamma_5
D^{-1}(y,x)
\gamma_5
\right)
\right\rangle
\\
+
&
\left\langle
\Tr\left(
D^{-1}(x,x)
\gamma_5
\right)
\Tr
\left(
D^{-1}(y,y)
\gamma_5
\right)
\right\rangle
\end{split}
\end{equation}
which has a disconnected contribution with a double trace in addition
to the single-trace connected contribution. The diagrammatic
representation is shown in the right panel of
fig.~\ref{fig:meprops}.

\subsection{Stochastic evaluation of the path integral}
\label{sect:num}

Having set up the framework of lattice QCD, we can now proceed to
stochastically evaluate the path integral eq.~\ref{eq:avgdef},
computing expectation values of target observables. It is clear from
eq.~\ref{eq:avgdef} that the expectation value of the target
observable $O$ is just a weighted average over the observable computed
on all possible gauge field backgrounds $O[U]$ with a weight
$\exp(-S_U)$ and the effective action $S_U=S_G-\ln\det D[U]$. The most
straightforward stochastic evaluation of the path integral would
therefore consist of producing random gauge configurations, computing
the effective action $S_U$ on them and taking the weighted
average. This procedure is very inefficient though because most of the
configurations will be exponentially suppressed.

A more promising approach is known as importance sampling. Instead of
generating the gauge fields with a uniform random weight, we can
produce them with a weight $\propto\exp(-S_U)$. It is important to
note that this is only possible if $S_U$ is real which in turn
requires the fermion determinant $\det D[U]$ to be real and
positive. We have seen in sect.~\ref{sect:ferm} that
$\gamma_5$-hermiticity implies a real fermion determinant. In addition,
the eigenmodes of all massless fermion operators we have covered possess a
nonnegative real part and consequently $\det D[U]$ is positive
definite for any positive bare mass. For all fermions retaining a
remnant of chiral symmetry, i.e. naive, staggered and Ginsparg-Wilson
fermions, there is no further subtlety. For fermions that explicitly
break chiral symmetry however, like Wilson-type fermions do, an
additive mass renormalization is required that typically renders the
bare mass negative. The positivity of the fermion determinant has then
to be ensured a posteriori and we will discuss in
sect.~\ref{sect:checks} how to check this important property in the
numerical treatment.

Assuming that we may use an importance sampling technique, we label
the gauge fields obtained with a weight $\propto\exp(-S_U)$ as
$U_i$. The expectation value of an observable is then given by a
straight, unweighted average
\begin{equation}
\langle 
O
\rangle
=
\lim_{N\rightarrow\infty}
\frac{1}{N}
\sum_{i=1}^N
O[U_i]
\end{equation}
Truncating the sum after a finite number of gauge configurations, we
obtain an estimate of the observable
\begin{equation}
\hat{O}
=
\frac{1}{N}
\sum_{i=1}^N
O[U_i]
=\langle O\rangle+\mathcal{O}\left(\frac{1}{\sqrt N}\right)
\end{equation}
which is affected by a standard statistical error of order $1/\sqrt
N$. Interpreting $\exp(-S_U)$ as a Boltzmann weight, the importance
sampling technique might also be viewed as generating microstates of a
thermodynamic system with the correct equilibrium distribution.

Except for simple cases of noninteracting theories however, it is
usually not straightforward to generate gauge configurations with a
weight proportional to $\propto\exp(-S_U)$. Typically, update
algorithms are used that generate a gauge configuration based on a
previous one using a stochastic technique. The simplest of these, the
Metropolis algorithm \cite{Metropolis:1953am}, proceeds in the
following steps: Starting with an initial gauge configuration $U_0$,
one iterates through the following steps
\begin{enumerate}
\item
  Generate $U_{k}$ from $U_{k-1}$ by a small random change
\item
  Measure the change in the action $\Delta S=S_U[U_{k}]-S_U[U_{k-1}]$
\item
  Accept the change if $\Delta S\le 0$
\item
  Accept the change with a probability $e^{-\Delta S}$ if $\Delta S>0$
\end{enumerate}
The resulting Markov chain of gauge configurations $U_i$ will
asymptotically (for large $i$) contain gauge configurations with the
correct weight distribution\footnote{for more details and a proof of
  this statement see e.g. \cite{Fodor:2012gf}}. There are some caveats
however that need to be realized. First, consecutive gauge
configurations are not independent. The ``time''-series $U_i$ will
therefore have some autocorrelation, which has to be taken into
account. As a consequence, the system will also not reach thermal
equilibrium instantly and a number of initial configurations will have
to be discarded because they suffer from thermalization
effects. Finally, one not only needs to make sure that the
configurations produced have the correct relative weight, but also
that any possible configuration can be reached by the algorithm with a
finite probability. This property is known as ergodicity. In practice,
some critical observables are typically monitored to ensure the system
has sensible autocorrelation times, is thermalized and ergodic. We
will come back to this point in sect.~\ref{sect:checks}.

The algorithm most widely used today for evaluating the partition
function of lattice QCD is the hybrid Monte-Carlo (HMC) algorithm
\cite{Callaway:1982eb,Callaway:1983ee,Polonyi:1983tm,Batrouni:1985jn,Duane:1985ym,Duane:1985hz,Duane:1986iw,Duane:1987de}. It is an essential extension of the Metropolis
algorithm that replaces the small random change of the first step,
which is very inefficient in full QCD, by a more global modification
of the gauge field. This global modification proceeds through first
reinterpreting the fermion determinant $\det D[U]$ as the contribution
to the partition function of an auxiliary scalar pseudofermion field
$\Phi$ via \cite{Weingarten:1980hx}
\begin{equation}
\label{eq:psf}
\det D[U]=\int \prod_{x}
[d\Phi^\dag(x)][d\Phi(x)]
e^{-\Phi^\dag\left(D^\dag[U]D[U]\right)^{-1/2}\Phi}
\end{equation}
and then evolving the resulting system classically in a fictitious
time with a Hamiltonian
\begin{equation}
\mathcal{H}=\frac{1}{2}\Pi^2+S
\qquad
S=S_G+\Phi^\dag\left(D^\dag[U]D[U]\right)^{-1/2}\Phi
\end{equation}
where $\Pi$ are randomly initialized conjugate momenta. This
procedure guarantees that as long as the classical evolution part was
carried out with sufficient accuracy the change in the action $\Delta
S$ will be moderate despite the global nature of the change in the
gauge field. It will also provide the value of $\Delta S$, which might
otherwise require substantial effort to determine. For further
details on the HMC algorithm, the reader is referred to the
introductory literature \cite{Montvay:1994cy,Gupta:1997nd,Smit:2002ug,Rothe:2005nw,DeGrand:2006zz,Gattringer:2010zz,Fodor:2012gf}.

In more general terms however, it should be clear at this point that
independent of the specifics of the update algorithm the numerically
difficult part of lattice QCD are the fermion fields. The change in
the gauge action upon modification of a single link e.g. is easy to
compute. One only needs to compute the adjacent plaquettes i.e. take
some products and traces of $3\times 3$ matrices. Even the gauge
action of the entire system can be computed by $\mathcal{O}(N)$ such
operations, where $N$ is the number of lattice points. Typical values
of $N$ used today range from $N\sim 10^6$ for a $32^4$ lattice to
$N\sim 10^8$ for a $96^4$ lattice, which results in a manageable
computational effort.

For the fermion fields on the other hand one needs to compute
determinants or functions like inverse square roots (see
eq.~\ref{eq:psf}) of matrices that in the case of staggered fermions
are $3N\times 3N$ and $12N\times 12N$ for other fermion
formulations. It is not possible to significantly reduce the number of
lattice points $N$ either. Lattices have to be large enough in size to
accomodate the relevant physics - typically at least a few fm in
each direction. They have to be fine enough on the other hand that the
lattice spacing itself is firmly within the perturbative regime so
that the nonperturbative physics is reliably captures. Additionally,
carrying out the continuum limit requires having a range of lattice
spacings. Typically, they are chosen to be in a range $a\sim 0.05-0.1
\text{fm}$.

As a result, the computational cost of generating lattice QCD
ensembles arises almost entirely from the fermions. It has therefore
been customary in the early days of lattice QCD to eliminate this cost
entirely by a mean field approximation. This can be achieved by simply
ignoring the fermion determinant $\det D[U]$ in the path integral
eq.~\ref{eq:pfint}. In the lattice literature, this is referred to as
the quenched approximation. Although it has worked surprisingly well
in some cases, it has been largely phased out nowadays due to its
uncontrolled nature.

\subsection{Staggered rooting}

For staggered fermions, the numerical evaluation of the path integral
poses one additional problem. Since doubler fermions were not entirely
eliminated but merely the fermion multiplicity reduced to $2^{D/2}$ or
$4$ in $4$ dimensions, the effective action term $-\ln\det D[U]$
describes four fermion species instead of one. Very early on it was
suggested by Marinari, Parisi and Rebbi \cite{Marinari:1981qf} that
one could just divide this effective action by $2^{D/2}$, which
corresponds to taking the ${2^{D/2}}^\text{th}$ root of the fermion
determinant. Whether this is a valid procedure has been widely
discussed in the literature since. For the free case, Adams
\cite{Adams:2004mf} has proven that the procedure is valid for any
$m>0$. He could show that the free staggered operator can be
decomposed into 4 single flavour operators that have an identical
spectrum.

\begin{figure}[htb]
\centerline{
\includegraphics[width=12cm]{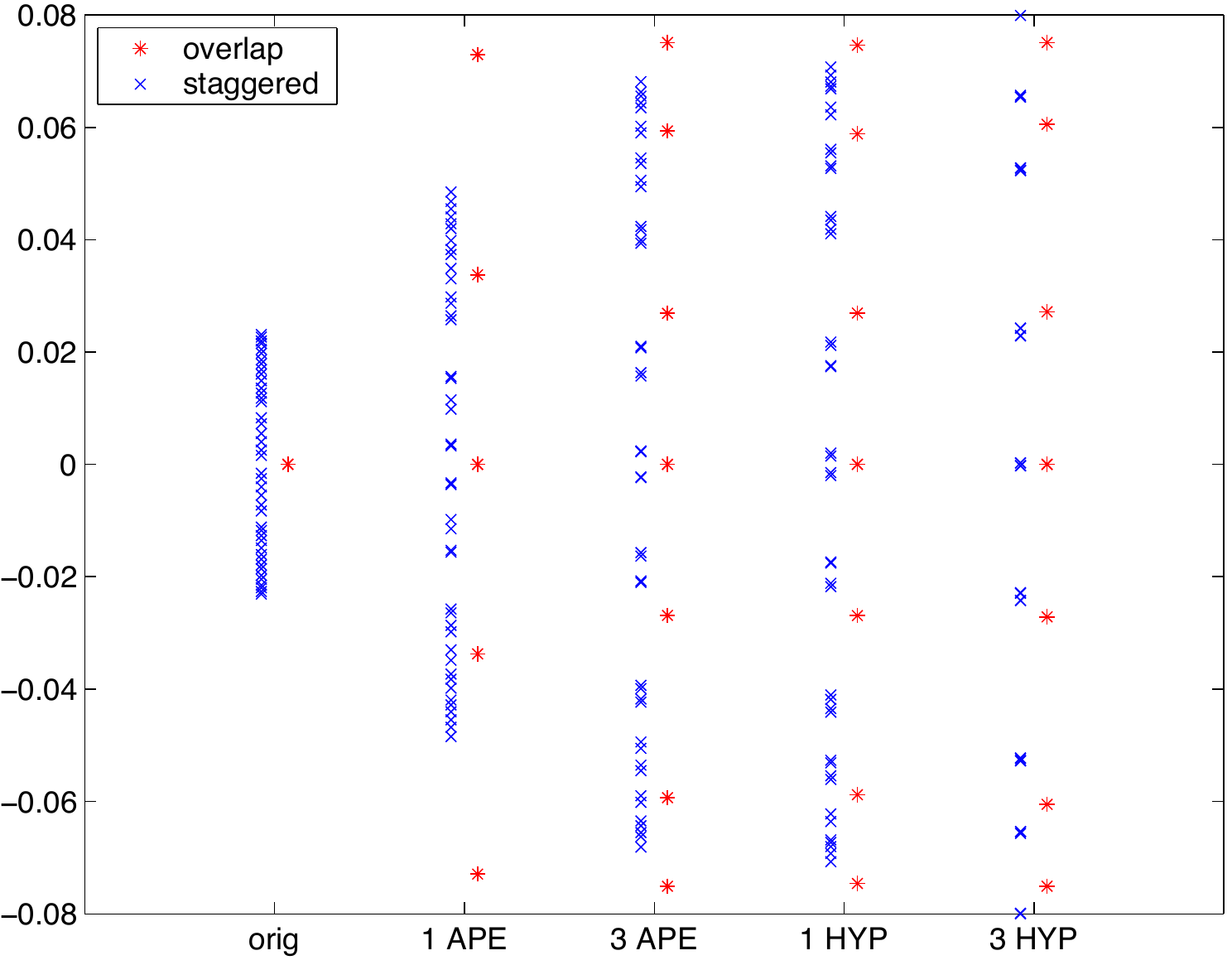}}
\caption{Comparison of low lying eigenmodes of the staggered and
  overlap operators (after chiral rotation eq.~\ref{eq:gwtild}) with
  different smearing levels on a single gauge configuration according
  to \cite{Durr:2004as}. As one can see, after sufficient smearing the
  eigenmodes of the staggered operator form approximate quadruples
  that correspond to a single overlap eigenmode up to a
  renormalization factor.}
\label{fig:stagovev}
\end{figure}

In the interacting case, it is instructive to again look at the
eigenmode spectrum of the Dirac operator. In fig.~\ref{fig:stagovev}
the physically relevant part of the eigenmode spectrum of the
staggered operator is plotted on a single gauge configuration and for
different smearing levels. As a comparison the corresponding
eigenmodes of the fully chirally symmetric overlap operators are
plotted, where the field transformation eq.~\ref{eq:gwtild} to
continuum chirality has already been performed so that the eigenmodes
lie along the imaginary axis. While at low smearing level there seems
to be no resemblance whatsoever between the spectra, one can see at
high smearing levels that an approximate 4-to-1 correspondence pattern
emerges between staggered and overlap eigenmodes (up to a
renormalization factor). This seems to suggest, that in the continuum
limit the staggered fermion determinant may indeed decompose into 4
degenerate single flavour determinants and that there are only small
corrections at finite lattice spacing if one properly suppresses the
coupling between the flavours. More evidence for this point of view is
presented in \cite{Durr:2004as,Durr:2004ta}.

\begin{figure}[htb]
\centerline{
\includegraphics[width=12cm]{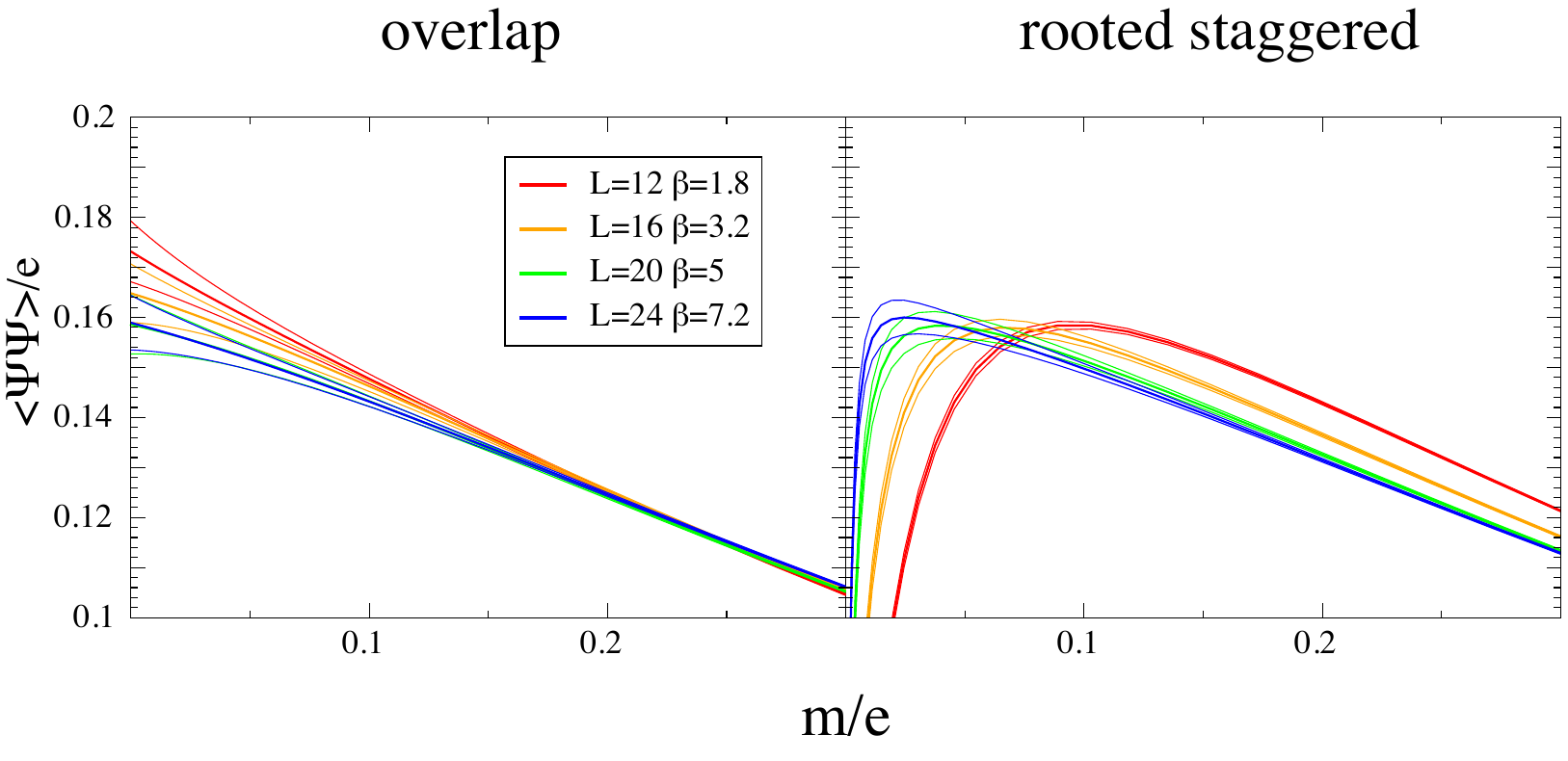}}
\caption{Chiral condensate of the 1-flavour Schwinger model for
  overlap and rooted staggered fermions versus fermion mass for
  different lattice spacings from \cite{Durr:2004ta}. While for
  overlap fermions the continuum and chiral limits commute, one has to
first perform the continuum limit at large enough mass for staggered
fermions before going to the chiral limit.}
\label{fig:stagovcond}
\end{figure}

There is one caveat to this argument however. While the
near-degeneracy may be good, it is not exact at finite lattice spacing
in the interacting theory. And since fermionic lattice observables
generically involve the inverse of the fermion matrix
(cf. eq.~\ref{eq:pfint}), there is potentially a huge difference
between an approximate and a true zero mode if the mass is small. An
observable that is especially sensitive to this effect, the one
flavour chiral condensate in the Schwinger model, is plotted in
fig.~\ref{fig:stagovcond}. As one can clearly see, the behaviour of
staggered and overlap fermions, while similar at high masses, is
dramatically different at low masses. Specifically, the continuum
limit at zero mass of the staggered theory is wrong. One does however
obtain the correct $m=0$ result with staggered fermions when the
continuum limit is first taken at finite mass and the chiral limit
afterwards. This subtle behaviour has to be kept in mind when dealing
with staggered fermions. When one avoids this dangerous region
however, there is substantial evidence that rooted staggered fermions
produce correct results
\cite{Bernard:2006zw,Bernard:2007qf,Prelovsek:2005rf,Durr:2004as,Durr:2003xs,Davies:2003ik,Davies:2004hc,Aubin:2004fs,Follana:2007uv,Bazavov:2010ru,Bazavov:2009bb,Bernard:2006ee,Giedt:2006ib,Bernard:2007ma,Shamir:2006nj,Bernard:2004ab,Durr:2006ze}
although there are some dissenting opinions
\cite{Creutz:2007yg,Creutz:2007pr}.

\subsection{Some important crosschecks}
\label{sect:checks}

\begin{figure}[htb]
\centerline{
\includegraphics[width=6cm]{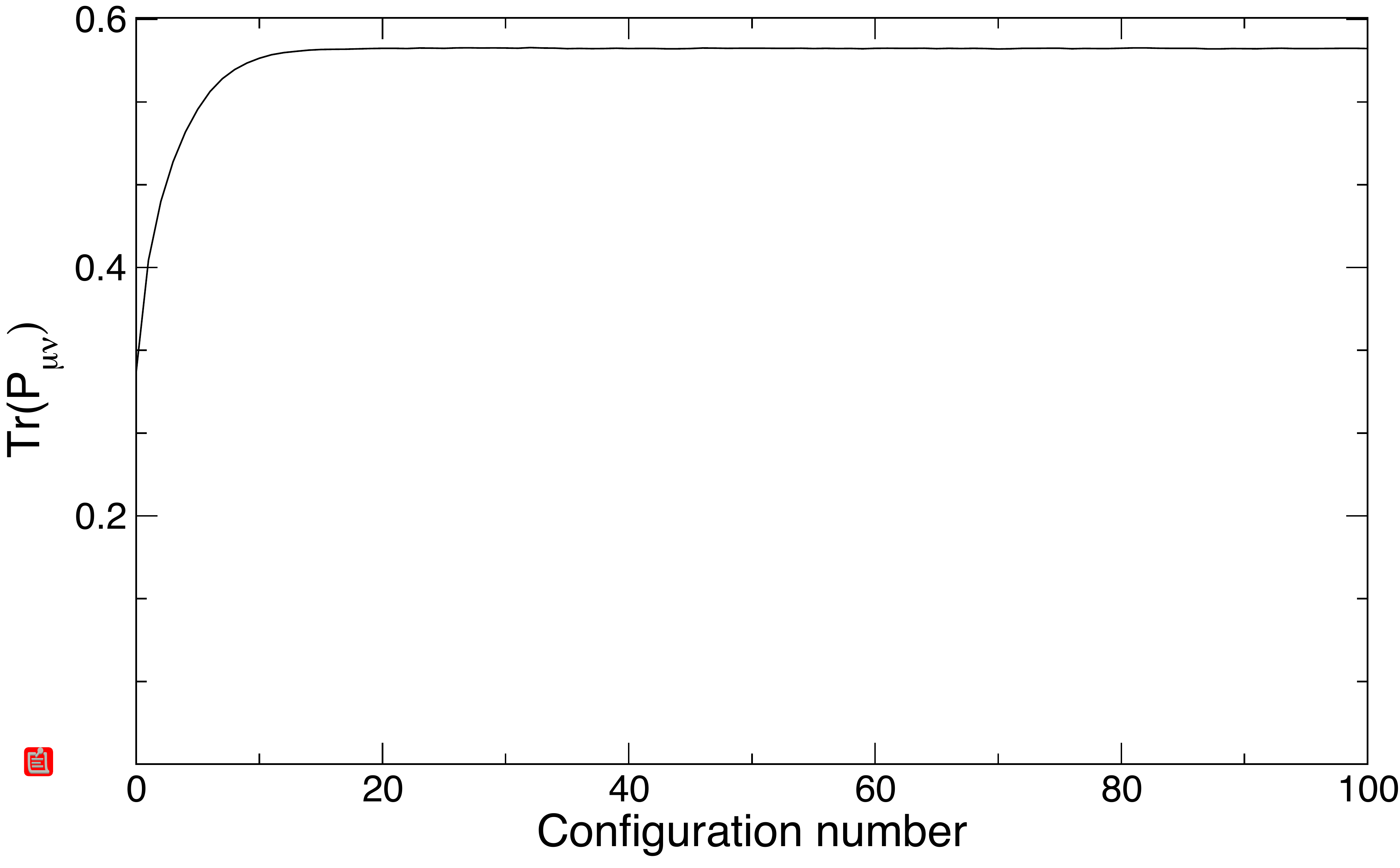}}
\caption{Illustration of the evolution of the plaquette in simulation time.}
\label{fig:plts}
\end{figure}

As already mentioned in section~\ref{sect:num}, it is important to
monitor the behaviour of the update algorithm to ensure a correct
sampling of configuration space. The most straightforward technique is
to monitor a simple observable such as e.g. the
plaquette. Fig.~\ref{fig:plts} shows a simple example where the
initial thermalization is clearly visible.

\begin{figure}[htb]
\centerline{
\includegraphics[width=12cm]{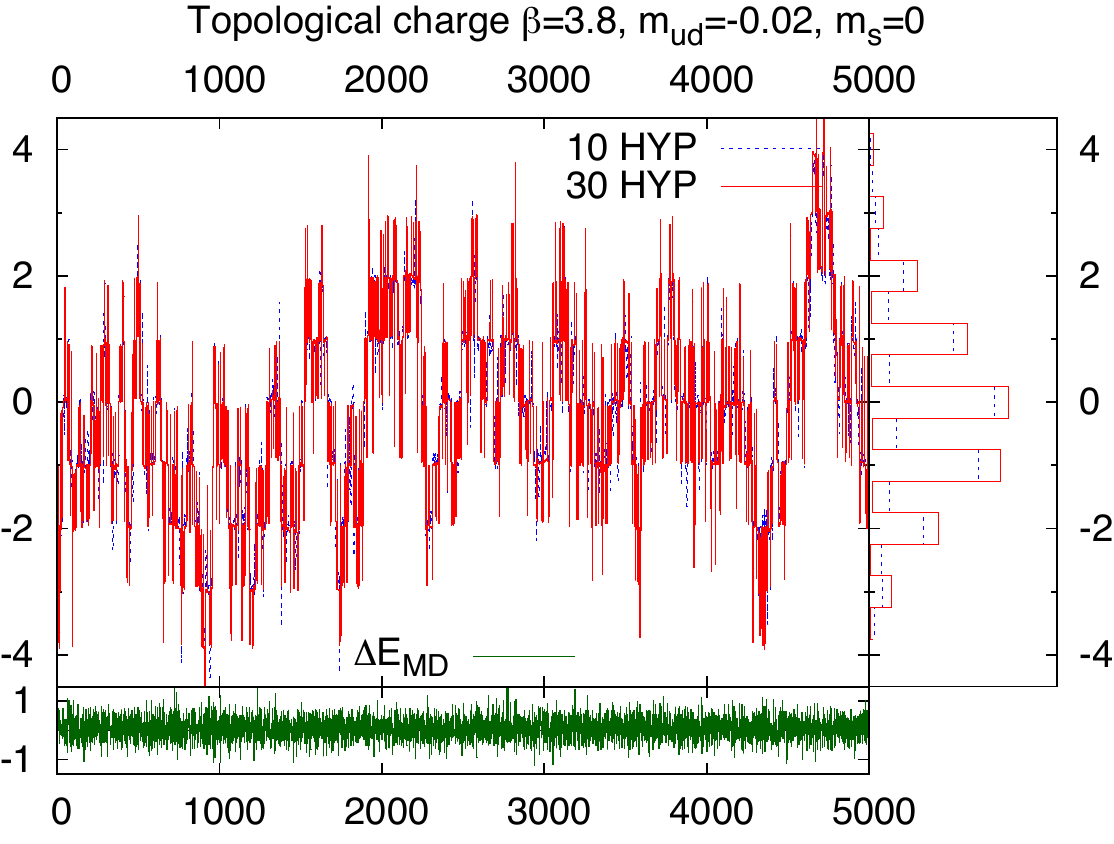}}
\caption{History and histogram of topological charge from a recent
  large scale simulation \cite{Durr:2010aw}. The relevant
  autocorrelation time of $|Q|$ in this instance was determined to be
  $27.3\pm7.4$}
\label{fig:khist}
\end{figure}

The plaquette however is a rather well behaved observable that
thermalizes and decorrelates relatively quickly. This property is
connected with the plaquette being a very local observable and one
might underestimate the true autocorrelation and thermalization time
of the system by looking at it exclusively. It has therefore become
customary to monitor other, more global observables of the system,
too. One example of such an observable that is in wide use today is
the topological charge. It is a global property of the gauge field and
the fermion operator that can only change in integer steps on typical
hypertorrodial geometries. In the continuum the
topological charge is defined as
\begin{equation}
Q=\frac{g^2}{32\pi}\int d^4x G^*_{\mu\nu}G_{\mu\nu}
\end{equation}
which can be easily generalized on the lattice \cite{Durr:2006ky}. Its
relation via the index theorem \cite{Atiyah:1967ih} to the zero modes
of a chiral fermion operator allows for an alternative extraction
method which however is much more demanding computationally. An
example from a recent work is displayed in fig.~\ref{fig:khist}.

\begin{figure}[htb]
\centerline{
\includegraphics[width=12cm]{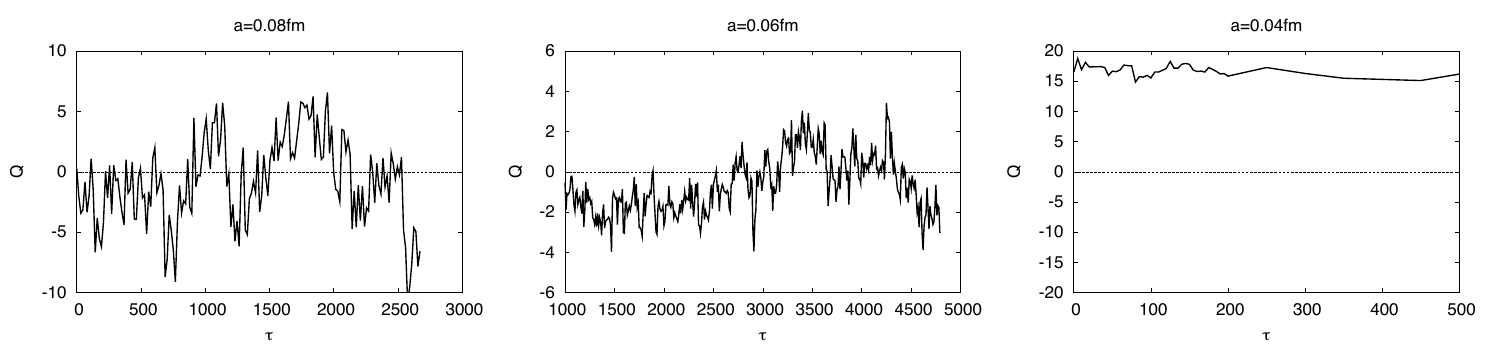}}
\caption{History of topological charge for for different lattice
  spacings from \cite{Schaefer:2009xx}.}
\label{fig:tophop}
\end{figure}

Studying the topological charge autocorrelation as one goes to the
continuum limit, Schaefer, Sommer and Virotta \cite{Schaefer:2009xx}
have pointed out a potentially severe problem. Their results are
displayed in fig.~\ref{fig:tophop}. As one can clearly see, the
topological charge does not tunnel anymore at the finest lattice
investigated for the entire Markov chain but instead remains frozen at
a certain value. The details of this behaviour are of course dependent
on the specific action and algorithm used, but there is a physical
cause for it which is again maid clear by looking at the eigenmode
spectrum of the fermion operator.

Remember that for Wilson type operators topological modes lie
along the real axis. The number of modes in the physical branch
determines the topological sector (cf. fig.~\ref{fig:smearsp}). To
change topological sector these modes therefore need to either appear
or disappear. The only possibility for this to happen is however that
either a pair of complex conjugate modes approaches the real axis, mix
and split into a physical and doubler chiral mode or the other way
round. As one approaches the continuum however, this is not easily
possible in a continuos manner. Chiral symmetry is restored and there
is a gap developing between the physical and the doubler branches. The
fermion operator will therefore have to develop a discontinuity in the
underlying gauge field at the boundary of a topological sector.

This behaviour should therefore be more pronounced the better the
fermions discretization realizes chiral symmetry. In fact, it has
been observed earlier that for Ginsparg-Wilson fermions changing the
topology is a far greater challenge already on relatively coarse
lattices \cite{Fodor:2003bh,DeGrand:2004nq,Cundy:2005pi}.

Several suggestions have been made over the years on how to deal with
this problem. For Ginsparg-Wilson fermions, special update algorithms
have been proposed \cite{Egri:2005cx}. An alternative point of view is
that fixing a topological sector is only a finite volume effect that
can be corrected for \cite{Hashimoto:2006rb}. Ultimately, for large
enough volumes, subvolumes will decorrelate and reproduce the correct
fluctuation pattern even if the overall topological charge is
fixed. Along similar lines, it was suggested to use open boundary
conditions \cite{Luscher:2011kk} for which topological charge is not
an integer. Here too the open boundary results in additional finite
volume effects that can ultimately be eliminated by a proper infinite
volume limit. For current lattice calculations however, the
potentially long autocorrelation times in the continuum limit are not
a limiting factor yet. For fermion discretizations that do not have
exact chiral symmetry, these effects become relevant only for lattices
finer than $a\sim0.05 \text{fm}$.

\begin{figure}[htb]
\centerline{
\includegraphics[width=12cm]{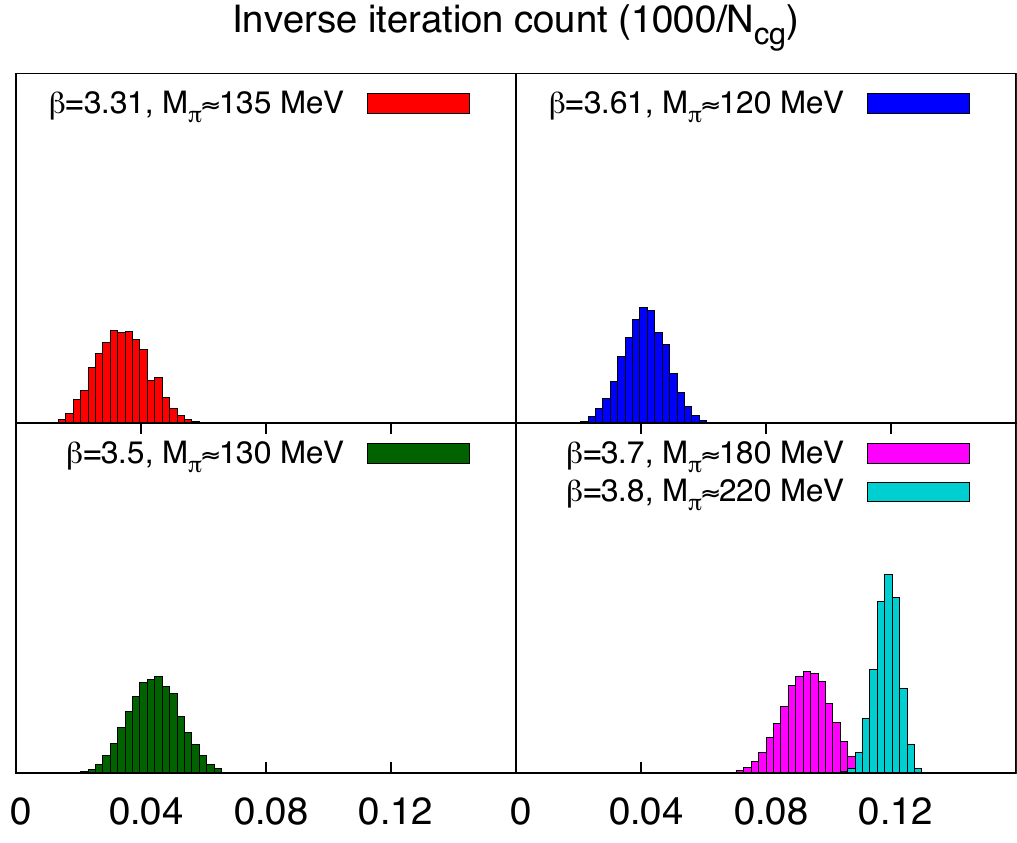}}
\caption{Histogram of the inverse iteration count in the inversion of
  the fermion operator from a recent large scale simulation
  \cite{Durr:2010aw}.}
\label{fig:inviter}
\end{figure}

For Wilson-type fermions one needs to perform another crucial
check. Because of the additive mass renormalization, the fermion
determinant was not guaranteed to be positive definite. In terms of
the eigenmode spectrum, a negative fermion determinant can only appear
when an odd number of real modes is negative after the additive mass
renormalization has been applied (cf. fig.~\ref{fig:smearsp}). In
principle, one should therefore monitor the real modes of the fermion
operator. Since this is computationally expensive, another quantity is
typically monitored that is closely related but also directly
obtainable from the simulation itself. Remember, that in every update
step a fermion matrix has to be inverted (on a source vector) in order
to construct the pseudofermion action eq.~\ref{eq:psf}. This inversion
is performed iteratively and the numer of iterations is very sensitive
to the condition number of the matrix, i.e., the ratio of its largest
to smallest eigenvalue. If during the classical evolution in the
pseudo-time a real mode would come close to the origin, it would
immediately be recognizable as an increase in the iteration count of
the inverter. In the limiting case of a zero mode, the inverter would
not converge and the corresponding gauge configuration is called
exceptional. One can therefore plot the iteration count or, as in
fig.~\ref{fig:inviter} the histogram of the inverse iteration
count. If the tail of the distribution has a safe distance from 0, one
can conclude that no real mode has crossed over to the negative side
and therefore the fermion determinant is indeed positive.

Some further checks of algorithm stability and efficiency that are
routinely done include monitoring the acceptance rate and the forces
in the classical evolution of the fermion field. Sometimes hystereses
are recorded with respect to varying fermion masses or the gauge
coupling when one suspects the proximity of an unphysical phase
transition due to lattice artefacts. In general one can say that
lattice QCD has a large set of tools for checking the integrity of the
simulation algorithms.

\section{An example calculation: hadron masses}
\label{sec:had}

\subsection{Skeleton of a lattice calculation}
\label{sect:skeleton}

With the basic techniques established, the next step is to actually
make physical predictions using lattice QCD. We will look at the
computation of ground state light hadron masses
\cite{Durr:2008zz,Borsanyi:2014jba} as a prototypical example. In
principle, the strategy is straightforward: we want to go to the
physical point and read off the target observables. But the physical
point can of course never be reached directly. One always has to
extrapolate to the continuum limit and to infinite volume. In
addition, the physical values of the parameters of the lattice QCD
action, namely the gauge coupling and the quark masses, are
unknown. Hence it is necessary to define the physical point through a
set of quantities that can be measured both experimentally and on the
lattice and to interpolate or extrapolate lattice results to the
physical point thus defined.

Typical lattice QCD calculations currently include two flavours of
degenerate light quarks, a strange quark and possibly a charm
quark. In lattice terminology, these setups are referred to as $2+1$
resp. $2+1+1$. Isospin splitting is usually treated as a perturbation
while the effects of b and t quarks can generally be ignored at the
current level of precision. In such a setup, each lattice calculation
has 3 or 4 parameters: the gauge coupling $g$ and the masses of the
quarks $m_{ud}$, $m_s$ and possibly $m_c$. Through dimensional
transmutation, the gauge coupling is closely linked to the scale of
the theory, i.e. the lattice spacing. Light and strange quark masses
on the other hand are related to the masses of the pseudoscalar
mesons. To leading order, this relation reads \cite{GellMann:1968rz}
\begin{equation}
\label{eq:gmor}
M_\pi^2\propto 2 m_{ud}
\qquad
M_K^2\propto m_s+m_{ud}
\end{equation}
One can therefore characterise an ensemble of gauge configurations by
lattice spacing and the observable quantities $M_\pi$ and
$\sqrt{2M_K^2-M_\pi^2}$ instead of $g$, $m_{ud}$ and $m_s$. In
addition, the size of the lattice is a relevant parameter.

\begin{figure}[htb]
\includegraphics[width=\textwidth]{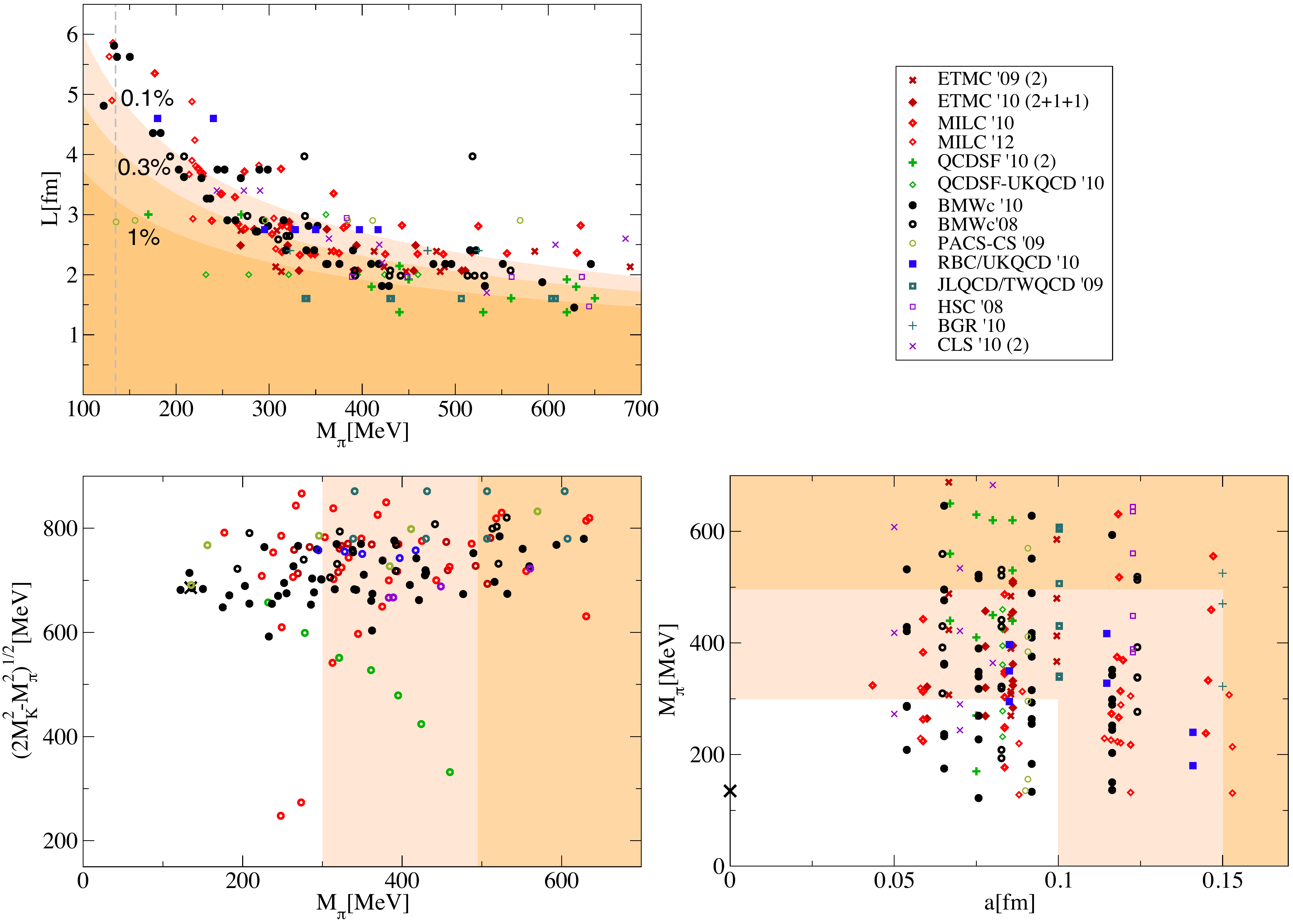}
\caption{\label{fig:landscape} Plot of simulation parameters of some
  recent lattice calculations with dynamical fermions following
  \cite{Hoelbling:2011kk}. Data are from ETMC'09(2)
  \cite{Blossier:2009bx}, ETMC'10(2+1+1) \cite{Baron:2011sf}, MILC'10
  \cite{Bazavov:2009bb}, QCDSF'10(2) \cite{Schierholz:2010xx},
  QCDSF-UKQCD'10 \cite{Bietenholz:2010si}, BMWc'08 \cite{Durr:2008zz},
  BMWc'10 \cite{Durr:2010aw}, PACS-CS'09
  \cite{Aoki:2009ix,Aoki:2008sm}, RBC-UKQCD'10
  \cite{Aoki:2010dy,Mawhinney:2010xx}, JLQCD/TWQCD'09
  \cite{Noaki:2009sk}, HSC'10 \cite{Lin:2008pr}, BGR'10(2)
  \cite{Engel:2010my} and CLS'10(2) \cite{Brandt:2010ed}. The cross in
  the lower left and right hand plots denote the physical point. The
  percent marks in the upper left hand plot indicate the estimated
  relative finite volume corrections on the pion mass according to
  \cite{Colangelo:2005gd}.}
\end{figure}

Fig.~\ref{fig:landscape} displays these simulation parameters for some
recent lattice QCD calculations. Since it is substantially less
demanding numerically, a lot of calculations are still performed at
large pion masses, relatively coarse lattice spacings and small
volumes. One can however reach physical pion masses today at multiple
lattice spacings and volumes as large as $(6 \text{fm})^3$, which
allows a controlled continuum and infinite volume extrapolation as
well as an interpolation to physical $M_\pi$ to be performed.

\subsection{Extraction of hadron masses}

As a first step towards a physical prediction, we need to actually
measure the hadron masses on our ensembles of gauge configurations. 
We first choose two (not necessarily different) operators $O_{1/2}$
that couple to the target hadron $|h\rangle$
\begin{equation}
\langle h|O_{1/2}|0\rangle\ne 0
\end{equation}
We then compute the correlator
\begin{equation}
G(t,0)=\langle 0|O^\dag_2(t)O_1(0)|0\rangle=
\langle 0|e^{\mathcal{H}t}O^\dag_2e^{-\mathcal{H}t}O_1|0\rangle
\end{equation}
as described in sect.~\ref{sect:obs}. Inserting a complete set of
eigenstates 
\begin{equation}
\mathbbm{1}=\sum_n\frac{1}{2E_n}|n\rangle\langle n|
\qquad
\mathcal{H}|n\rangle=E_n|n\rangle
\quad
E_0=0
\end{equation}
in the standard fashion, we obtain
\begin{equation}
\label{eq:propst}
G(t,0)=\sum_n\frac{
\langle 0|O^\dag_2|n\rangle
\langle n|O_1|0\rangle
}{2E_n}e^{-E_nt}
\end{equation}
If our target state $|h\rangle$ happens to be the ground state, we can
extract its mass $M=E_h$ by simply going to asymptotic times
\begin{equation}
\label{eq:fwc}
G(t,0)\stackrel{t\rightarrow\infty}{\longrightarrow}\frac{
\langle 0|O^\dag_2|h\rangle
\langle h|O_1|0\rangle
}{2M}e^{-Mt}
\end{equation}
and measuring the exponent in the decay of the propagator with time
separation. One can also define an effective mass
\begin{equation}
M_\text{eff}=\ln\frac{G(t,0)}{G(t+1,0)}\stackrel{t\rightarrow\infty}{\longrightarrow}M
\end{equation}
that will signal when the regime has been reached where excited state
contributions are negligible.

On a lattice of finite time extent $T$ it is of course not possible to
go to asymptotic times. In fact, due to the periodicity of the lattice
in time direction, the propagator $G(t,0)$ will typically be dominated
by backward propagating states for $t>T/2$. In fact, for $T\gg T-t$ we
can find
\begin{equation}
\label{eq:bwc}
G(t,0)\stackrel{T-t\rightarrow\infty}{\longrightarrow}\frac{
\langle 0|O^\dag_1|\tilde{h}\rangle
\langle \tilde{h}|O_2|0\rangle
}{2E_{\tilde{h}}}e^{-E_{\tilde{h}}(T-t)}
\end{equation}
where $\tilde{h}$ is the lowest energy state coupling to the adjoint
of the source operators
\begin{equation}
\langle \tilde{h}|O^\dag_{1/2}|0\rangle\ne 0
\end{equation}
In principle, there are also contributions to the propagator from
multiple windings around the time direction. Each additional winding
in forward direction e.g. gives an additional factor $e^{-MT}$, which
however only gives a tiny correction to the prefactor and is therefore
irrelevant for extracting masses.

As a typical example, the charged pion mass can be extracted using a source
operator
\begin{equation}
O_1=(\bar\psi_u\gamma_5\psi_d)_{\vec{x}}
\end{equation}
Using a sink operator $O_2$ of the same form, we obtain the observable
discussed in eq.~\ref{eq:mespropg5h}, which can be computed by
inverting the fermion matrices of the u and d $D_\text{u}$ and
$D_\text{d}$ for the source point $(0,\vec{x})$ only. One can see from
eq.~\ref{eq:mespropg5h}, that no additional inversions are required to
go to an arbitrary sink point. In fact, we can sum over all sink
points in a given time slice and thus project the final state to
$\vec{p}=0$ without any substantial additional cost. It is therefore
customary to use as a sink operator
\begin{equation}
O_2=\sum_{\vec{y}}(\bar\psi_u\gamma_5\psi_d)_{\vec{y}}
\end{equation}

There is a wealth of additional techniques to construct efficient
operators which is beyond the scope of these notes to cover and I
refer the interested reader to the introductory literature for further
details \cite{DeGrand:2006zz,Gattringer:2010zz,Fodor:2012gf}.

\begin{figure}[htb]
\centerline{
\includegraphics[width=12cm]{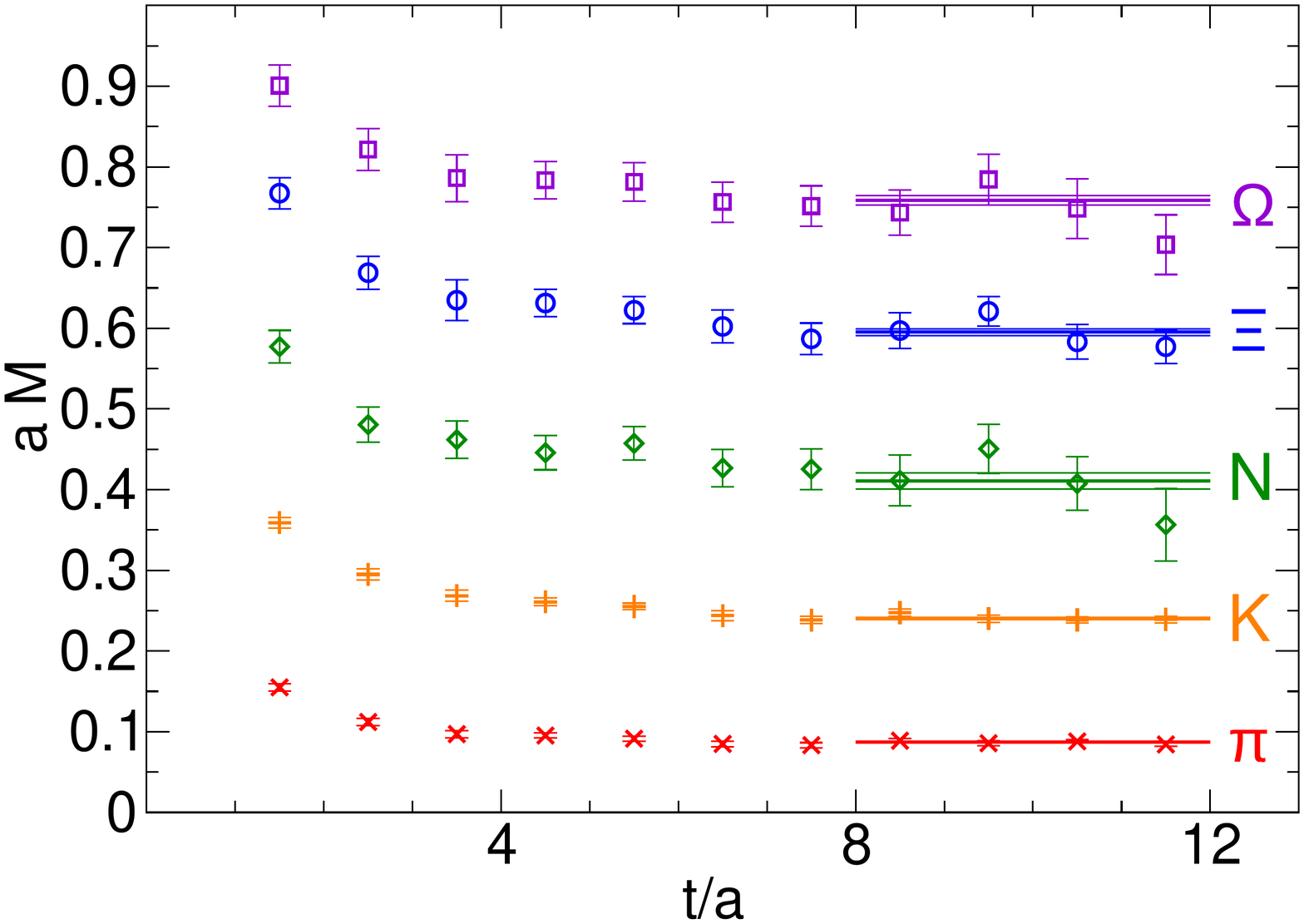}}
\caption{Plateaus of the effective mass and fit ranges for some hadron
  channels from \cite{Durr:2008zz}.}
\label{fig:platmef}
\end{figure}

For our specific example, we have $|\tilde{h}\rangle=|h\rangle$ and
also $\langle 0|O^\dag_1|\tilde{h}\rangle \langle
\tilde{h}|O_2|0\rangle=\langle 0|O^\dag_2|h\rangle \langle
h|O_1|0\rangle$ so that the prefactors as well as the masses are the
same in the forward (eq.~\ref{eq:fwc}) and backward (eq.~\ref{eq:bwc})
contributions. In a region where excited state contributions are
irrelevant we therefore obtain
\begin{equation}
\label{eq:cosh}
G(t,0)
\propto
e^{-Mt}+e^{-M(T-t)}
\propto
\cosh M(T/2-t)
\end{equation}
We can use eq.~\ref{eq:cosh} as a fit ansatz to extract $M$ from
$G(t,0)$ or solve it with respect to $M$ for two time slices to obtain
an effective mass.

Fig.~\ref{fig:platmef} gives an example of an effective mass plot and
corresponding fit ranges for several hadronic channels. As one can
see, it is not entirely clear what is an optimal fit range to
choose. It is therefore essential to perform fully correlated fits and
monitor the fit quality. It is also good practice to perform the
analysis with multiple fit ranges that seem sensible and let the
corresponding spread of the results enter the systematic error.

\begin{figure}[htb]
\centerline{
\includegraphics[width=12cm]{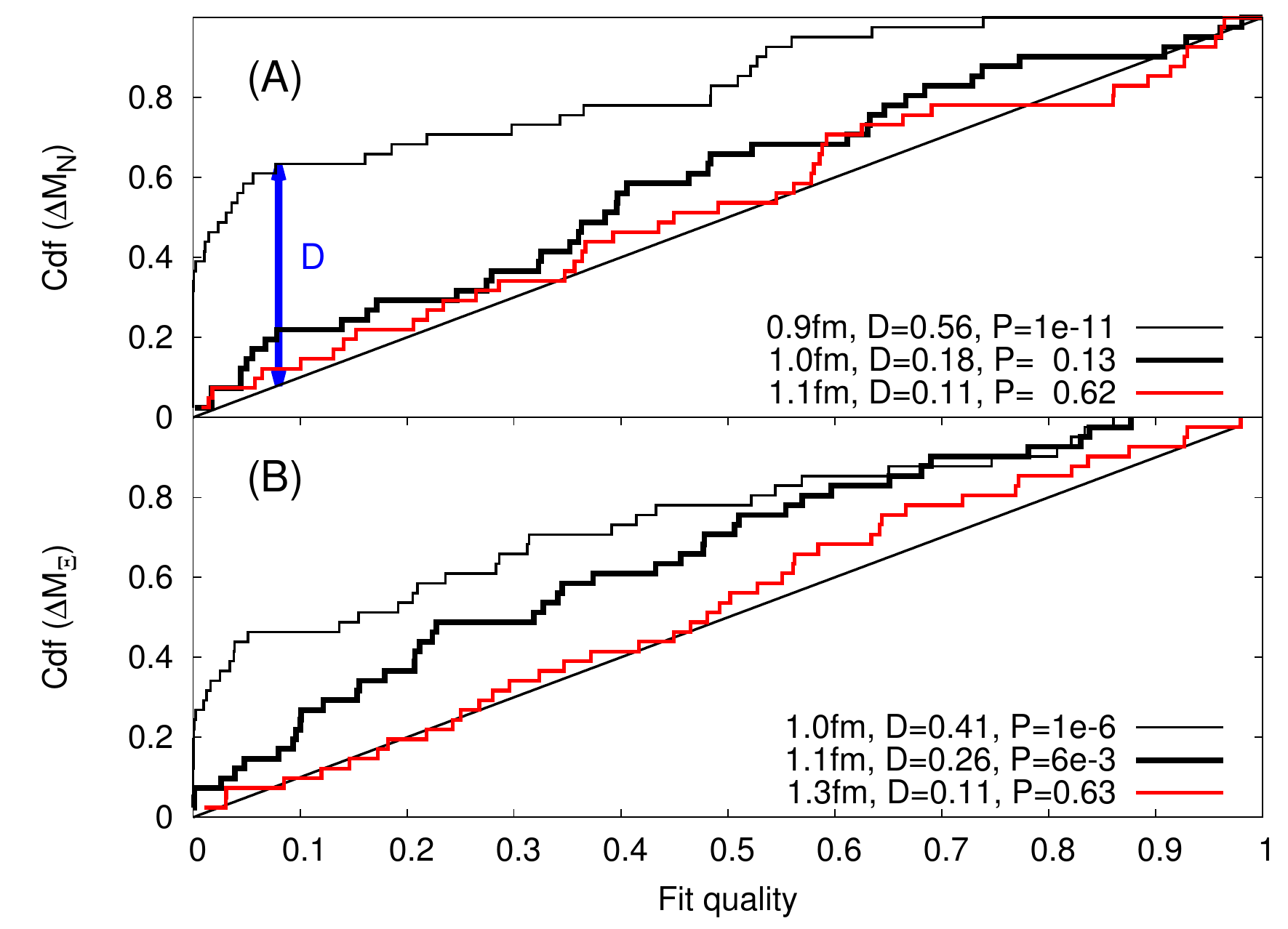}
} 
\caption{\label{fig:ks} Cumulative histogram of fit quality over 48 ensembles for different
initial times and two observables from \cite{Borsanyi:2014jba}. For
each case, the legend gives the maximum distance $D$ between expected
and measured CDF as well as the probability $p$ that they are
identical according to the Kolmogorov-Smirnov test.
}
\end{figure}

A further check for a sensible fit range is possible over a set of
ensembles. If there is no excited state contribution, the fit quality
$Q$ is expected to be randomly fluctuating between 0 and 1. One can
plot the CDF of the fit quality and check with a Kolmogorov-Smirnov
test whether it is compatible with the expected linear rise (see
fig.~\ref{fig:ks}).

Extracting excited state masses from eq.~\ref{eq:propst} is much more
difficult as the contribution of excited states decays exponentially
with separation. It is in principle possible to make an ansatz
\begin{equation}
G(t,0)
=
a_0 e^{-M_0t}+a_1 e^{-M_1t}
\end{equation}
and fit for the ground state mass $M_0$, the excited state mass $M_1$
and the two prefactors $a_0$ and $a_1$. In practice however, these
fits tend to be unstable and have a limited accuracy. It is preferable
in these cases to use a larger operator basis $O_i$  and measure the
full cross-correlator between its elements
\cite{Michael:1985ne,Luscher:1990ck}
\begin{equation}
G_{ij}(t,0)=\langle 0|O^\dag_i(t)O_j(0)|0\rangle
\end{equation}
From two time slices an effective mass matrix 
\begin{equation}
M(t,t_0)=G(t,0)G^{-1}(t_0,0)
\end{equation}
can now be computed and its eigenvalues will asymptotically give the
energies of the lowest lying states
\begin{equation}
\lambda_n\rightarrow e^{-E_n(t-t_0)}
\end{equation}
This so-called variational method may in fact be advantageous for
extracting ground state masses. The reason is that after
diagonalization the contamination of the ground state from the $N-1$
lowest excited states has been removed where $N$ is the size of the
operator basis.

For the variational method to work, it is essential that each of the
target states has good overlap with at least one operator of the
basis. This is a nontrivial requirement and the classical example is
the scalar operator
\begin{equation}
O_s=(\bar\psi_u\psi_u)
\end{equation}
Although $O_s$ should in principle couple to two-pion states, this
coupling is practically zero. In order to be sensitive to two-pion
states, four fermion operators like
\begin{equation}
O_4=(\bar\psi_u\gamma_5\psi_d) (\bar\psi_d\gamma_5\psi_u)
\end{equation}
need to be considered, too.

When extracting excited states, one also needs to keep in mind that
the relation between discrete energy levels extracted on a finite
volume lattice and the spectral density characterising an infinite
volume resonance is not straightforward. If one aims at predicting the
mass of a physical resonance, it is not sufficient to simply measure a
corresponding energy level on the lattice. We will briefly return to
this point in sect.~\ref{sect:fv}

\subsection{Scale setting}

Being able to compute hadron masses, we can now determine the
parameters of our lattice ensembles in terms of physical
quantities. As outlined in sect.~\ref{sect:skeleton}, we can use the
pseudoscalar meson masses $M_\pi$ and $M_K$ to locate the physical
light and strange quark masses. The coupling $g$ is related to the
lattice scale and we need one additional physical observable to fix
it. Setting the mass of the charm quark for cases where it is present
trivially follows along the same lines and will not be further
discussed.

The ideal scale setting observable should satisfy a few obvious
criteria. Most importantly, it should be known from experiment with a
high accuracy and it should be computable on the lattice with high
precision, too. In addition, it should not depend on quark masses
strongly. The most obvious choices are the masses of some heavy
hadrons. In the early days of lattice QCD, the mass of the $\rho$
meson was often used. This however is not an ideal choice, as the
$\rho$ is a broad resonance which makes its mass difficult to
determine precisely both for the experiment and on the lattice.

A quantity that is widely used for scale setting today is the mass of
the $\Omega$ baryon and to a lesser extent the cascade $\Xi$. Both of
them can be measured precisely in experiment and on the lattice and
both have little light quark mass dependence. Also in wide use today
are the pseudoscalar decay constants $F_\pi$ and $F_K$. While they can
easily be determined with high precision on the lattice, one has to
keep in mind that their physical value is not obtained directly from
experiment.

There is also a variety of intermediate scale setting variables that
are often used in lattice calculations today. They are not directly
related to any experimentally observable quantity but simple to
measure on the lattice. Their physical values have to be determined
initially however, which is usually done by a lattice calculation
using a scale setting observable that is experimentally accessible.

One group of scale setting observables that have been in use since the
early days of lattice QCD is based on the static quark potential. The
potential between static sources of color charge at a distance $R$ in
direction $x_i$ can be expressed in terms of a Wilson loop
$W_{0i}^{T\times R}$ (see sect.~\ref{sect:lr}) with a long extent in
time direction as
\begin{equation}
V(R)=-\lim_{T\rightarrow\infty}\frac{\ln W_{0i}^{T\times R}}{T}
\end{equation}
Historically, the large separation limit of the force or string
tension
\begin{equation}
\sigma=\lim_{R\rightarrow\infty}\frac{dV(R)}{dR}
\end{equation}
has been in wide use. More recently, the Sommer scale $r_{0/1}$
\cite{Sommer:1993ce} which is related to the force at a finite
distance
\begin{equation}
\left.R^2\frac{dV(R)}{dR}\right|_{R=r_{0/1}}=1.65/1
\end{equation}
has become a standard scale setting observable.

Even more recently, scale setting observables based on the gradient flow,
an infinitesimal form of the gauge field smearing procedure, have been
suggested \cite{Luscher:2010iy,Borsanyi:2012zs}. For a generic field
theory with fields $\phi$ and action $S$, the gradient flow of the
field $\phi$ is defined by
\begin{equation}
\label{eq:gflow}
  \dot{\phi}(x):=\frac{\partial\phi(x)}{\partial t}=-\frac{\delta S[\phi,\partial_\mu\phi]}{\delta\phi(x)}
\end{equation}
in a flow time $t$. Obviously, the field is driven towards the
classical solution for large flow time (see also
\cite{Narayanan:2006rf}). Applying this generic concept to a gauge
theory with Wilson plaquette action (eq.~\ref{eq:wga}), one obtains an
infinitesimal form of APE smearing (eq.~\ref{eq:apesmear}). A scale
can now be defined by integrating the flow equation to obtain the
smeared gauge fields $G_{\mu\nu}(t)$ at a finite flow time $t$ and
demanding that 
\begin{equation}
t_0^2\langle E(t_0)\rangle=0.3
\qquad
E(t)=-\frac{\Tr G_{\mu\nu}^2(t)}{4}
\end{equation}
According to eq.~\ref{eq:spread}, the effective smearing radius at
flow time $t$ is given by $\sqrt{8 t}$ and we can use $t_0$ to define
a lattice scale. An alternative method is to use $w_0$ defined via
\begin{equation}
\left. t\frac{d}{dt}\left(t^2\langle E(t_0)\rangle\right)\right|_{t=w_0}=0.3
\end{equation}

Both the static quark potential and the gradient flow are purely
gluonic quantities, which are much easier to measure than fermionic
ones. In addition, the gradient flow method does not require fitting
any data and is therefore very straightforward to implement.

\subsection{Finite volume effects}
\label{sect:fv}

As a last step before we can extrapolate the lattice results to the
physical point, we need to look at the effect of the finite lattice
volume. The good news is that for masses of hadrons that do not decay
via the strong interaction, QCD finite volume effects are typically
small. The reason for this is that QCD is a theory with a mass gap. A
hadron in a finite box will be affected by mirror charge effects,
i.e. it will interact with itself over a distance $L$, where L is the
spatial size of the box. Due to the mass gap of QCD, this interaction
will however be exponentially suppressed in the lightest particle
mass. Therefore, one generically expects finite volume effects to be
proportional to $e^{-M_\pi L}$.

\begin{figure}[htb]
\centerline{
\includegraphics[width=6cm]{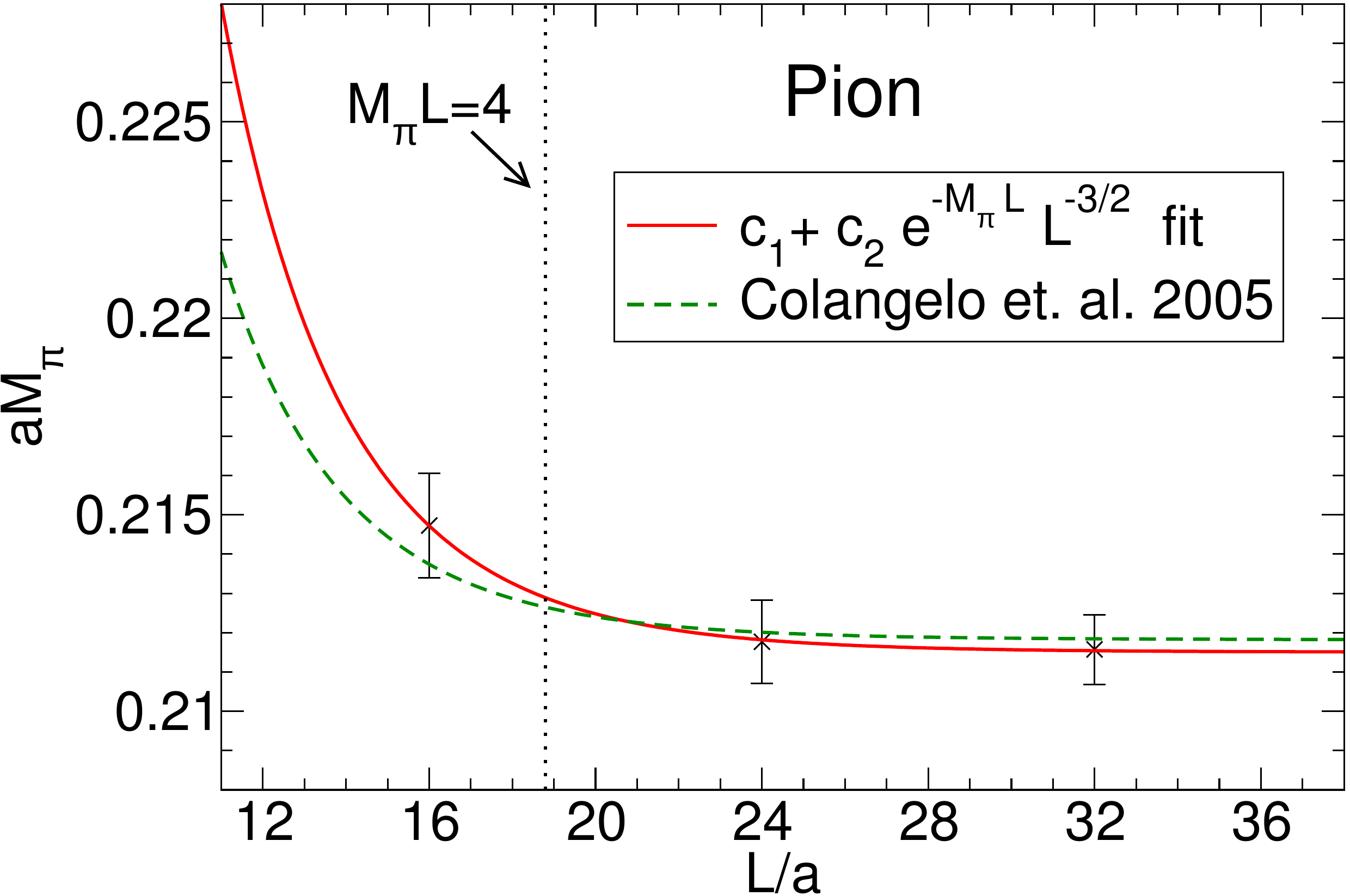}
\quad
\includegraphics[width=6cm]{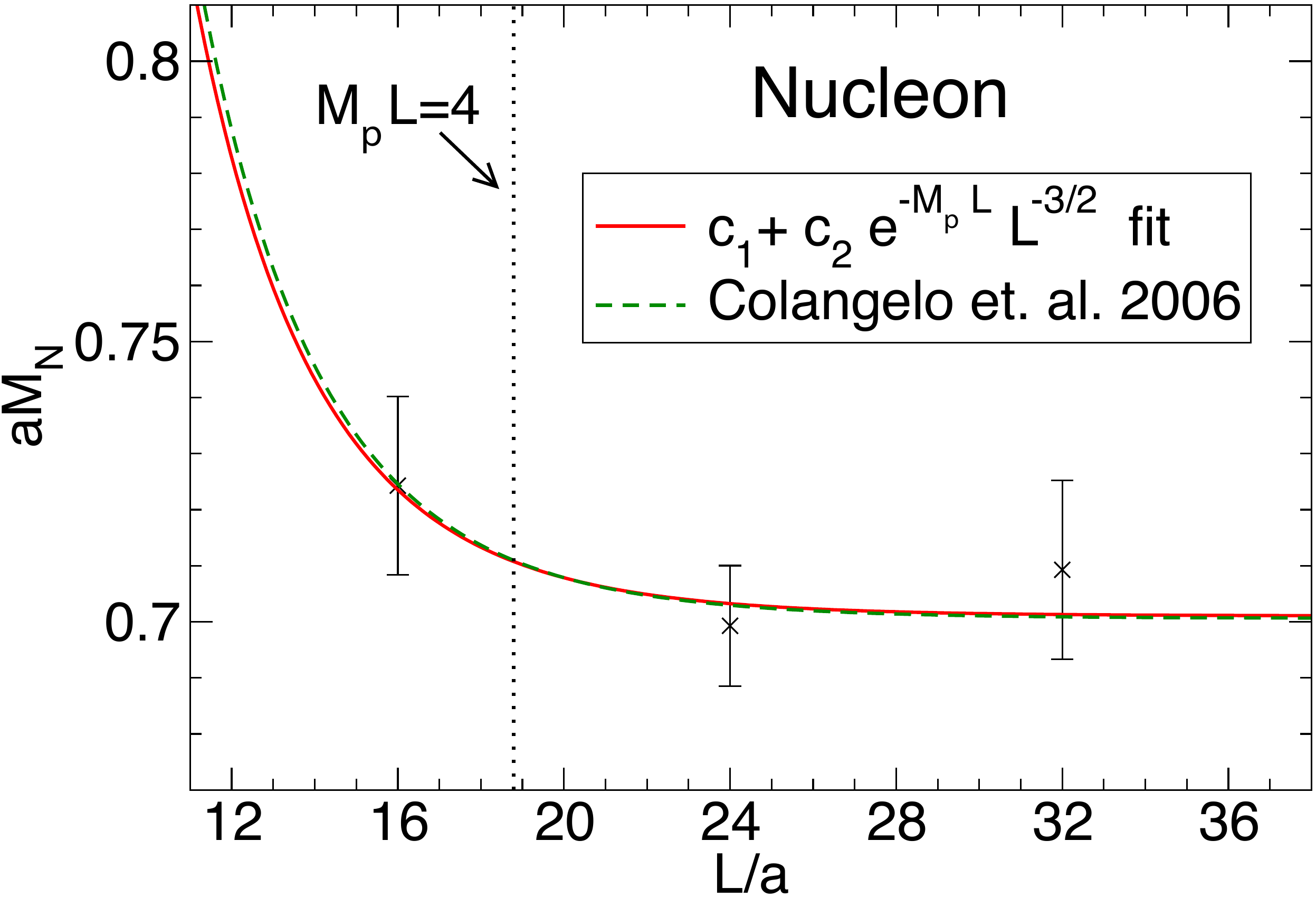}}
\caption{Pion (left panel) and nucleon (right panel) mass in lattice
  units versus lattice size. Lattice data \cite{Durr:2008zz} agree
  very well with the theoretical prediction
  \cite{Colangelo:2005gd,Colangelo:2005cg}.}
\label{fig:fv}
\end{figure}

Corrections to this leading order behaviour for mesons
\cite{Luscher:1985dn,Gasser:1986vb,Gasser:1987ah,Gasser:1987zq,Colangelo:2003hf,Colangelo:2005gd}
and, to a lesser extent, baryons
\cite{Colangelo:2005cg,Colangelo:2010ba} have been computed. As
demonstrated in fig.~\ref{fig:fv}, they describe lattice data very
well. As a rule of thumb, lattices with $M_\pi L\ge 4$ generate small
finite volume corrections for non-resonant particle masses where the
condition can be somewhat relaxed for lower $M_\pi$.

\begin{figure}[htb]
\centerline{
\includegraphics[width=10cm]{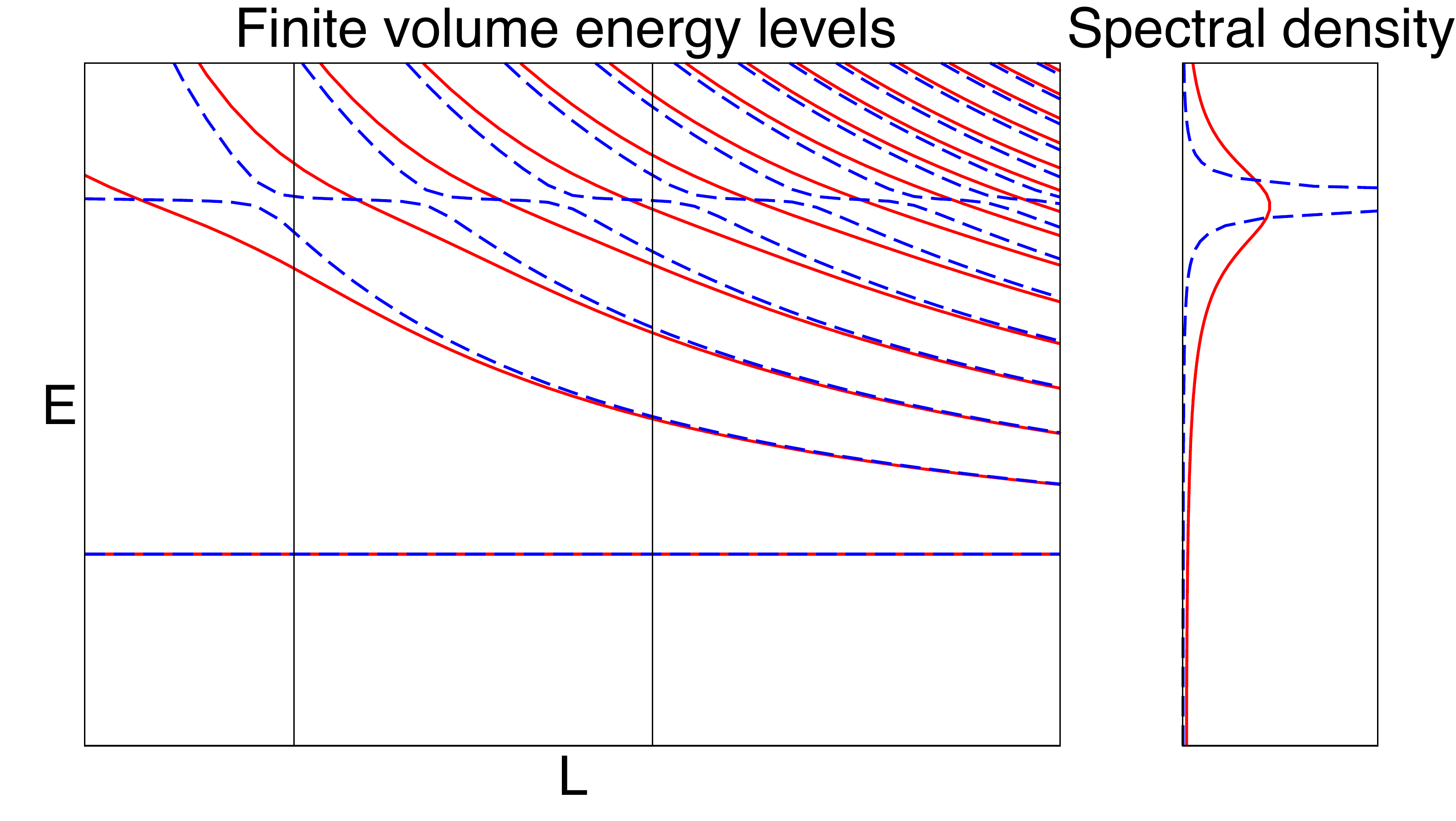}}
\caption{Illustration of the infinite volume energy density (right
  panel) and the corresponding finite volume energy levels (left
  panel) plotted versus box size $L$ for a narrow (dashed line) and a
  broad (full line) resonance. The physical $\rho$ resonance
would be much closer to the broad resonance case.}
\label{fig:fvr}
\end{figure}

For resonances, finite volume effects are more pronounced however. In
the continuum, a resonance is characterized by an increase in the
spectral density (see fig.~\ref{fig:fvr}). In finite volume however,
there are only discrete energy levels that correspond to some linear
combination of the resonance and its decay products. In order to make
an infinite volume prediction for the mass of the resonant states it
is therefore necessary to disentangle these effects.

\begin{figure}[htb]
\centerline{
\includegraphics[width=4cm]{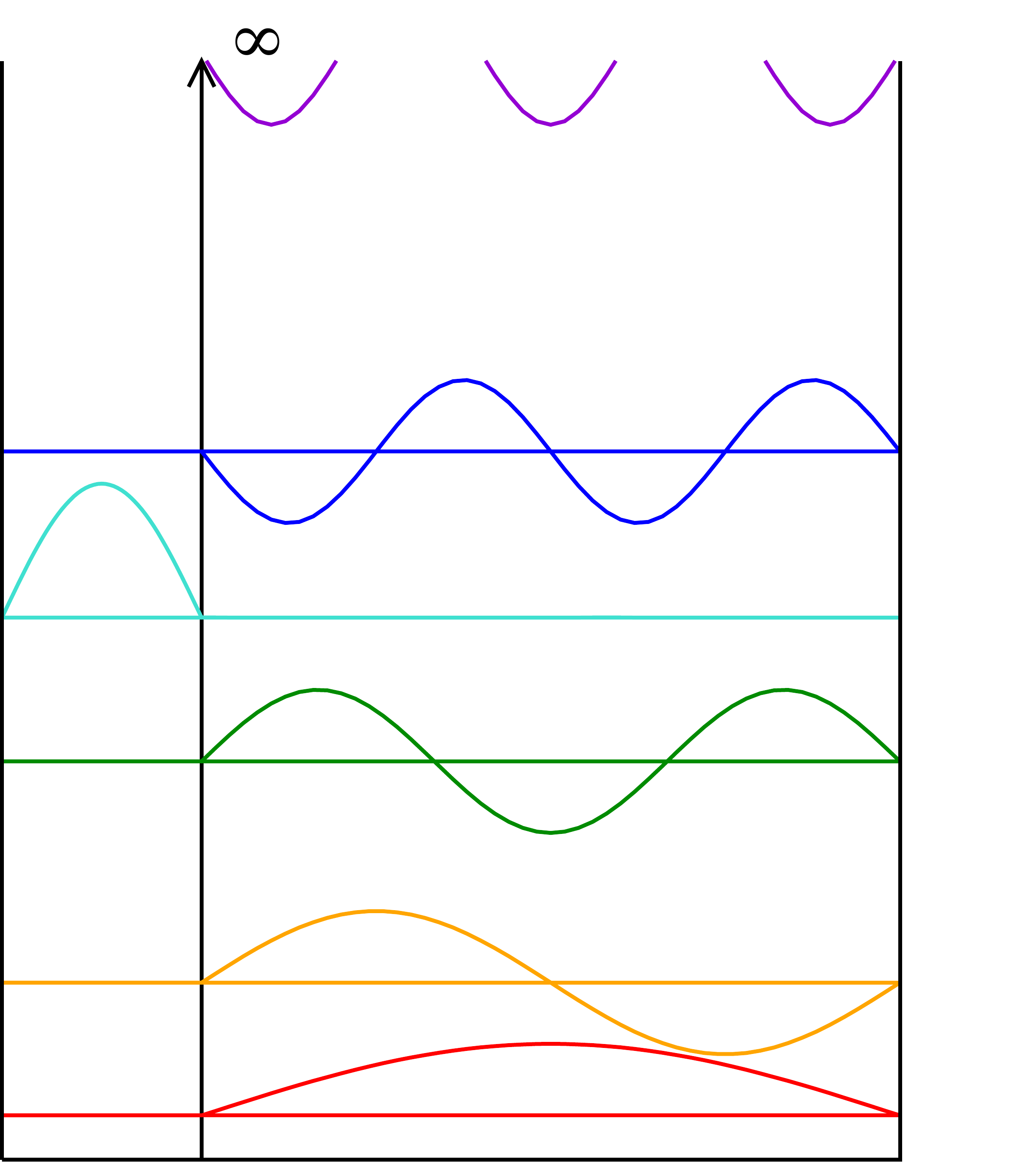}
\quad
\includegraphics[width=4cm]{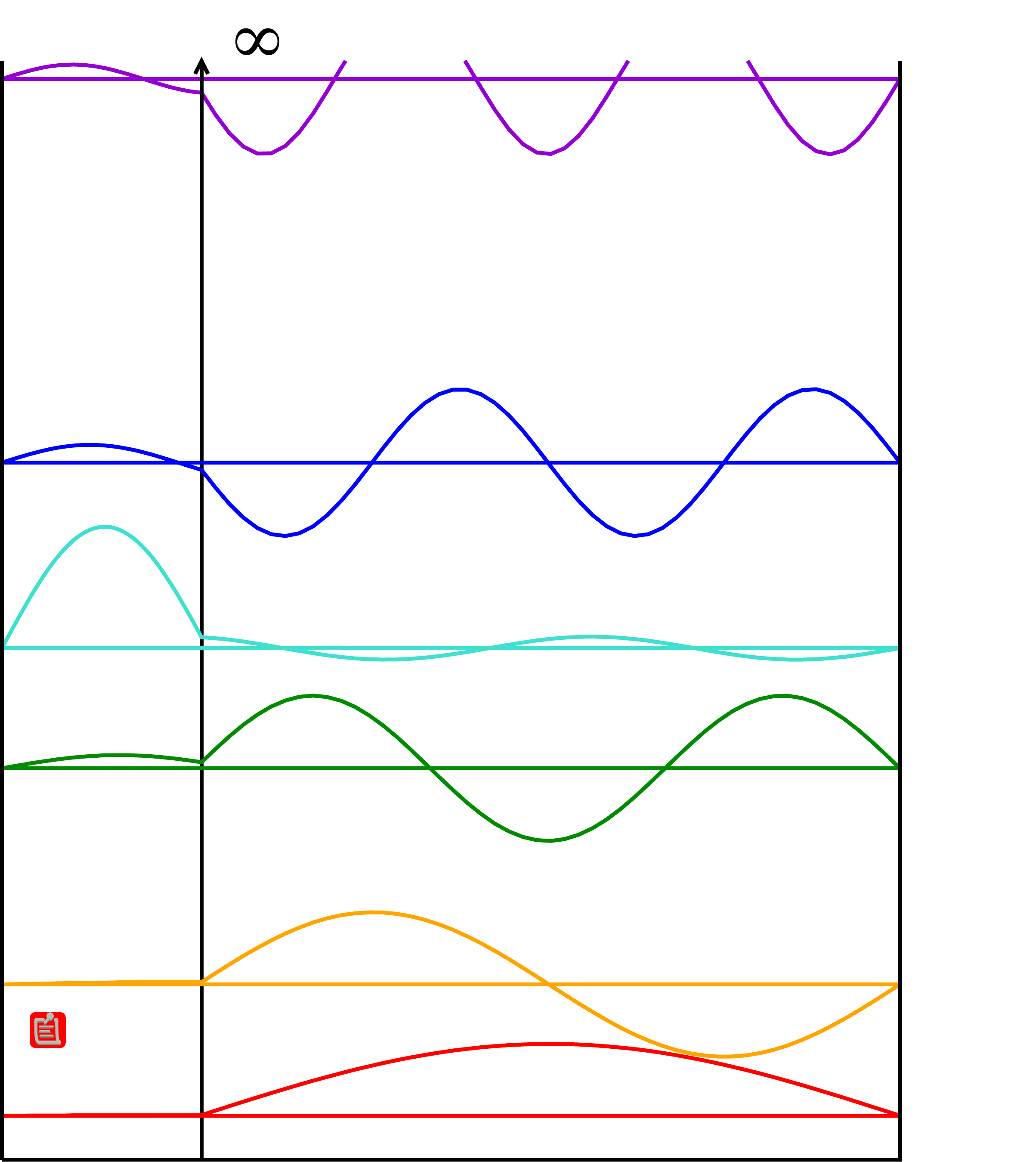}
\quad
\includegraphics[width=4cm]{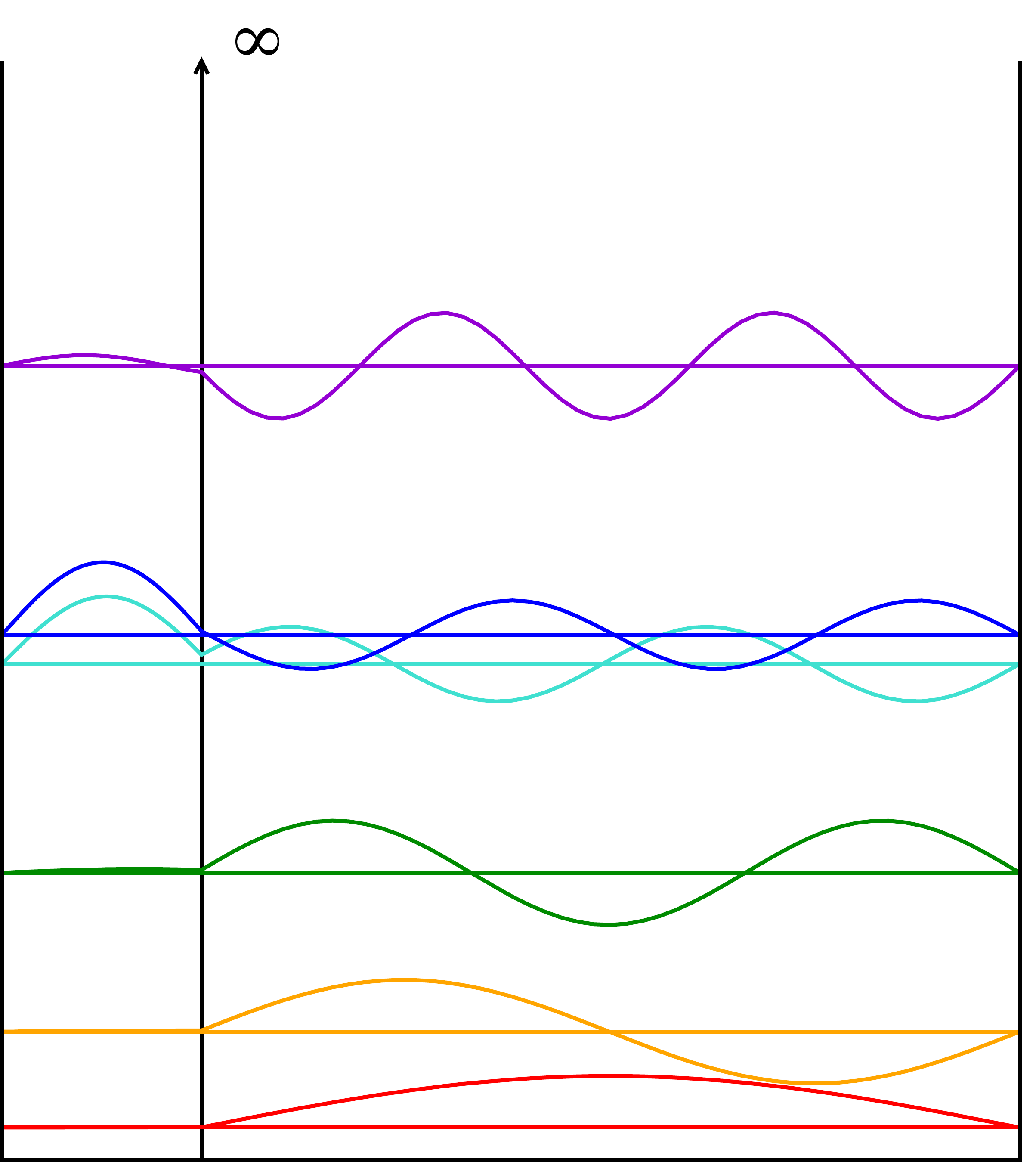}}
\caption{
Energy levels and wave functions of the double well potential model.
}
\label{fig:qmfv}
\end{figure}

A very simple quantum mechanical model can nicely illustrate the
underlying physics. We consider a particle in a double potential well
\begin{equation}
V(x)=K\delta(r)
\qquad
x\in [0,L]
\quad
0<r\ll L
\end{equation}
For an infinite separation between the two wells $K\rightarrow\infty$,
the energy levels of the small left well $x<r$, representing the
resonance, and the large right one $x>r$, representing the scattering
states, are independent as illustrated in the left panel of
fig.~\ref{fig:qmfv}. We now introduce a coupling of the resonance by
making the bareer height $K$ finite, which causes the left and right
hand side modes to mix (middle panel). Changing the volume of the box
will then change the mixing pattern (right panel) and the phenomenon
of level repulsion occurs. Depending on the box size $L$, different
modes will be in the vicinity of the uncoupled resonance energy and
overlap with the wave function of the uncoupled resonance
state. Measuring these energy levels for different box sizes, we can
infer the energy of the uncoupled resonance and the bareer height or
coupling.

In a very similar manner, it is possible to treat the problem of
resonances in QCD with scattering theory
\cite{Luscher:1986pf,Luscher:1990ux,Luscher:1991cf,Rummukainen:1995vs,Durr:2008zz}.
The energy of the uncoupled scattering states is known - in the case
of the $\rho$ it is simply the energy of a two pion system at the
relative momenta $\vec{p}$ allowed by the volume - and one can
therefore extract the mass and the coupling of the resonance by
measuring the energy levels of the system at various finite volumes.

It should also be clear from the model consideration that not all
energy levels are equally sensitive to the resonance mass. In order to
have large sensitivity, the level must be close to the resonance mass
itself.

\subsection{Extrapolating to the physical point}

We finally have all ingredients ready to extrapolate our target hadron
mass $M_X$ to the physical point and make a physical prediction. We
have measured $M_X$ on each ensemble and in addition $M_\pi$, $M_K$
and the scale setting observable $M_\Xi$ or $M_\Omega$. As explained in
sect.~\ref{sect:skeleton}, we traded the bere parameters of our theory
for $M_\pi$, $M_K$ and $M_{\Xi/\Omega}$, so we need the correct
functional dependence of $M_X$ on these (and on the lattice size) to
extrapolate to the physical point.

Generically, we can expand any heavy hadron mass $M$ in terms of the
quark masses
\begin{equation}
M=M^{(0)}+\hat{\alpha}m_{ud}+\hat{\beta}m_s+\ldots
\end{equation}
which, according to eq.~\ref{eq:gmor} translates into
\begin{equation}
M=M^{(0)}+\alpha M_\pi^2+\beta M_K^2+\ldots
\end{equation}
Depending on the precision of our data, we may need to add higher
order terms to this expansion. The specific form of these terms
depends on the expansion point we choose. For an optimal convergence
radius, a Taylor expansion around finite $M_\pi^2$ and $M_K^2$ seems
to be a good choice. Alternatively, one may perform an asymptotic
expansion around $M_\pi^2=0$ for which chiral perturbation theory
generically gives a term $\propto M_\pi^3$ as the next higher order in
$M_\pi^2$ \cite{Langacker:1974bc}. Since the lattice data are not
sensitive to further terms, it is a good idea to use both ans\"atze
\begin{equation}
\label{eq:mdep}
M=M^{(0)}+\alpha M_\pi^2+\beta M_K^2+\gamma\left\{
M_\pi^4
\atop
M_\pi^3
\right.
\end{equation}
for extrapolating/interpolating to the physical point and let the
spread between the results contribute towards the systematic error.

Sensible ans\"atze for the infinite volume behaviour have been
discussed in sect.~\ref{sect:fv}. We can therefore perform a combined
fit of the scale setting observable $M_\Xi$ or $M_\Omega$ versus
$M_\pi$, $M_K$ and the lattice size $L$, introducing one fit parameter
$a_\beta$ for each of the bare couplings $\beta=6/g^2$ in our
ensembles. Requiring the fit to go through the physical point,
which can be defined by the physical value of the ratios
$M_\pi/M_{\Xi/\Omega}$ and $M_K/M_{\Xi/\Omega}$, the lattice spacings
are determined.

We can then make an ansatz for the continuum limit of $M_X$ guided by
the scaling behaviour of the action (see sect.~\ref{sec:lqcd}). For
an unimproved Wilson action, one would e.g. choose
\begin{equation}
\label{eq:scala}
M=M^{(0)}+\eta a+\ldots
\end{equation}
while for a nonperturbatively improved clover action
\begin{equation}
\label{eq:scala2}
M=M^{(0)}+\eta a^2+\ldots
\end{equation}
might be more appropriate. The specific example calculation I am
following used a perturbatively improved smeared clover
action. Formally, its scaling is $\mathcal{O}(\alpha_s a)$, which is in
between eq.~\ref{eq:scala} and eq.~\ref{eq:scala2}. Numerically, even the
leading scaling term is barely relevant. It is therefore a very
conservative choice to use both eq.~\ref{eq:scala} and
eq.~\ref{eq:scala2} and again add the spread between the results thus
obtained to the systematic error.

\begin{figure}[htb]
\centerline{
\includegraphics[width=12cm]{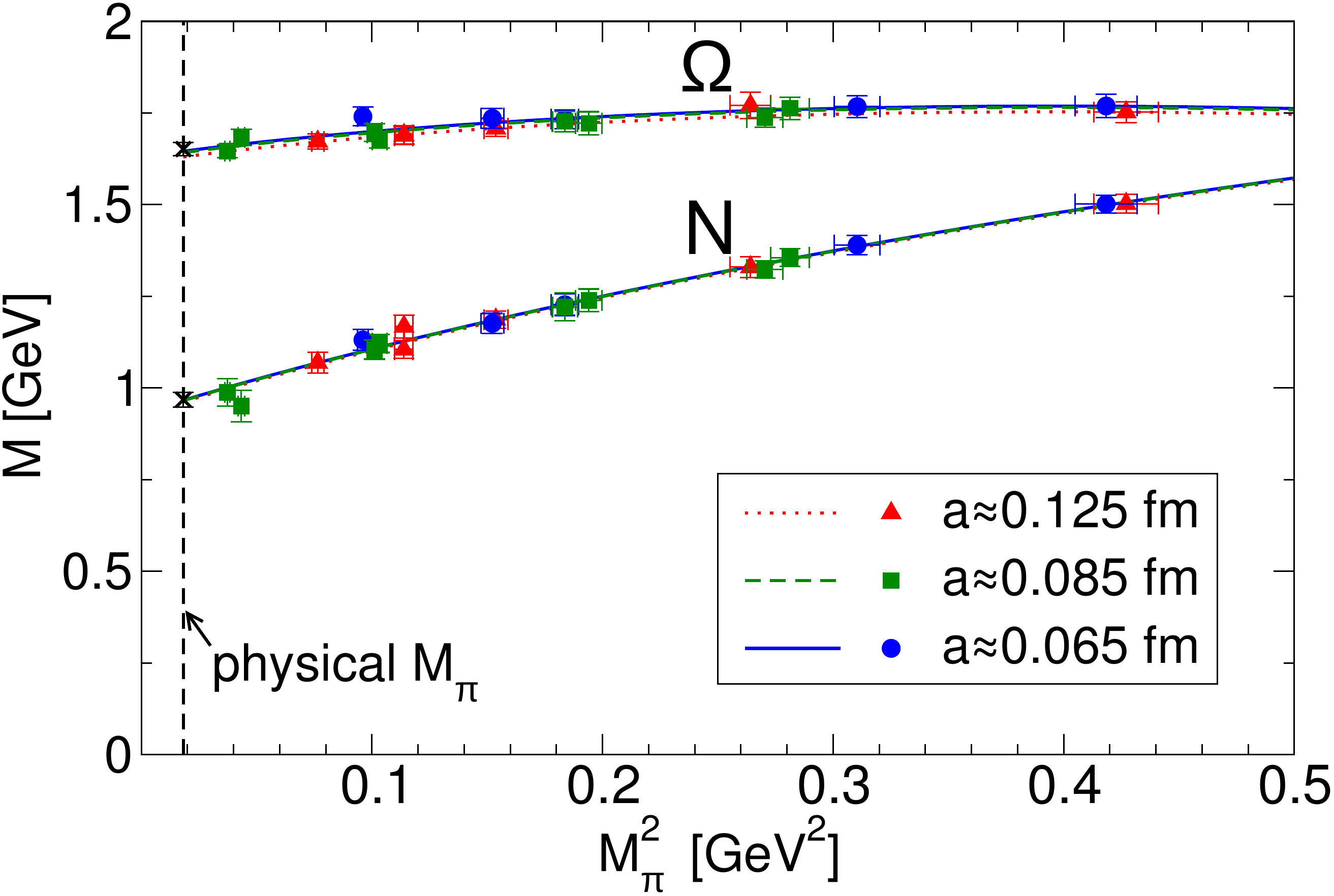}
}
\caption{\label{fig:fit}
Nucleon and $\Omega$ mas vs. $M_\pi^2$ from \cite{Durr:2008zz}.
Points represent lattice data shifted with the combined fit function
to be at physical $M_K^2$ and corrected for finite volume
effects. Curves represent the fit for 3 different lattice spacings. It
is evident that the extrapolation to both the continuum limit and the
physical $M_\pi^2$ is very mild.
}
\end{figure}

Performing a combined continuum, infinite volume, $M_\pi$ and $M_K$
fit, we thus obtain a prediction for $M_X$ at the physical point. An
example of such a fit for the $\Omega$ and nucleon masses with the scale
set by $M_\Xi$ is displayed in fig.~\ref{fig:fit}

\begin{figure}[htb]
\centerline{
\includegraphics[width=12cm]{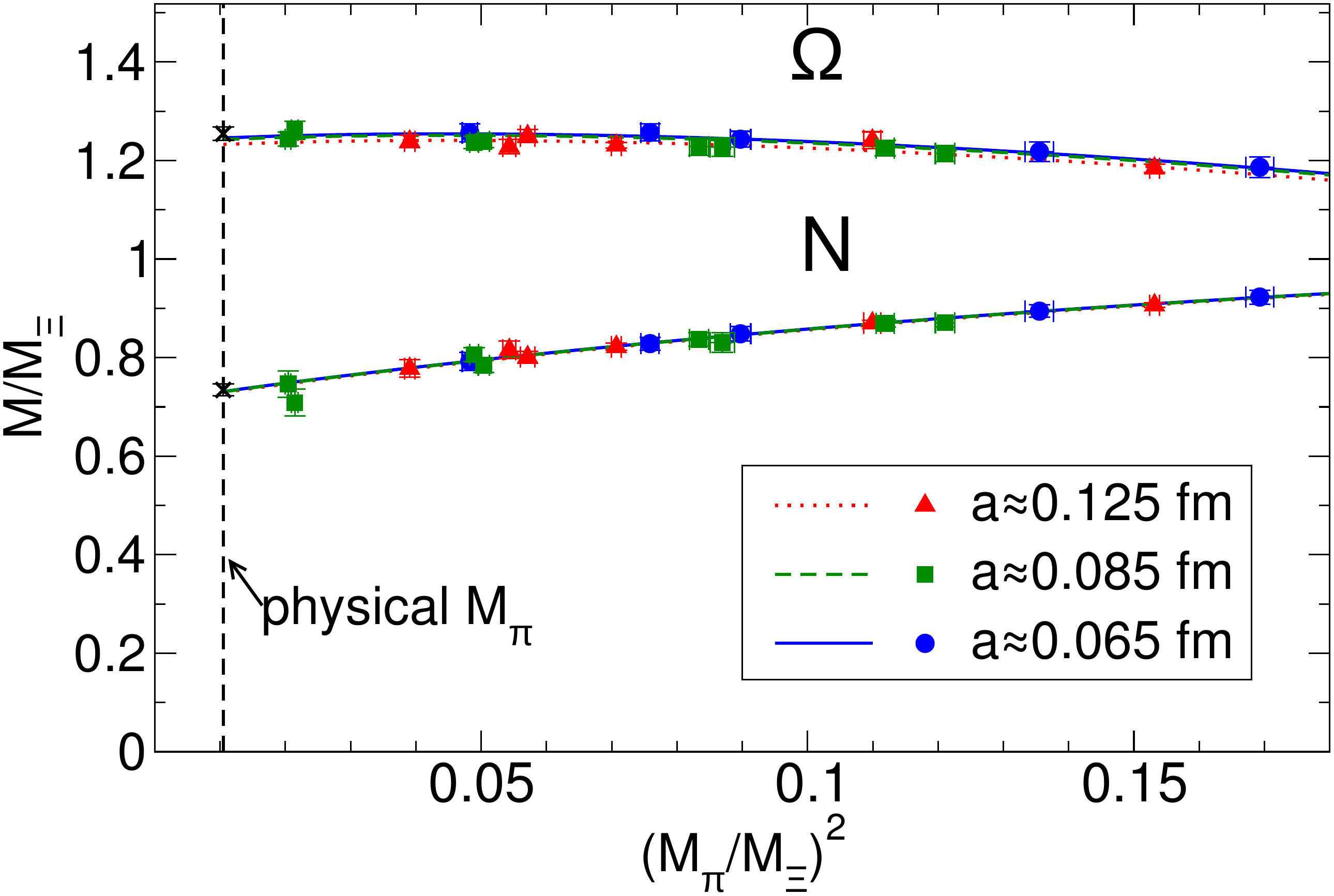}
}
\caption{\label{fig:fitr}
Nucleon and $\Omega$ mas vs. $r_\pi^2$ from \cite{Durr:2008zz}.
Points represent lattice data shifted with the combined fit function
to be at physical $r_K^2$ and corrected for finite volume
effects. Curves represent the fit for 3 different lattice spacings.
}
\end{figure}

It is interesting to note, that in the derivation of the fit function
we have so far made a choice that the lattice spacing depends only on
$\beta$ among the bare lattice parameters, which is known as mass
independent scale setting. Since the lattice spacing is ill defined
outside the physical point (i.e. different definitions may lead to
different values) we may actually choose any other procedure as long
as it coincides at the physical point. One might e.g. assume that
outside the physical point the scale setting observable keeps the same
physical value. An alternative way to eq.~\ref{eq:mdep} for
parametrizing the quark mass dependence would therefore be
\begin{equation}
\label{eq:rdep}
M=M^{(0)}+\alpha r_\pi^2+\beta r_K^2+\gamma\left\{
r_\pi^4
\atop
r_\pi^3
\right.
\end{equation}
where $r_X:=M_X/M_{\Xi/\Omega}$ is the ratio of $M_X$ to the scale
setting mass. A fit with this ratio method is displayed in
fig.~\ref{fig:fitr}.

\subsection{Systematic errors}

Having extrapolated the target observable to the physical point, we
have a prediction with a statistical error. In fact, we have many
predictions of the same observable - hundreds or even thousands are
not uncommon. During the analysis we had many points where a number of
different procedures were reasonable - the time interval for
extracting baryon masses, the scale setting observable or various fit
forms - and we said that we will include the corresponding spread into
the systematic error. This is what we will do now.

It is important to note that there is no uniquely correct procedure to
compute the systematic error. In fact, unlike the statistical error,
the systematic error can not be computed. The systematic error tries
to quantify the effects which we are not able to control - e.g. the
size of the terms in a taylor expansion that we truncated because our
data are no more sensitive to it. It is therefore a guess or at best
an estimate. The most important point about estimating a systematic
error therefore is not to omit any relevant part of it. A
sophisticated estimate of the error in the $M_\pi$ extrapolation
e.g. is useless if all data were obtained at a single lattice spacing
or the fit window for extracting the bare masses was not varied.

\begin{figure}[htb]
\centerline{
\includegraphics[width=10cm]{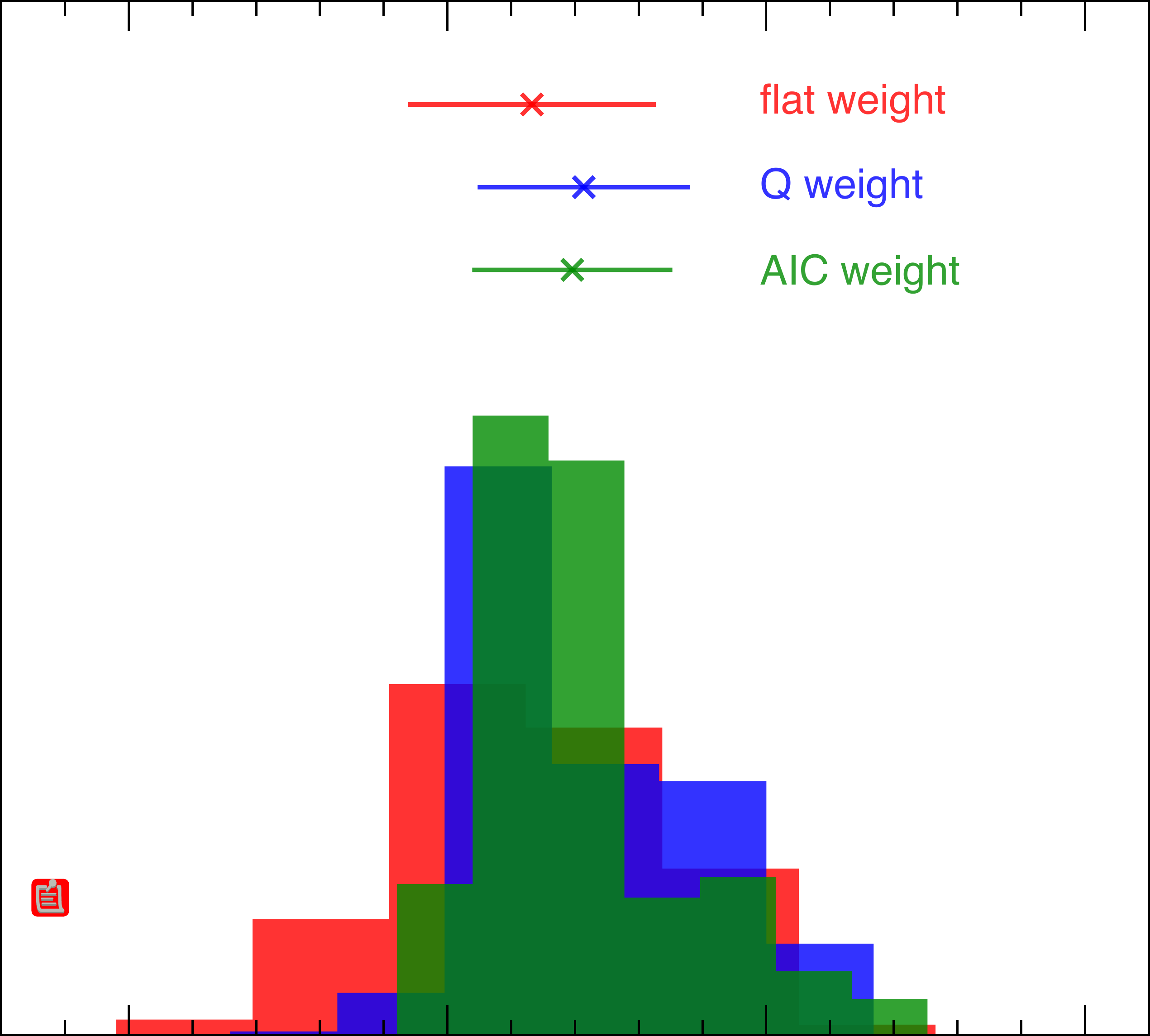}
}
\caption{\label{fig:syscom}
Comparison of three methods to compute the systematic error.
}
\end{figure}

Provided that we have varied our analysis procedure to cover all
relevant effects, A simple procedure for estimating the systematic
error would then be to take all the results and compute the
spread. The average or mean of the distribution can serve as the
central value. I label this procedure as flat weight.

One might be worried that with this procedure an analysis that did not
describe the data well will have the same weight as one that did. One
might therefore put a weight to each analysis when computing the spread
and the average. A reasonable weight would e.g. be the fit quality
$Q$.

Another weight that is motivated by information theory is the Akaike
information criterion (AIC) \cite{Akaike:1974aa}. It estimates the
information contained in a specific fit $m$ by computing the information
cross-entropy $J_m$ of the given fit with the best one in the
sample. For a large number of points, the cross-entropy is then given by
\begin{equation}
\label{eq:AIC}
J_m=-\frac{\chi^2}{2}-p_m
\end{equation}
where $p_m$ is the number of parameters of the fit and $\chi^2_m$ is
given by the least square fit. The probability that a fit is correct
is proportional to the exponential of the cross-entropy $\exp
J_m$. The AIC punishes fits with a large number of parameters, since
according to eq.~\ref{eq:AIC}, $\chi^2$ has to decrease by 2 for every
new parameter to even achieve the same weight.

\begin{figure}[htb]
\centerline{
\includegraphics[width=12cm]{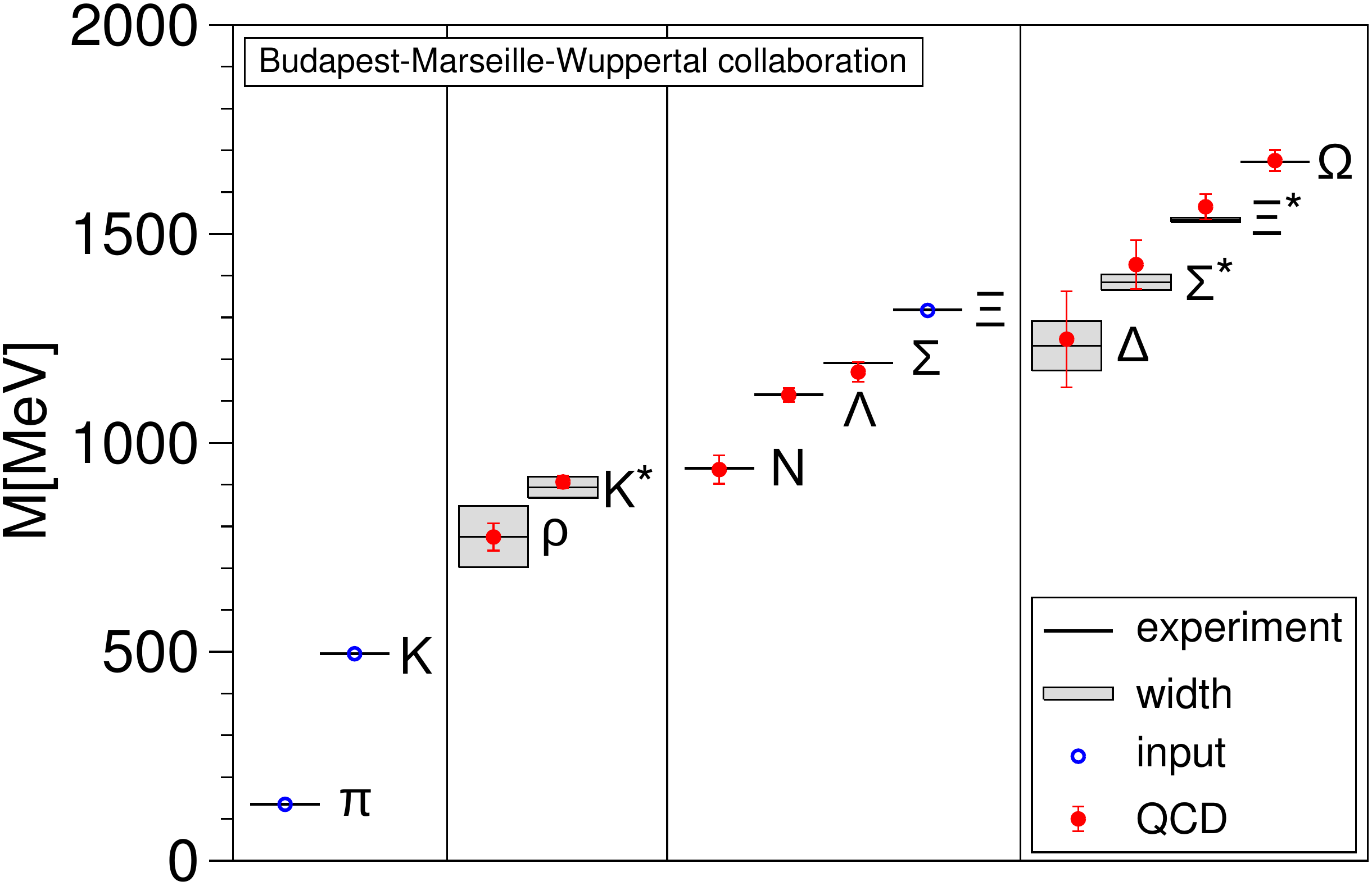}
}
\caption{\label{fig:spect}
 The ground state light hadron spectrum from
\cite{Durr:2008zz}. Blue points denote observables that were used as
input quantities in the lattice calculation, while red points are
lattice predictions with combined statistical and systematic
error. Experimental values are isospin averaged and boxed denote the
width of resonances.
}
\end{figure}

 Although the AIC might seem to be an optimal method, it is important
to note that it only gives relative weights among fits that were
chosen beforehand or, in other words, that we have provided as an
input a certain measure in the space of all possible fits that the AIC
only modified. Since we have no a priori knowledge about a proper
measure, even the AIC weighted systematic error is just a guess.

From a practical point of view, it is important to note, that all
sensible estimates of the systematic error should give compatible
values (see fig.~\ref{fig:syscom}). In fact, this agreement is a
valuable crosscheck for the entire analysis procedure.

Adding the statistical and systematic errors in quadrature, the ground
state light hadron spectrum can now be computed. The result is
displayed in fig.~\ref{fig:spect}. 

\subsection{QED and strong isospin splitting}

We now turn our attention to the fine structure of the hadron
spectrum. The electromagnetic and strong isospin splitting effects in
the hadron spectrum are typically a few MeV.  After having made
physical predictions in QCD, it might seem trivial to include these
effects in our lattice calculations. For the case of strong isospin
splitting, this is true at least conceptually - one only needs to
introduce independent u and d quark masses. From a computational point
of view it is however extremely demanding to accommodate for the very
light u quark mass. Both u and d quarks are light, but $m_u/m_d\sim
0.5$ resulting in a much worse conditioned fermion matrix. For Wilson
type fermions, the probability of encountering an exceptional
configuration is drastically increased, while for staggered fermions
one approaches the dangerous low mass regime at finite lattice
spacing. Apart from these technical difficulties and the need for an
additional observable to fix $m_u-m_d$ however the
introduction of strong isospin splitting is straightforward.

QED on the other hand has a number of features that make its ab-initio
treatment more difficult. First of all, the QED coupling constant
$\alpha$ has a pole in the UV \cite{Landau:1955aa} so we are dealing
with an effective theory. Secondly, all electrically charged particles
are no more gauge invariant. Computing propagators with the methods
described in sect.~\ref{sec:cpi} will trivially give $0$ unless one
fixes a gauge or inserts appropriate gauge links in between source and
sink to points. Furthermore, for Wilson-type fermions there will be an
additional additive mass renormalization that will be different for up
and down type quarks due to their different electrical
charge. Finally, QED does not have a mass gap. It therefore features
power law finite volume effects, in contrast to QCD. One might also
think that adding QED will necessitate adding electrons to the lattice
theory, which would be difficult because of their very small
mass. Their contribution however only appears at
$\mathcal{O}(\alpha^2)$ compared to $\mathcal{O}(\alpha\alpha_s)$ for
quarks, so they may be neglected.

One advantage of QED though is that it is an Abelian theory. However,
compactifying the photon field $A_\mu$ via gauge links $U_\mu$ as we
did for the gluon field in sect.~\ref{sect:lr} would introduce
spurious self couplings. It is therefore reasonable to use a
non-compact photon action, e.g.
\begin{equation}
\label{eq:sgamma}
S_\gamma=\frac{1}{2V_4}\sum_{k,\mu}|\hat{k}|^2
|A_\mu^k|^2
\qquad
\hat{k}=\frac{e^{iak_\mu}-1}{ia}
\end{equation}
in Feynman gauge momentum representation. Note that in
eq.~\ref{eq:sgamma} the prefactor of the $k=0$ mode is $0$. Therefore,
$A_\mu^0$ is not constrained and may freely fluctuate. It is also easy
to check that it is both gauge invariant and not contributing to the
field strength $F_{\mu\nu}=\partial_\mu A_\nu-\partial_\nu A_\mu$. Its
only effect is to create a potential difference when winding around
the lattice nontrivially and returning to the same point. This is a
pure lattice artefact appearing at finite volume and one should
subtract it \cite{Duncan:1996xy}. Due to the $1/2V_4$ prefactor in
eq.~\ref{eq:sgamma}, the theory thus obtained has the same infinite
volume limit.

There is a problem with this simple subtraction scheme
however. Subtracting the $k=0$ mode is equivalent to adding a term
\begin{equation}
\label{eq:sub0}
\xi\sum_\mu\left(\sum_x a^4
A_\mu(x)
\right)^2
\end{equation}
to the action and letting the Lagrange multiplier
$\xi\rightarrow\infty$. Evidently this term spoils reflection
positivity as it connects all field components at points on arbitrary
time slices with each other and the resulting theory is not guaranteed
to possess a well-defined Hamiltonian.

This deficiency can be cured by making the Lagrange multiplier in
eq.~\ref{eq:sub0} time dependent
\begin{equation}
\label{eq:subhu}
\sum_t \eta(t)\sum_\mu\left(\sum_{\vec{x}} a^4
A_\mu(t,\vec{x})
\right)^2
\end{equation}
With $\eta(t)\rightarrow\infty$, this is equivalent to subtracting all
modes $\vec{k}=0$ from the action, a procedure first proposed by
Hayakawa and Uno \cite{Hayakawa:2008an}. Using the Hayakawa-Uno (HU)
subtraction in Coulomb gauge results in a theory that is reflection
positive \cite{Borsanyi:2014jba}. Additionally, the HU subtraction is
a pure finite volume effect. Finite volume terms are universal up to
$O(1/L^2)$ and $O(1/L^3)$ terms do not diverge for infinite volume or
time extent \cite{Davoudi:2014qua,Borsanyi:2014jba}.

It is interesting to note that although the HU subtraction
is not gauge invariant, one can define a slightly modified subtraction
that is gauge invariant and coincides with the HU
subtraction in temporal gauge. We can define this scheme by adding to
the action
\begin{equation}
\label{eq:submy}
\xi\left(\sum_x a^4
A_0(x)
\right)^2
+
\sum_t \eta(t)\sum_i\left(\sum_{\vec{x}} a^4
A_i(t,\vec{x})
\right)^2
\end{equation}
with $\xi,\eta(t)\rightarrow\infty$. It thus removes from the action
the components $A_0^0$ and $\vec{A}^{(k_0,\vec{0})}$ for all
$k_0$. With the additional $A_0^k=0$ in temporal gauge, this new
scheme is identical to HU in that gauge and therefore seems to fulfill
reflection positivity, too.

QED and strong isospin splitting have introduced two new parameters
$\alpha$ and $m_u-m_d$ to our lattice theory that need to be
extrapolated to the physical point. We again would like to find two
experimentally accessible observables that are strongly dependent on
$\alpha$ and $m_u-m_d$ and not so on other parameters. One such
observable is the mass difference between charged and neutral kaons
$M_{K^\pm}^2-M_{K^0}^2$. As a second observable, one can take the
value of the renormalized QED coupling $\alpha$ itself because in
contrast to $\alpha_s$ of QCD it is very small at low energies. The
renormalization scheme best suited to obtain $\alpha$ is in fact provided
by the gradient flow (see eq.~\ref{eq:gflow}) for photon fields.

\begin{figure}[htb]
\centerline{
\includegraphics[width=10cm]{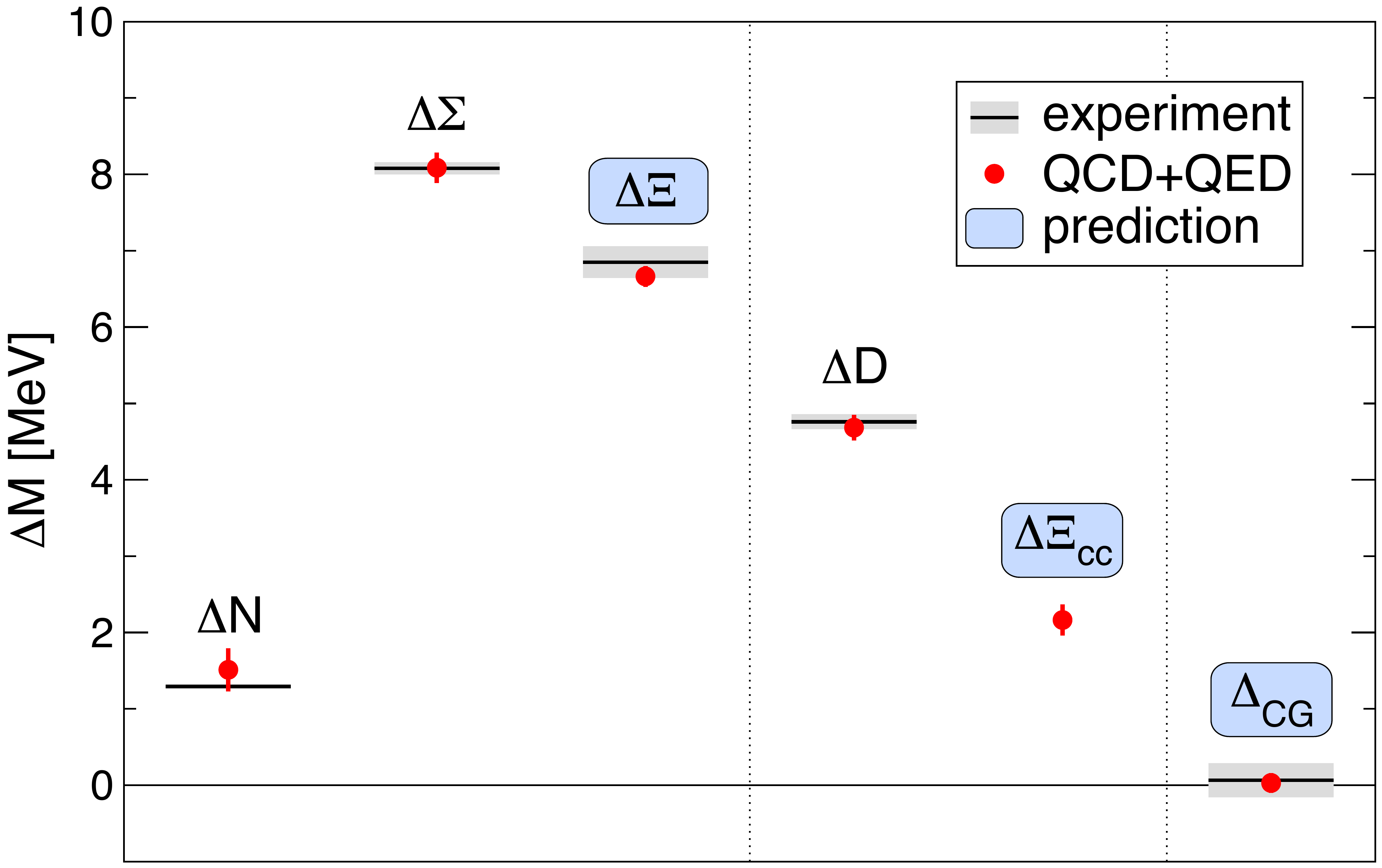}
}
\caption{Isospin splitting of some ground state hadrons 
  from \cite{Borsanyi:2014jba}. In the left panel, grey boxes
  represent experimental uncertainties and red dots are lattice
  predictions. $\Delta_\text{CG}$
  denotes the violation of the Coleman-Glashow
  relation\cite{PhysRevLett.6.423,Zweig:1981pd}, a quark model relation predicting
  $\Delta\Sigma-\Delta\Xi-\Delta N=0$. }
\label{fig:iso}
\end{figure}

Having defined the physical point, the target observables, which in
this case are hadron isospin splittings, can be extrapolated
there. Skipping further technical details that can be found in
\cite{Borsanyi:2014jba}, the lattice predictions for some hadronic
isospin splittings are displayed in the left panel of
fig.~\ref{fig:iso}.

\section{QCD thermodynamics}
\label{sec:therm}

\subsection{Formulation of QCD at finite temperature}

We now turn our attention to finite temperature QCD. Introducing
finite temperature into an Euclidean quantum field theory is
straightforward. For a generic QFT with fields $\Phi$ and Lagrangian
$L$, the path integral is given as the vacuum to vacuum
transition amplitude
\begin{equation}
\mathcal{Z}=\int D\Phi e^{-\int_{-\infty}^\infty dt  L}=\langle 0|0\rangle
\end{equation}
At finite Euclidean time extent ${\mathcal T}$ and with corresponding
periodic/antiperiodic boundary conditions for bosons/fermions, one
instead obtains
\begin{equation}
\mathcal{Z}_{\mathcal T}=\int D\Phi e^{-\int_0^{\mathcal T} dt
  L}=\sum_i\langle i|e^{-E_i{\mathcal T}}|i\rangle
=\Tr(e^{-{\mathcal H \mathcal T}})
\end{equation}
which describes a thermal ensemble at a temperature $T=1/\mathcal{T}$
given by the inverse time extent of the system.

Introducing finite temperature into a lattice theory is therefore
achieved by simply reducing its time extent $T=a N_t$. One can vary
$T$ in steps of $a$ which, especially at high temperatures, can be
very coarse. As an alternative, one may keep the number of lattice
points in the time direction $N_t$ fixed and vary the lattice spacing
$a$ instead which can be achieved by varying the gauge coupling
$\beta=6/g^2$. The advantage of this method is that the temperature
might be varied continuously, but one has to keep in mind that all
other quantities vary with $\beta$, too. The spatial volume is
directly affected by a change in $a$, and so is the relation between
bare quark masses in lattice units and renormalized physical quark
masses. In order to ensure that by changing $\beta$ one does not
change the parameters of the theory as well, we have to determine the
physical point at each $\beta$ as outlined in
sect.~\ref{sect:skeleton}. The resulting path through the parameter space
of our theory connects parameter sets describing the same physical
situation as the cutoff is varied. In the thermodynamics literature,
this is known as the line of constant physics (LCP).

The LCP will of course depend on the specific observables chosen to
identify the physical point. The difficulty of determining it also
depends largely on the action used. For the $2+1$ flavour staggered
action that is in broad use today, one of the two ratios necessary to
fix the LCP is simply given by the ratio of bare quark masses
$m_s/m_{ud}$. Although it is not an experimentally accessible
quantity, it is known to a good enough accuracy and its use reduces
the number of parameters that need to be independently tuned by
one. For Wilson-type fermions on the other hand, the additive quark
mass renormalization renders the search for an LCP much more
difficult so that it is preferable to vary the temperature by varying
$N_t$.

\begin{figure}[htb]
\centerline{
\includegraphics[width=12cm]{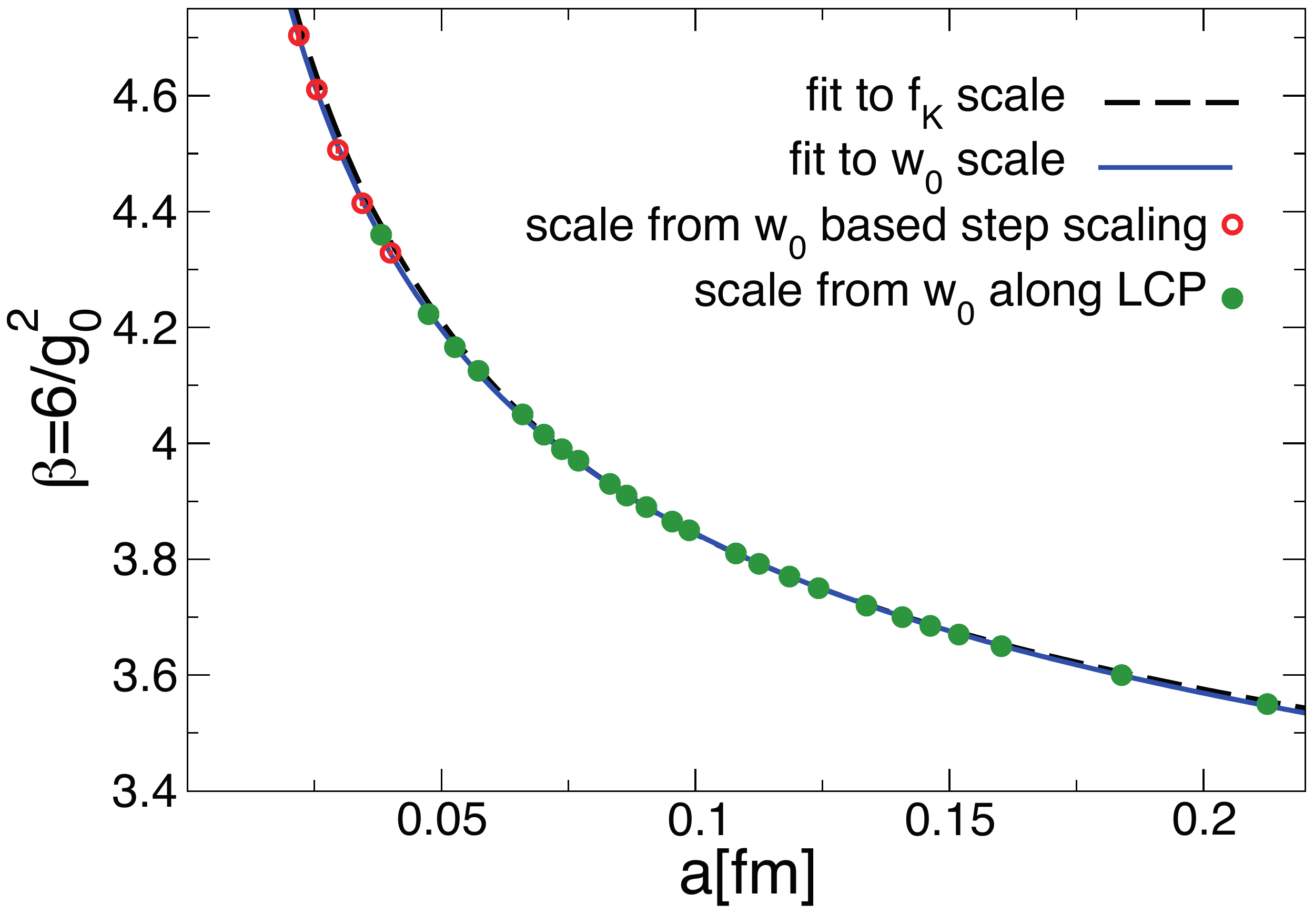}
}
\caption{Relation between bare gauge coupling and lattice spacing from
\cite{Borsanyi:2013bia}.}
\label{fig:lcp}
\end{figure}

As an example, fig.~\ref{fig:lcp} displays a determination of the
relation between lattice spacing $a$ and bare gauge coupling $\beta$
along a LCP from a recent calculation with $2+1$ flavour stout-smeared
staggered fermions. It is important to note, that one can define LCPs
which are not physical. One can e.g. set the quark mass ratio
$m_s/m_{ud}$ to unphysically small values. The results obtained will
be consistent, but will not describe the real physical situation.

\subsection{Identifying phase transitions}

One of the main motivations for dealing with thermodynamics is the
exploration of the phase structure of a theory. Generically, phase
transitions appear only in infinite volume. Studying their emergence
at finite volume is best achieved by finite size scaling techniques.

\begin{figure}[htb]
\centerline{
\includegraphics[width=12cm]{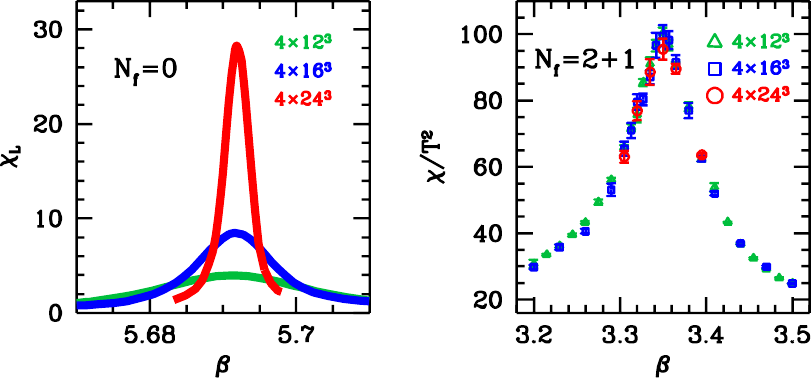}
}
\caption{
Polyakov loop susceptibility in pure $SU(3)$ gauge theory (left panel) and
chiral susceptibility of $2+1$ flavour QCD (right panel) versus
$\beta$ from
\cite{Fodor:2009ax,Aoki:2006we}}
\label{fig:susp}
\end{figure}

 In QCD, we can define the chiral susceptibility as the second
derivative of the partition function with respect to the (light) quark
mass
\begin{equation}
\label{eq:chisus}
\chi_{\bar\psi\psi}=\frac{T}{V}\frac{\partial^2 {\mathcal Z}}{\partial m^2}=\frac{\partial\langle\bar\psi\psi\rangle}{\partial m}
\end{equation}
The corresponding quantity in pure gauge theory, the Polyakov loop
susceptibility, is plotted in the left panel of fig.~\ref{fig:susp}
versus $\beta$ for three different lattice volumes. It is clearly
visible that the peak height scales with volume, which is a sign for a
phase transition that develops in the infinite volume limit. The
exponent with which the peak height diverges with volume depends on
the universality class. In the present case, the height scales
$\propto V$, which is characteristic of a first order phase
transition.

In one of the landmark calculations of lattice QCD \cite{Aoki:2006we},
it was demonstrated that at physical quark masses (and vanishing
chemical potential) QCD does not exhibit a phase transition but rather
a crossover. In the right panel of fig.~\ref{fig:susp}, the chiral
susceptibility is plotted versus $\beta$ for the same three lattice
volumes as in the pure gauge theory case. In contrast to pure gauge
theory, the peak does not show an increase with volume though.

\begin{figure}[htb]
\centerline{
\includegraphics[width=12cm]{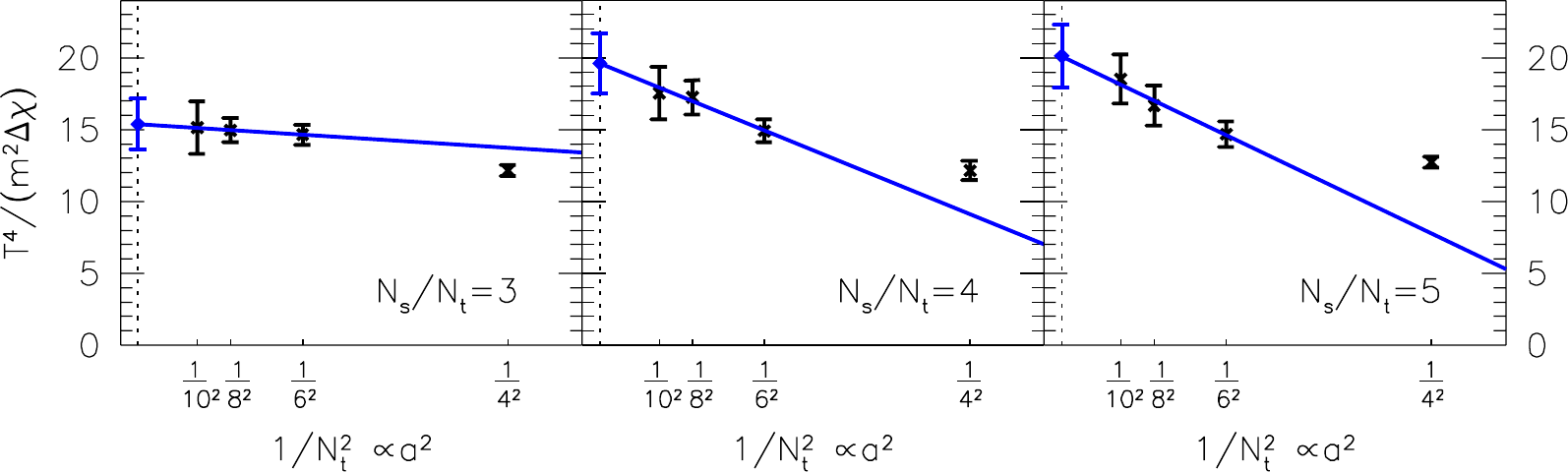}
}
\centerline{
\includegraphics[width=8cm]{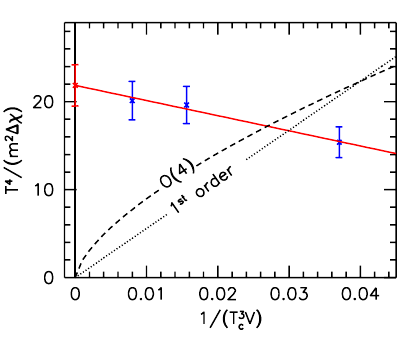}
}
\caption{
Continuum extrapolation of the peak height for 3 different lattice
volumes (upper panel) and infinite volume behaviour of the inverse
peak height (lower panel) from
\cite{Aoki:2006we}}
\label{fig:opi}
\end{figure}

As clear as this evidence might seam, it is not conclusive yet because
it lacks a proper continuum limit. The calculation was therefore
repeated for 4 different values of $N_t$, which allowed for taking the
continuum limit of the peak height (upper panel of fig.~\ref{fig:opi})
before extrapolating it to infinite volume (lower panel of
fig.~\ref{fig:opi}). The result clearly shows that the peak height
does not diverge and that therefore QCD has no phase transition at
vanishing chemical potential.

\subsection{Critical and pseudocritical temperatures}
\label{sect:tc}

As we have seen now that QCD does not exhibit a phase transition, the
question about its critical temperature is moot: there simply is no
critical temperature. One might nonetheless be interested in finding
the temperature where the thermodynamic observables exhibit the
largest change, e.g. where the peak in the chiral susceptibility is
located even if it does not diverge.

\begin{figure}[htb]
\centerline{
\includegraphics[width=12cm]{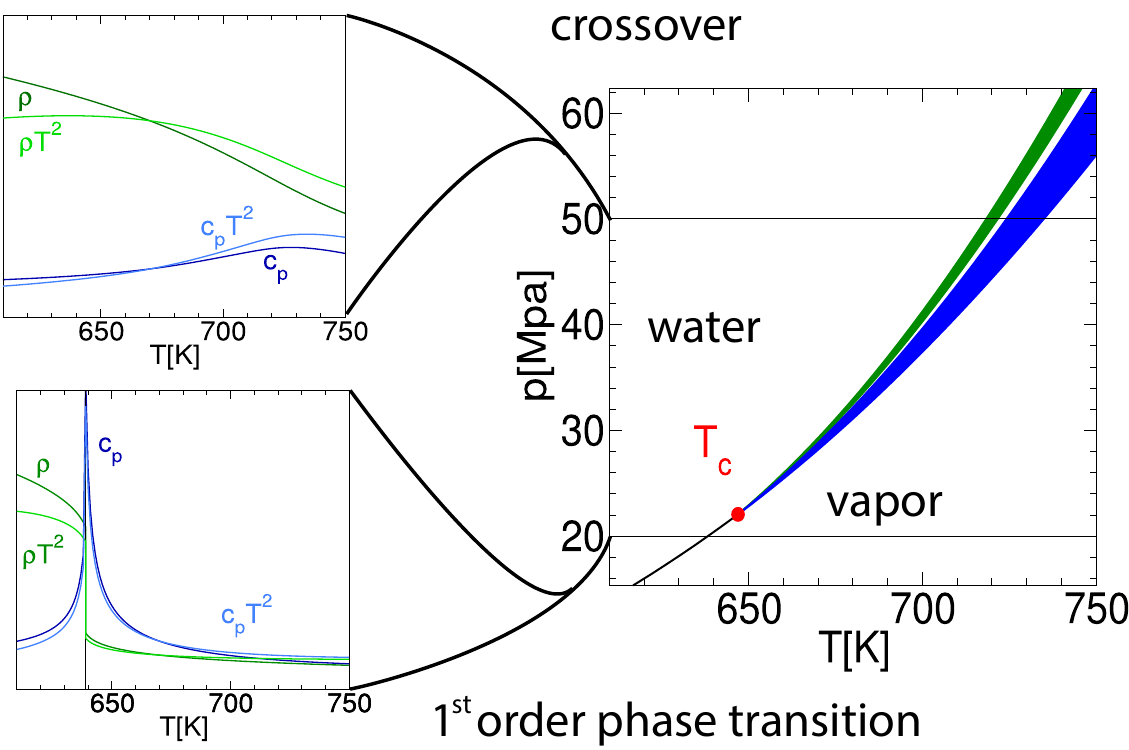}
}
\caption{Illustration of the phase diagram of water. In the right
  panel the phase structure is plotted in the $p$ vs. $T$ plane, the
  left panels show the temperature dependence of the density $\rho$
  and the specific heat $c_p$ vs. temperature at a line of constant
  pressure below and above the critical point.}
\label{fig:water}
\end{figure}

It is evident that there is no unique pseudocritical temperature. As a
simple illustration, we might look at the phase diagram of water
(fig.~\ref{fig:water}). Let us assume we want to determine the
transition temperature of water at constant pressure below the
critical point. We could e.g. measure the density $\rho$ or the
specific heat $c_p$ for different temperatures. The critical
temperature will be the unique point where $\rho$ is discontinuous and
$c_p$ diverges and redefining our observables by multiplying them with
a smooth function of $T$, e.g.  $T^2$ will not change the situation
(lower left panel of fig.~\ref{fig:water}).  If on the other hand the
pressure is above the critical value, not only will the peak of $c_p$
generically be at a different temperature than the largest change in
$\rho$, but redefining the observables by multiplying them with $T^2$
will shift those temperatures (upper left panel of
fig.~\ref{fig:water}).

In QCD, a pseudocritical temperature may also be computed for
different observables. In addition to the chiral susceptibility
(eq.~\ref{eq:chisus}), quark number susceptibilities
\begin{equation}
\label{eq:qsu}
\chi_2^q=\frac{T}{V}\left.\frac{\partial^2 {\mathcal Z}_\mu}{\partial \mu_q^2}\right|_{\mu_q=0}
\end{equation}
may be used. One may also use the renormalized chiral condensate
\begin{equation}
\label{eq:chico}
\langle\bar\psi\psi\rangle_R=
\frac{m_{ud}}{M_\pi^4}
\left(\langle\bar\psi\psi\rangle_{ud}-\langle\bar\psi\psi\rangle_{ud,T}\right)
\end{equation}
itself or a quantity called the strange subtracted chiral condensate
that is defined as
\begin{equation}
\label{eq:strange}
\Delta_{l,s}=\frac{
\langle\bar\psi\psi\rangle_{ud,T}
-
\frac{m_{ud}}{m_s}
\langle\bar\psi\psi\rangle_{s,T}
}{
\langle\bar\psi\psi\rangle_{ud}
-
\frac{m_{ud}}{m_s}
\langle\bar\psi\psi\rangle_{s}
}
\end{equation}
where $\langle\bar\psi\psi\rangle_{q,T}$ is the chiral condensate for
quark flavour $q$ at temperature $T$ and
$\langle\bar\psi\psi\rangle_{q}$ the corresponding condensate at zero
temperature.

\begin{figure}[htb]
\centerline{
\includegraphics[height=4cm]{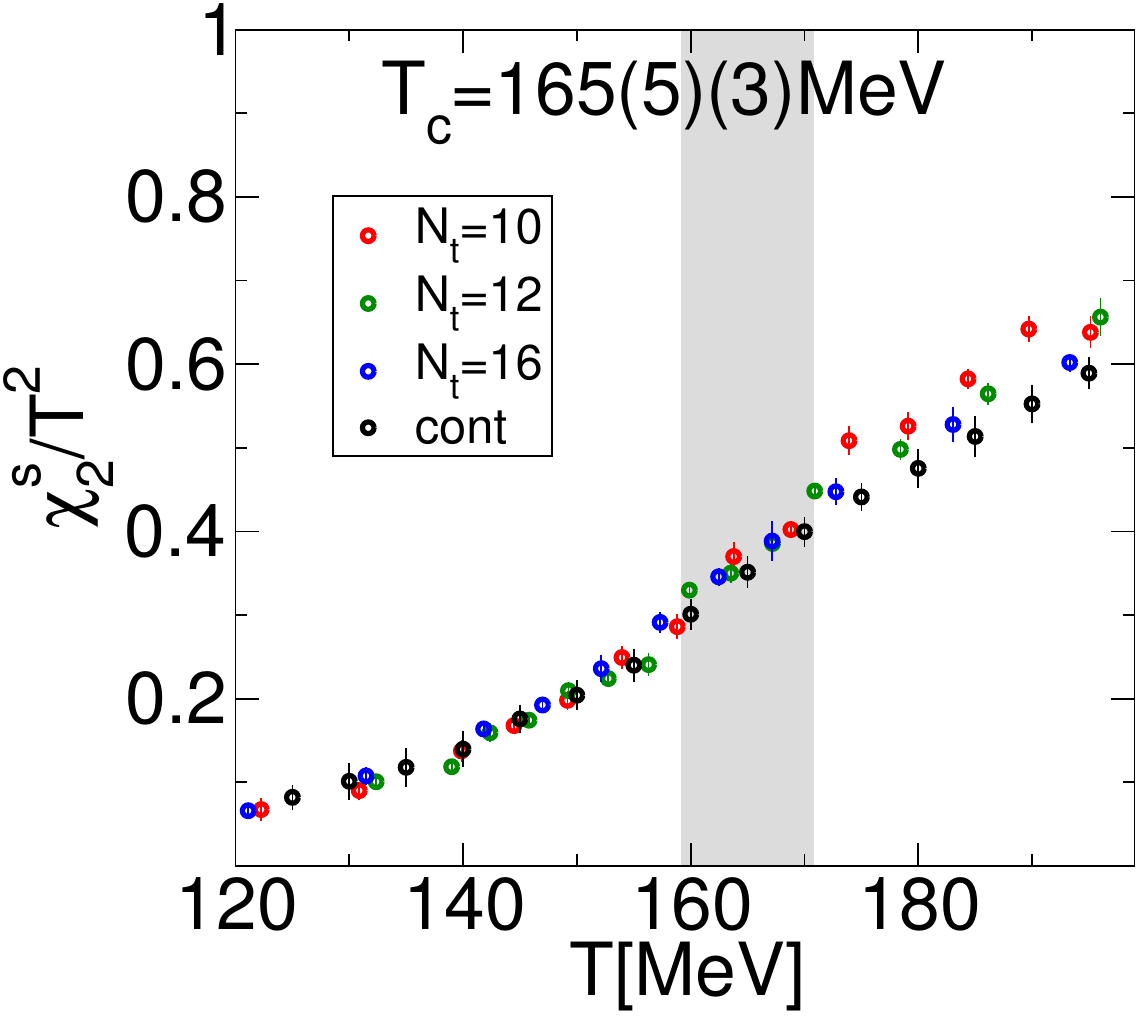}
\includegraphics[height=4cm]{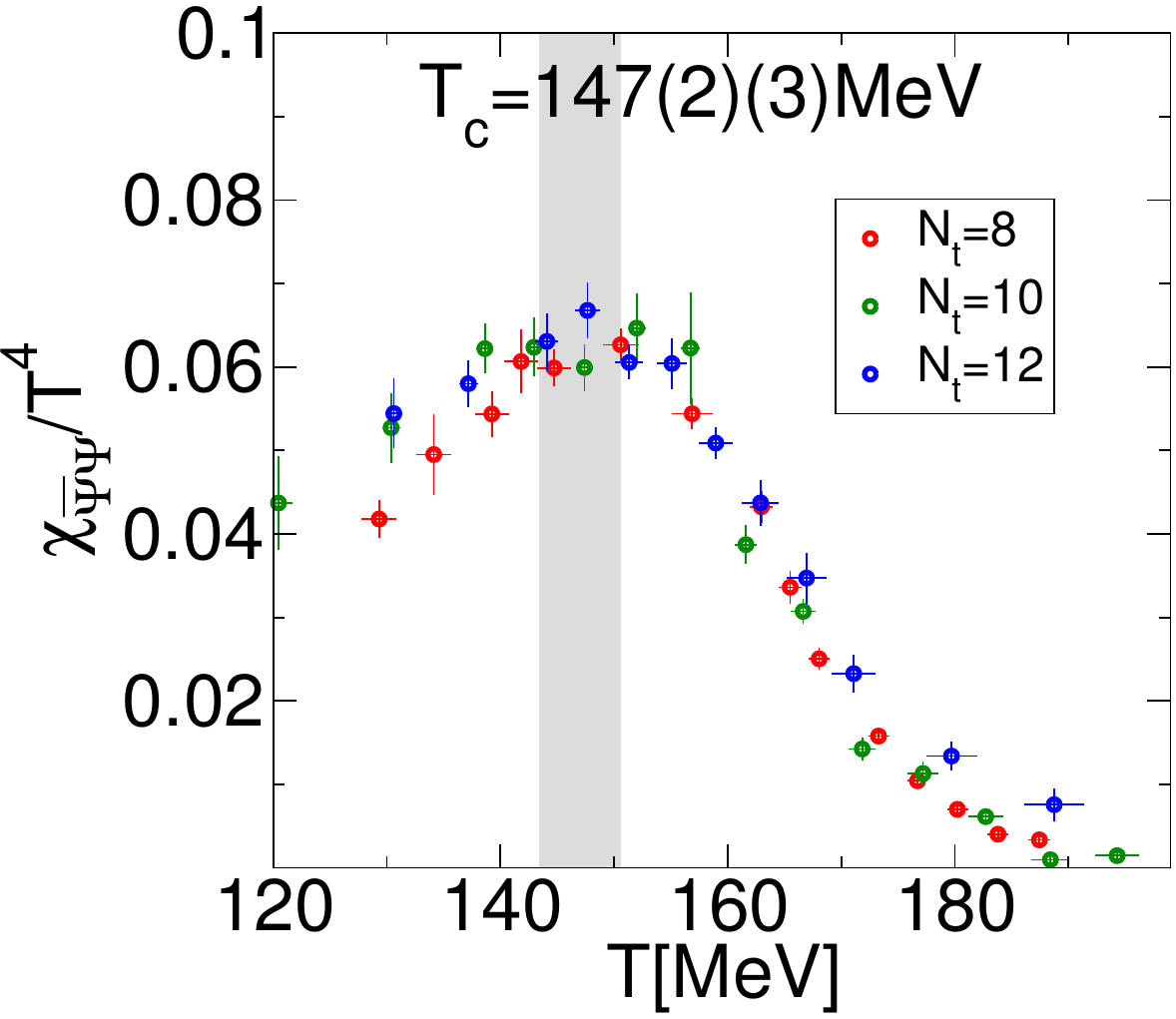}
\includegraphics[height=4cm]{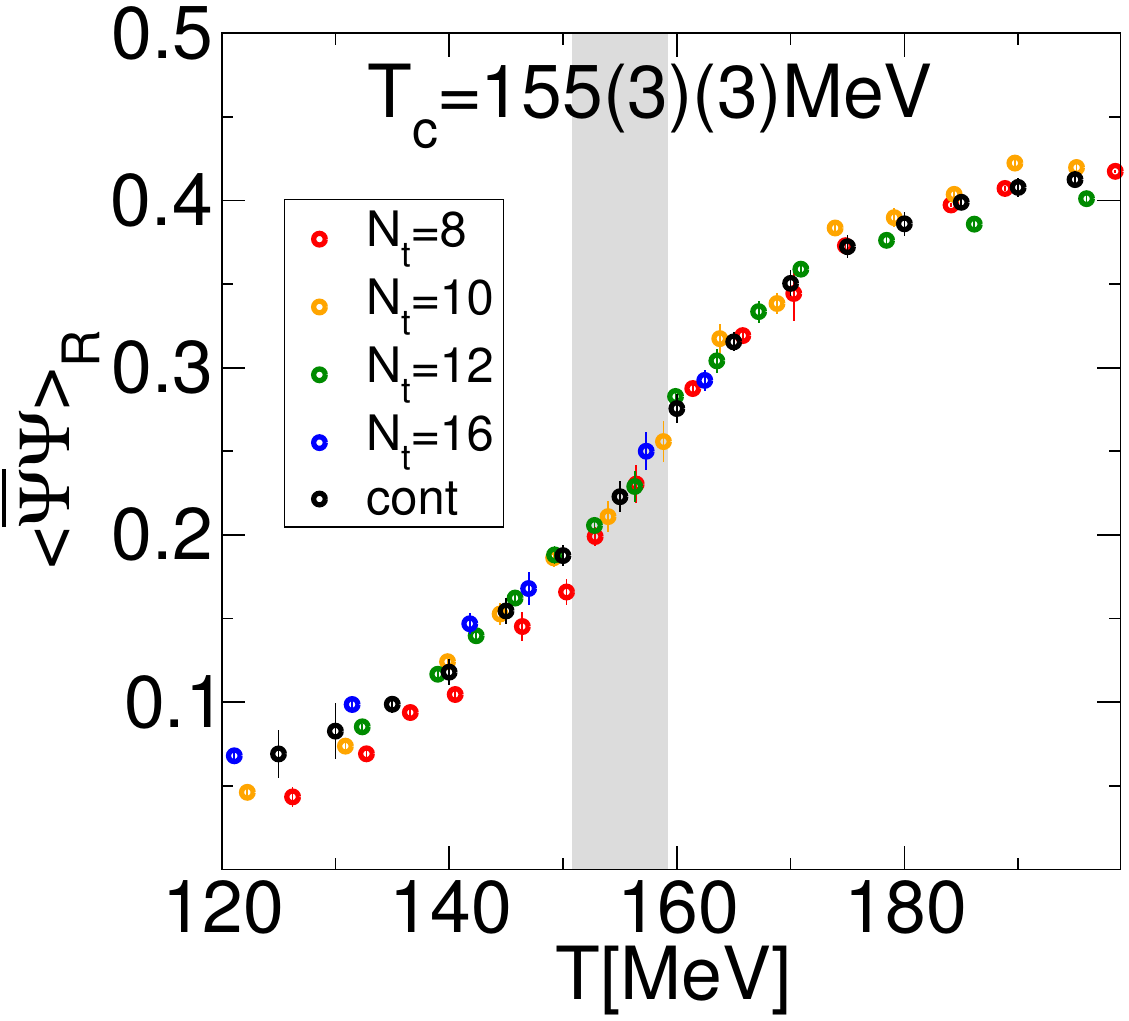}
}
\caption{
Determination of the pseudocritical temperature of QCD from 3
different observables from
\cite{Borsanyi:2010bp}.}
\label{fig:tc}
\end{figure}

\begin{figure}[htb]
\centerline{
\includegraphics[width=\textwidth]{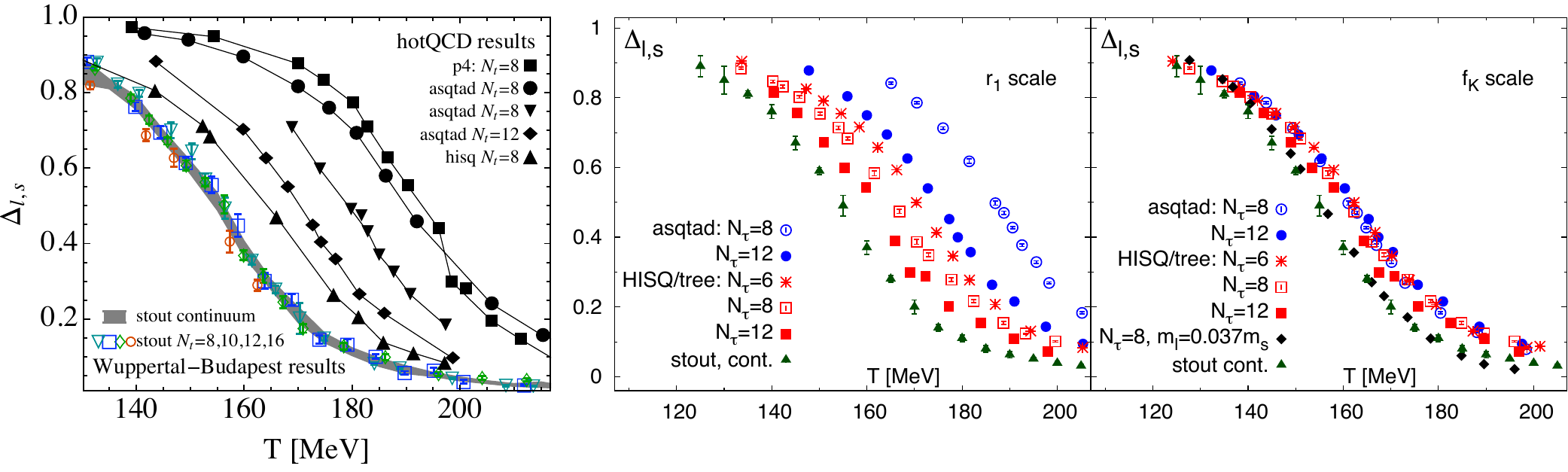}
}
\caption{ Comparison of the strange subtracted chiral condensate of
  the Wupertal-Budapest collaboration (stout) and the hotQCD
  collaboration (p4, asqtad and hisq actions) as presented by the
  Wuppertal-Budapest collaborations (left panel,
  \cite{Borsanyi:2010bp}) and the hotQCD collaboration (central and
  right panels, \cite{Bazavov:2011nk}). Note that unless otherwise
  noted hotQCD results are for an unphysical quark mass ratio
  $m_s/m_{ud}=20$.  }
\label{fig:hot}
\end{figure}

In fig.~\ref{fig:tc}, the behaviour of three of these observables is
plotted. The pseudocritical temperature extracted is in the range
$T_c\sim 145-165 \text{MeV}$. These results are in agreement with
older results from the Budapest-Wuppertal collaboration
\cite{Aoki:2006br,Aoki:2009sc}. They are also in agreement with recent
results of the hotQCD collaboration \cite{Bazavov:2011nk} that had
previously quoted substantially higher numbers \cite{Cheng:2006qk}
(see fig.~\ref{fig:hot}).

\begin{figure}[htb]
\centerline{
\includegraphics[width=\textwidth]{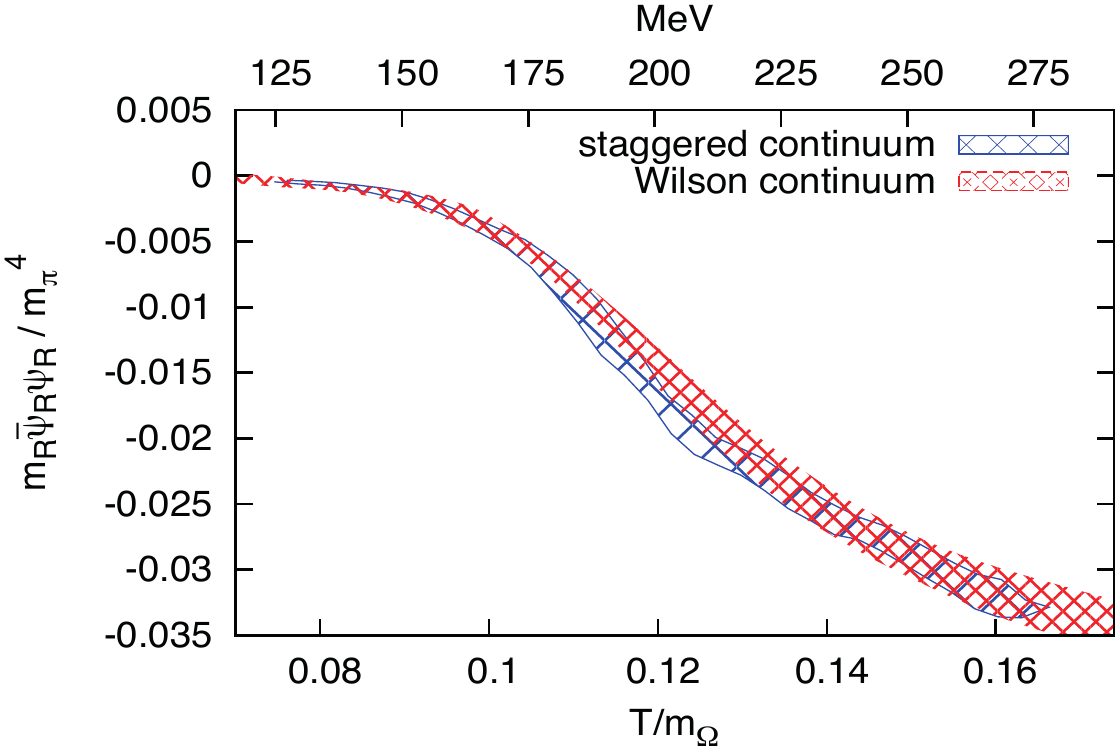}
}
\caption{
Comparison of the continuum extracted renormalized chiral condensate
(eq.~\ref{eq:chico}) obtained from staggered and Wilson fermions at an
unphysically lagre $M_\pi\approx 545 \text{MeV}$
\cite{Borsanyi:2012uq}}
\label{fig:whot}
\end{figure}

Some valuable crosschecks of the results on the phase transition are
beginning to emerge from lattice calculations with alternative fermion
formulations. Although continuum results at physical quark masses are
currently only available for staggered fermions due to their
relatively low computational cost, there are results at larger quark
masses from Wilson and overlap fermions
\cite{Umeda:2012er,Borsanyi:2012uq,Borsanyi:2012xf}. As an example,
fig.~\ref{fig:whot} shows a comparison of the staggered and Wilson
chiral condensate at an unphysically large pion mass. As one can see,
the continuum results are in nice agreement.

\subsection{Equation of state}

The final lattice calculation I want to briefly discuss is the QCD
equation of state. From standard thermodynamic relations we find that
up to finite volume corrections the pressure is given by
\begin{equation}
\label{eq:pres}
p^\text{lat}(\beta,m_q)=\frac{Ta^4}{V}\ln{\mathcal Z}(\beta,m_q)
\end{equation}
Since lattice QCD does not give us the normalization of the partition
function, eq.~\ref{eq:pres} can only be used to compute pressure differences
\begin{equation}
\label{eq:pdif}
p^\text{lat}({\beta},{m_q})-p^\text{lat}({\beta^0},{m_q^0})=
\frac{Ta^4}{V}
\int_{({\beta^0},{m_q^0})}^{({\beta},{m_q})}
{\frac{\partial\ln{\mathcal Z}}{\partial\beta^\prime}d\beta^\prime}
{\frac{\partial\ln{\mathcal Z}}{\partial m_q^\prime}dm_q^\prime}
\end{equation}
It is important to note that the pressure difference in
eq.~\ref{eq:pdif} is independent of the integration path, which allows
one to choose optimal integration paths dependent on the problem.
The derivatives of the partition function occurring in
eq.~\ref{eq:pdif} are the gauge action and the chiral condensate
\begin{equation}
{\frac{\partial\ln{\mathcal Z}}{\partial\beta}=-\langle S_G\rangle}
\qquad
{\frac{\partial\ln{\mathcal Z}}{\partial m_q}=\langle\bar\psi\psi\rangle_q}
\end{equation}
Like in the previous section, a $T=0$ subtraction has to be performed
on them to remove divergences.

We can now integrate the pressure starting from a reference
point. One straightforward method to do so is to compute the
derivative of the pressure with respect to temperature along the
LCP. The temperature derivative of the pressure is related to the
trace anomaly
\begin{equation}
I=\Theta^{\mu\mu}=\epsilon-3p
\end{equation}
via the relation
\begin{equation}
\label{eq:ta}
\frac{I}{T^4}=T\frac{\partial}{\partial T}\frac{p(T)}{T^4}
\end{equation}
Alternatively, one can construct the pressure as a function of $\beta$
and the quark masses, constraining its form by computing derivatives
with respect to all these parameters \cite{Borsanyi:2010cj}.

\begin{figure}[htb]
\centerline{
\includegraphics[width=6cm]{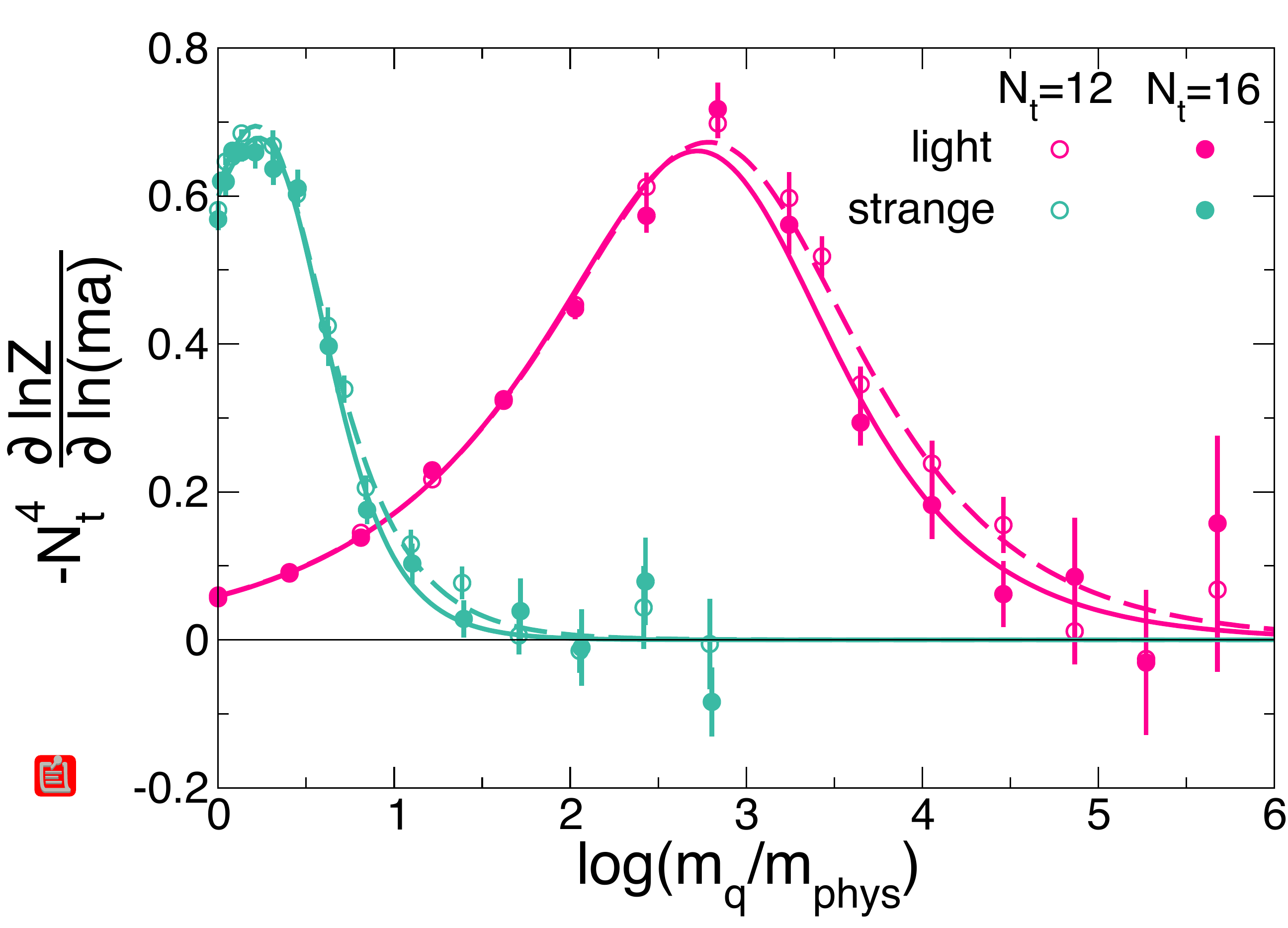}
\quad
\includegraphics[width=6cm]{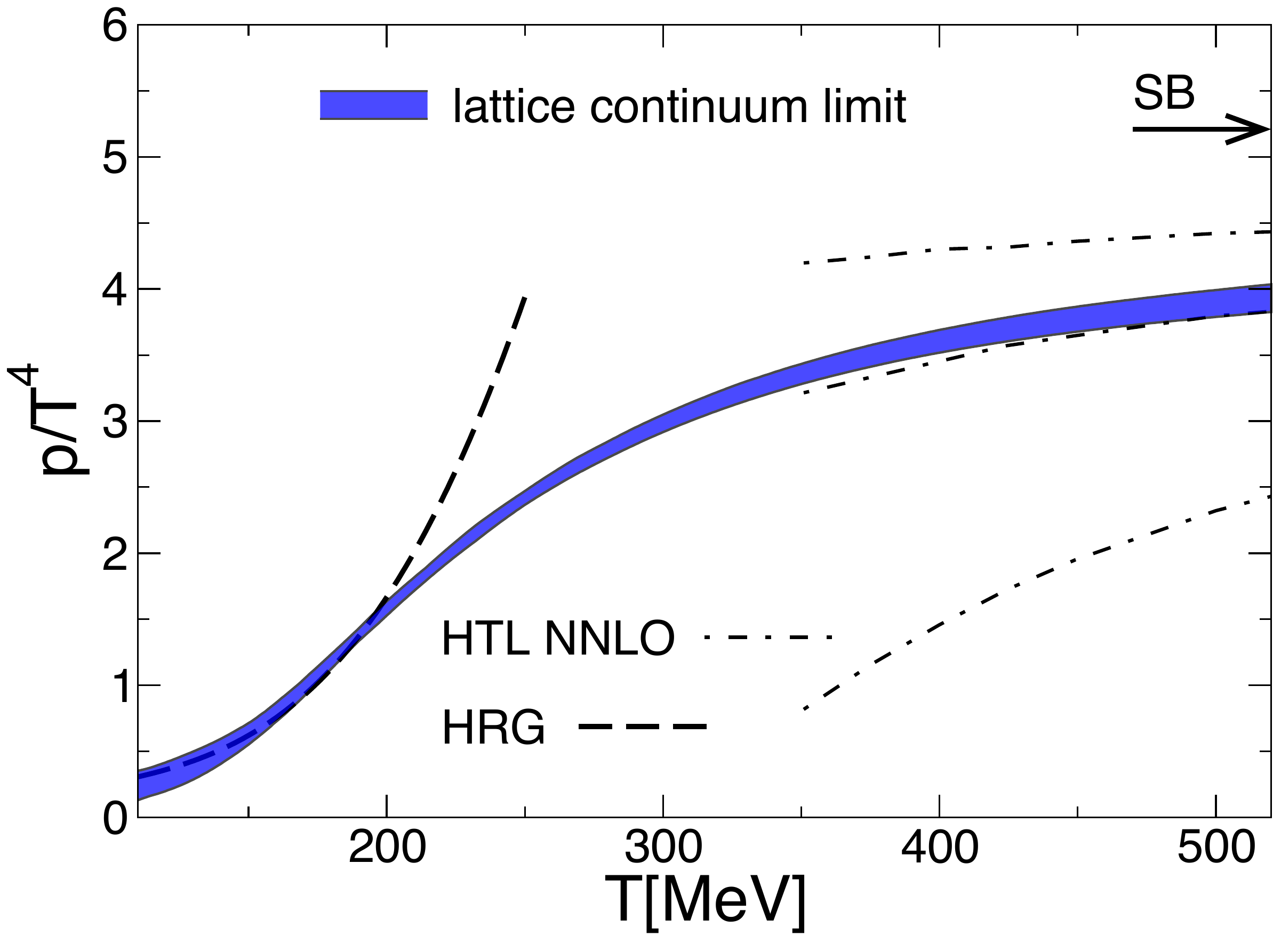}}
\caption{
Pressure integration from a reference point at high bare quark mass
for two $N_t$ (left panel) and the continuum extrapolated pressure as
a function of temperature (right panel) from \cite{Borsanyi:2013bia}.
The $\beta$ of the reference point corresponds to $T\sim 214
\text{MeV}$ at the physical point. In the right panel, HRG denotes the
hadron resonance gas result and HTL refers to the hard thermal loop
result of \cite{Andersen:2011sf}.}
\label{fig:pres}
\end{figure}

There are also different possible choices for a reference
point. Ideally, the pressure should be negligible at the reference
point. One natural choice is e.g. a physical point al low
temperature. The pressure will be low at low temperature and in
addition it can be estimated rather accurately in the hadron resonance
gas model \cite{Karsch:2003vd,Karsch:2003zq,Tawfik:2004sw}. However,
in a fixed $N_t$ approach, the lattice spacing $a=N_t/T$ increases
dramatically at low $T$ and there are potentially large discretization
effects. Alternatively, one may choose a reference point that is not
on the LCP. Increasing the bare quark masses at a fixed $\beta$ leads
to an unphysical theory deeply in the confined phase where the
pressure is almost zero which can serve as an ideal reference point
\cite{Borsanyi:2013bia}. In the left panel of fig.~\ref{fig:pres} the
result of a pressure integration from such a reference point are
displayed for two lattice spacings.

In the right panel of fig.~\ref{fig:pres}, the continuum extrapolated
pressure is plotted vs. temperature and compared to the hadron
resonance gas prediction, the hard thermal loop prediction and the
Stefan-Boltzmann limit. Further thermodynamic quantities can be
obtained from $p(T)$. Apart from the trace anomaly (eq.~\ref{eq:ta}),
we can compute the energy density $\epsilon$, the entropy density $s$
and the speed of sound $c_s$ via
\begin{equation}
\epsilon=I+3p
\qquad
s=\frac{1}{T}(\epsilon+p)
\qquad
c_s^2=\frac{\partial p}{\partial\epsilon}
\end{equation}

\begin{figure}[htb]
\centerline{
\includegraphics[width=10cm]{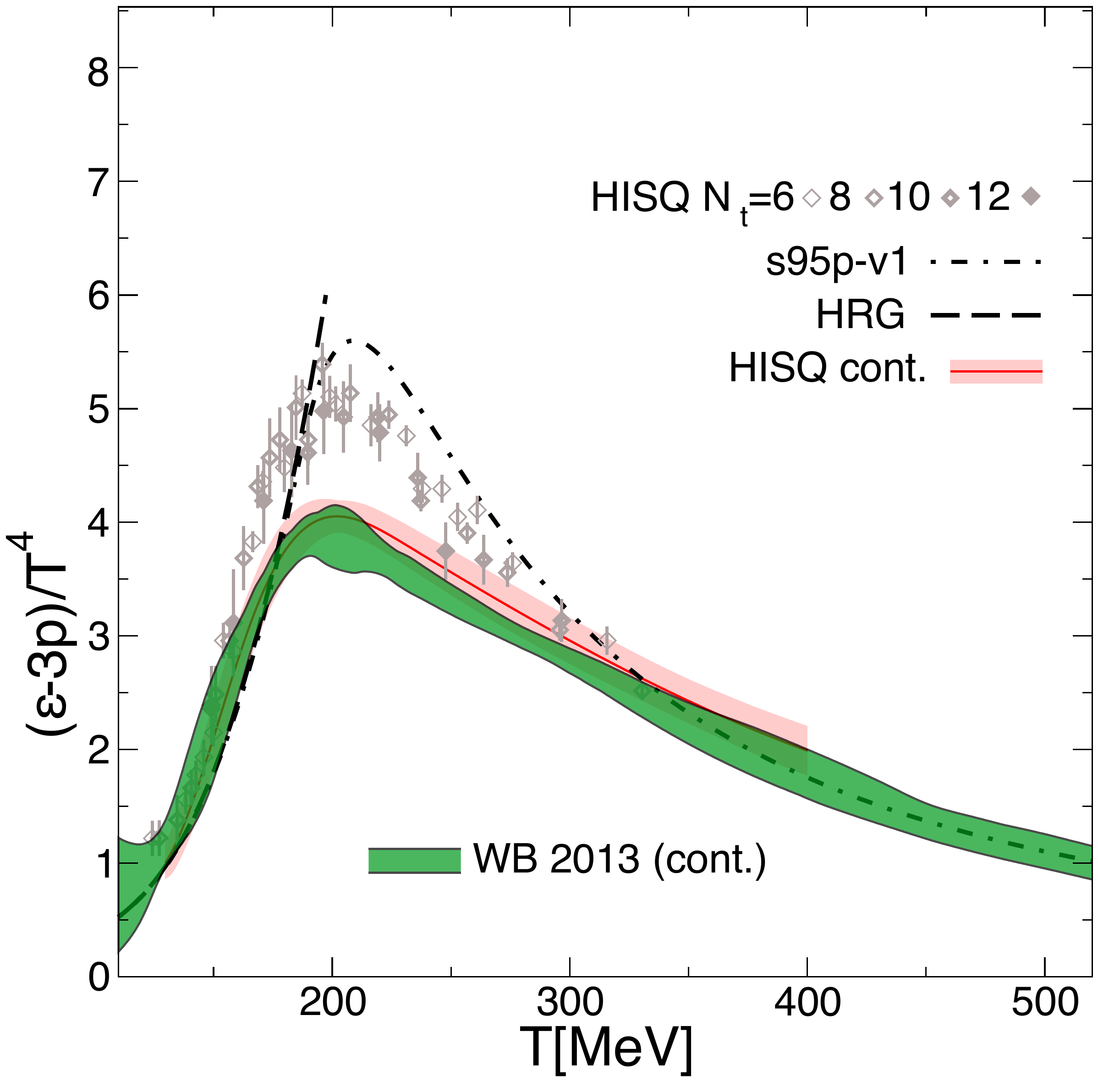}}
\caption{Trace anomaly vs. temperature from the Wuppertal-Budapest
  collaboration \cite{Borsanyi:2013bia} and the hotQCD collaboration
  \cite{Bazavov:2014pvz}. The ``s95p-v1'' parametrisation
  \cite{Huovinen:2009yb} and a previous hotQCD result (HISQ, 
  \cite{Petreczky:2012gi}) is plotted for comparison.}
\label{fig:anomal}
\end{figure}

Similarly to the case of the pseudocritical temperature discussed in
sect.~\ref{sect:tc}, there has been until very recently a marked
discrepancy in the literature regarding the equation of state as
obtained by the two major collaborations computing it. The peak height
of the trace anomaly reported by the hotQCD collaboration
\cite{Bazavov:2009zn,Petreczky:2012gi} was substantially larger than
that reported by the Wuppertal-Budapest collaboration
\cite{Borsanyi:2010cj,Borsanyi:2013bia}. With the latest results of
the hotQCD collaboration \cite{Bazavov:2014pvz} this discrepancy has
disappeared and there is now consensus on the QCD equation of
state at vanishing chemical potential (see fig.~\ref{fig:anomal}).

\section*{Acknowledgements}

I would like to thank Stephan D\"urr, Zoltan Fodor, Sandor Katz,
Laurent Lellouch and Kalman Szabo for discussions and support in
preparing the lectures.

\bibliography{references}{}
\bibliographystyle{aipnum4-1}

\end{document}